\documentclass[11pt,a4paper]{article}
\pdfoutput=1
\usepackage{jheppub}
\usepackage{braket}
\usepackage[english]{babel}
\usepackage[utf8]{inputenc}
\usepackage{amsmath}
\usepackage{array, url}
\usepackage{fancyhdr}
\usepackage{amsfonts}
\usepackage{amsthm}
\usepackage{mathtools}
\usepackage{bm}
\usepackage{titlesec}
\usepackage{graphicx}
\graphicspath{{./figures/}}
\usepackage{color}
\usepackage{tensor}
\usepackage{mathrsfs}
\usepackage{caption}
\usepackage{subcaption}
\usepackage{etoolbox}
\usepackage{tikz}
\usetikzlibrary{positioning}
\usepackage{comment}
\usepackage{tensor}
\usepackage[title]{appendix}

\numberwithin{equation}{section}

\usepackage{amsmath}

\DeclareMathOperator\vol{Vol}

\theoremstyle{stylename}
\usepackage{ae,aecompl}
\usepackage{xcolor}
\renewcommand*\thesection{\arabic{section}}
\definecolor{mathematica1}{rgb}{0.368417, 0.506779, 0.709798}
\definecolor{mathematica2}{rgb}{0.880722, 0.611041, 0.142051}
\definecolor{mathematica3}{rgb}{0.560181, 0.691569, 0.194885}
\definecolor{mathematica4}{rgb}{0.922526, 0.385626, 0.209179}
\definecolor{mathematica6}{rgb}{0.772079, 0.431554, 0.102387}
\definecolor{pink}{rgb}{1, 0.5, 0.5}

\title{Flavored ABJM theory on the sphere  and holographic $F$-functions }

\author[\flat, \natural]{Niko~Jokela,}
\author[\natural,\dagger]{Jani~Kastikainen,}
\author[\dagger,\sharp]{Elias~Kiritsis,}
\author[\dagger]{Francesco~Nitti}

\affiliation[\flat]{Department of Physics, P.O.Box 64, FIN-00014 University of Helsinki, Finland}
\affiliation[\natural]{Helsinki Institute of Physics, P.O.Box 64, FIN-00014 University of Helsinki, Finland}
\affiliation[\dagger]{Université de Paris, CNRS, Astroparticule et Cosmologie, F-75013 Paris, France}
\affiliation[\sharp]{Crete Center for Theoretical Physics, Institute for Theoretical and
Computational Physics, Department of Physics
University of Crete, Heraklion, Greece}

\emailAdd{niko.jokela@helsinki.fi}
\emailAdd{jani.kastikainen@helsinki.fi}
\emailAdd{nitti@apc.in2p3.fr}

\preprint{
\begin{flushright}  
HIP-2021-48/TH\\
CCTP-2021-8
\end{flushright}
}

\abstract{We study strongly coupled ABJM theory on the 3-sphere with massive quenched flavor using the AdS/CFT correspondence.  The holographic dual consists of type IIA supergravity with probe D6-branes.  The flavor mass is a relevant deformation driving an RG flow whose IR endpoint is pure ABJM theory. At non-zero mass,  we find that the theory on the 3-sphere exhibits a quantum phase transition at a critical value of the sphere radius. The transition  corresponds to a topology change in the D6-brane embeddings whose dual interpretation is the meson-melting transition. We perform the holographic computation of the free energy on 3-sphere and we use it to construct various candidate $F$-functions. These were recently proposed  in the context of Einstein-scalar gravity to interpolate monotonically between the values of the sphere free energies of the UV and IR CFTs. We find that while the $F$-functions of the flavored ABJM theory have the correct UV and IR limits, they are not monotonic. We surmise that the non-monotonicity is related to the presence of the phase transition.}

\begin{document}
\maketitle

\section{Introduction}

UV-complete quantum field theories may be constructed as conformal field theories (CFTs) deformed by  a relevant operator,  which drives a renormalization group (RG) flow interpolating between high-energy (UV) and low-energy (IR) regimes. In  these two extreme limits, conformal invariance is recovered, so one can think of the RG-flow as interpolating between a UV and an IR CFT. The parameter along the RG-flow can be taken as the energy scale, and one defines running couplings in the Wilsonian sense as a way to evolve couplings with scale in order to leave observables unchanged. In this sense, running couplings are not observables, nor are $\beta$-functions.

When  the  UV CFT is defined on a curved spacetime,   an interesting interplay can arise between the curvature length scale\footnote{This can be the typical curvature scale of the geometry, or more specifically the curvature radius in case of constant-curvature spacetimes, which will be the subject of this paper.} $\alpha$  and the scale at the origin of the RG flow $m$ (corresponding to a dimensionful coupling): one can form a new dimensionless parameter $\alpha m$ on which physical quantities in general depend non-trivially. One is then led to consider {\em curvature}-RG flows of physical quantities, as a function of $\alpha m$ \cite{ghosh_holographic_2018}. The limit $\alpha m \to 0$ corresponds to  the UV, since for $m\to 0$ and fixed $\alpha$, one recovers the UV CFT. Equivalently, $\alpha  \to 0$ corresponds to large curvature (high energy) limit for fixed $m$.  Similarly, the IR  endpoint corresponds to the limit  $\alpha m \to \infty$.

Notice that both $\alpha$ and $m$ are defined in the UV CFT, therefore they are physical parameters defining the UV theory: a curvature-RG flow describes a relation between physical quantities in different theories.  It is therefore conceptually different from the usual concept of RG flow (of a single theory, with respect to energy scale), and it offers an {\em a priori} different interpolation between the UV and the IR: one expects that  in the  $\alpha m \to 0$ and   $\alpha m \to \infty$ limits one should recover the behavior of (suitably defined) observables in the UV and IR CFTs, respectively.

An important aspect of field theories on curved space-times concerns the conjectured existence of $c$-functions in odd dimensions. These are functions which decrease monotonically from the UV to the  IR, and assume specific values at the fixed points, which are characteristic of the UV and IR CFTs. In even dimensions,  the fixed-point values are combinations of the Weyl anomaly coefficients of the CFT. In odd dimensions, a similar role is taken by the value of the Euclidean CFT renormalized free energy on a sphere: this value (which one may call an $F$-value) is scheme-independent, independent of the sphere radius and, similarly  to central charges in even dimensions, it  ``measures'' the number of degrees of freedom of the CFT \cite{jafferis_towards_2011}.

The question has arisen whether, from the sphere free-energy of a non-conformal theory, one can construct  $F$-{\em functions},  which interpolate monotonically along a curvature RG-flow ({\emph{i.e.}}., as a function of the sphere radius) between the UV and IR $F$-values. Although it is known that a monotonic  $F$-function can be constructed using the entanglement entropy of a spherical region in {\em flat}-space \cite{liu_refinement_2013}, it is still an open problem whether the sphere free energy can play a similar role.

In the context of the gauge/gravity duality, the description of curvature RG-flows is relatively simple. The duality relates a $d$-dimensional CFT to a higher-dimensional  geometry (the {\em bulk}) containing an AdS$_{d+1}$ factor, and  AdS$_{d+1}$ admits slicing by curved $d$-dimensional hypersurfaces. The geometry of the slices corresponds to the geometry on which the CFT is defined \cite{graham_volume_1999}. This generalizes to curved holographic RG-flows of  non-conformal theories, in which the UV CFT is deformed by a relevant operator, and  the dual geometry is close to  AdS$_{d+1}$ only asymptotically,  but still admits a slicing by curved   $d$-dimensional hypersurfaces. The identification  of the geometry where the UV theory is defined is performed at the asymptotic near-AdS conformal boundary.

Recently, curved-slicing holographic RG flow solutions  have been systematically investigated in \cite{ghosh_holographic_2018} for  $d+1$-dimensional Einstein-scalar gravity, in the case of constant-curvature slicing, which corresponds to  the dual field theory being defined on a constant (positive or negative) curvature spacetime. The bulk theory in this case was constructed from the bottom-up, with a scalar potential which was left general, as long as it possessed extrema with negative cosmological constant, which are dual to RG flow fixed points. Under the holographic duality, having a single bulk scalar  corresponds to concentrating on a single relevant scalar operator driving the flow. The setup is simple but flexible enough to capture very interesting features, such as exotic RG flows \cite{Kiritsis:2016kog} and curvature-driven first-order phase transitions \cite{ghosh_holographic_2018}.

The holographic calculation of the free energy $F(\alpha)$ on a sphere of radius $\alpha$  is also relatively simple: in the large-$N$ limit,  it coincides with the classical bulk  action evaluated on the solution, after  divergences due to the infinite volume of AdS are subtracted. For CFTs the result is unambiguous, but for holographic RG flows it is scheme dependent, as one can add finite boundary counterterms to the action with arbitrary coefficients.  Moreover, whereas the UV CFT value is correctly  recovered by $F(0)$,  to obtain the IR $F$-value one cannot simply take $\alpha \to \infty$, as this quantity is typically divergent, simply because in this limit the volume of $d$-dimensional space becomes infinite. This IR divergence has nothing to do with the one associated to the infinite {\em radial} volume of AdS (which corresponds to UV divergences in the field theory) and it persists after holographic renormalization. This IR problem of the holographic free energy on the sphere   was observed, {\emph{e.g.}}, in \cite{taylor_holographic_2016}, in which it was taken as an obstruction to obtaining $F$-functions from the sphere free energy.

The obstruction was recently addressed in \cite{ghosh_holographic_2019} in  Einstein-scalar gravity, for the case of $d=3$. In that work  it was shown that,  remarkably, if one expands  $F(\alpha)$ around infinity, the  IR-finite contribution is  scheme-independent  {\em and} it  coincides with the correct IR  $F$-value one would compute directly in the IR fixed point CFT. This led the authors of \cite{ghosh_holographic_2019} to conjecture that one may construct $F$-functions from the IR-finite part of the free energy. To define $F$-functions,  one can  either choose a special renormalization scheme in which the IR divergences are canceled (and it  was shown that such a scheme always exists) by picking  specific values for the finite boundary counterterms; and/or one can act on $F(\alpha)$ by derivative operators to obtain a function with has a finite limit as both $\alpha \to 0$ and $\alpha \to \infty$.

The procedure outlined above results, for $d=3$,  in four different IR-finite versions of the sphere free energy  which, by construction, all interpolate between the UV  and IR $F$-values. It was conjectured in \cite{ghosh_holographic_2019} that some (or all) of these play the role of monotonic $F$-functions, independent of the Liu-Mezei entanglement entropy \cite{liu_refinement_2013} defined in flat space-time. A noteworthy ingredient
of the proposal, motivated by holography is that when the perturbing operator has $\Delta<{d\over 2}$ then the Legendre transform of the free energy is used instead. Numerical evidence for the validity of this conjecture was presented in \cite{ghosh_holographic_2019}, by analysing  simple but generic Einstein-scalar RG flows  interpolating between  two extrema of the scalar potential, for a wide range of (dual) operator dimensions.
Moreover, the case of a free massive scalar field that made previous attempts fail, now it was shown to work.

An interesting by-product of this analysis was that in \cite{ghosh_holographic_2019} it was shown that apart from the Euclidean free energy, there is also an entropy. This could be understood by rotating the $S^3$-theory to de Sitter, and then the entropy function was the entanglement entropy of the two space hemispheres in global coordinates. It becomes the thermal energy of the static patch if one chooses static patch coordinates.
Interestingly, it was shown that if we use this entropy to construct an $ F $-function, then the ones obtained coincide with those obtained from the free energy, \cite{ghosh_holographic_2019}.

These results raise the question whether monotonicity of the conjectured $F$-function holds in general. So  far the analysis was limited  to  bottom-up models  which are not, strictly speaking, guaranteed to make sense holographically (for a generic scalar potential they may lie in the swampland, after all).  Therefore, one important question is whether  the putative $F$-theorem holds in top-down constructions, in which the holographic setup is known to be robust. Two possibilities are:
\begin{enumerate}
	\item  test these $F$-theorems in known 4-dimensional Einstein-scalar models which uplift to type IIB or type IIA in 10 dimensions;
	\item  more interestingly, consider 10D D-brane constructions  which {\em do not} reduce  to  scalar fields canonically coupled to gravity.
\end{enumerate}
In this work we choose the second route, and consider the string theory dual of sphere RG-flows of 3-dimensional ABJM theory with massive flavor. Another possibility is to consider relevant deformations of ABJM theory on $ S^3 $ by real masses, \cite{freedman_holography_2014,bobev_mass_2019,bobev_higher-derivative_2021}, and study monotonicity of $ F $-functions in this context.\footnote{We thank N. Bobev for bringing the references \cite{freedman_holography_2014,bobev_mass_2019,bobev_higher-derivative_2021} to our attention and for suggesting this possibility.} We will not consider this setup in the current work.

Pure ABJM theory \cite{aharony_n6_2008} is a 3-dimensional $ \mathcal{N} = 6 $ superconformal Chern--Simons matter theory which is conjectured to be holographically dual to M-theory on $ \text{AdS}_{4}\times S^{7}\slash \mathbb{Z}_k$. At large number of colors $N$ and strong coupling, M-theory can also be approximated by type IIA supergravity on AdS$_4 \times  \mathbb{C}\mathbb{P}^{3}$. The holographic recipe for adding $ N_f $ massless flavor fields to ABJM theory was described in \cite{hohenegger_note_2009,hikida_abjm_2009,gaiotto_notes_2012,ammon_adding_2009}, and in the type IIA description, it corresponds to adding $ N_f $ AdS$ _4 $-filling D6-branes in the bulk.  Addition of massless flavor to the theory preserves conformal invariance. This has been explicitly checked in perturbation theory \cite{Bianchi:2009ja,Bianchi:2009rf}.

To abandon conformality, one has to introduce a relevant deformation which we do by giving the flavor a mass $ m $.
For $m\neq 0$, the flat-space theory undergoes an RG flow interpolating between two CFTs: ABJM with $N_f$ massless flavors in the UV, and pure ABJM in the IR, since the flavor is gapped and decouples. In the gravity dual description,  the RG flow  is realized as a non-trivial profile for the D6-brane embedding as a function of the holographic radial coordinate.  In our setup, the brane embeddings are parametrized by a single $ \mathbb{C}\mathbb{P}^{3} $ angle, the slipping mode $ \xi $, which is the bulk field dual to the flavor mass deformation \cite{hikida_abjm_2009}.

Entanglement entropy based $ F $-functions in ABJM theory with massive flavor on flat space have been considered in \cite{bea_unquenched_2013,balasubramanian_information_2019} where also backreaction generated by smeared D6-branes was taken into account (see also \cite{jokela_thermodynamics_2013} for calculations of thermal free energy). Those $ F $-functions were found to monotonically interpolate between $ F $-values in agreement with the proof of the $ F $-theorem by Casini and Huerta \cite{casini_rg_2012}.

Placing Euclidean ABJM theory on a 3-sphere simply means, holographically, that one should work with a spherical foliation of AdS$_4$. However, fully backreacted  (even smeared) D6-branes solutions with spherical foliations are not known.   To simplify the analysis we will consider the approximation where the flavor is quenched, {\emph{i.e.}}, $N_f \ll N$. In this case, the dominant flavor contribution to the free energy is of order $N_f$, and one can neglect the $\mathcal{O}(N_f^2/N^2)$ contribution which arises from flavor loops in diagrams involving the color degrees of freedom. In the bulk, the quenched flavor approximation corresponds to the probe-brane approximation where the backreaction of the branes is neglected: the branes are embedded as a seven-dimensional submanifold of the (unmodified) AdS$_4 \times  \mathbb{C}\mathbb{P}^{3}$ geometry. As a result, the leading flavor contribution to the free energy comes purely from the D6-brane on-shell action.

To leading order in $N_f/N$, the curvature-RG flow of the free energy as a function of $ \alpha m $ is captured completely by the D6-brane on-shell action. Hence our strategy for the computation of the free energy can be summarized as follows:
\begin{enumerate}
	\item Write the background  AdS$_4 \times  \mathbb{C}\mathbb{P}^{3}$ by choosing coordinates  such that AdS$_4$  is sliced by  3-spheres;
	\item Find the probe D6-brane embedding in AdS$_4 \times  \mathbb{C}\mathbb{P}^{3}$  as a function of $m$ and $\alpha$;
	\item Compute the free energy in the probe limit by calculating the D6-brane on-shell action (including appropriate counterterms and renormalization).
\end{enumerate}

The holographic calculation is reduced to solving for the dynamics of the single scalar field $\xi$ which, however, is not a canonically coupled scalar with a potential (except in the UV). Therefore this model, besides being top-down and therefore well motivated theoretically,  provides a new testing   ground,  beyond Einstein-scalar gravity, for the ideas put forward in \cite{ghosh_holographic_2018,ghosh_holographic_2019}.

For massless flavor, we  compute the flavor contribution to the $F$-value of pure ABJM theory to leading order in $ N_f $ and the result agrees with the field theory localization calculation in \cite{santamaria_unquenched_2011} and with the expansion of the backreacted calculation in \cite{conde_gravity_2011}.

One may expect that massive flavors will gently flow from the UV and become irrelevant in the IR, as in flat space. However, things are more interesting: we find that  at a critical value of the dimensionless parameter $ \alpha m $, the theory undergoes a first order  quantum phase transition which is holographically realized as a topology-changing transition of the D6-brane embedding. The field theory interpretation of the transition is that meson bound states are broken by quantum fluctuations due to increasing zero-point energy as the radius of the sphere decreases. Meson breaking transitions have been observed first in $ \mathcal{N} = 4 $ super Yang--Mills theory at finite temperature \cite{mateos_holographic_2006,mateos_thermodynamics_2007} (meson melting) and later in holographic flavored $\mathcal{N}=4$ SYM on curved backgrounds \cite{karch_chiral_2006,karch_critical_2009,karch_precision_2015,karch_supersymmetric_2015},  as in our setup. Curvature-driven transitions were also found in the mass deformed ABJM theory by using supersymmetric localization in \cite{nosaka_large_2016,nosaka_mass_2017,honda_supersymmetry_2019}, in certain bottom-up models in \cite{ghosh_holographic_2018} and in \cite{Buchel:2011cc}.

The presence of the phase transition is both a blessing and a curse. On one hand, the phase transition is crucial to obtain  the correct value of the IR-finite part of the flavor on-shell action, which matches the $F$-value one expects from the IR theory, namely zero (since the flavors decouple). Therefore, we find that the {\em weak} version  of the $F$-theorem is reproduced by the curvature RG-flow, {\emph{i.e.}}, the holographic free energy computes the correct fixed-point $F$-values.

On the other hand, the interpolating $F$-functions constructed with the same strategy as in \cite{ghosh_holographic_2019} this time are {\em not} monotonic: in particular, they have the wrong monotonicity in the small-curvature phase, all the way up to the phase transition. However, three of the $F$-functions  have the correct monotonicity in the {\em high} curvature phase, all the way down to the phase transition. This seems to leave the possibility open that the phase transition is what causes the $F$-theorem to fail in this case. It would be interesting to test these questions further in top-down models dual to holographic RG flows which do not display phase transitions, for example those considered in \cite{Ahn:2000aq,Ahn:2000mf,Ahn:2001by,Ahn:2002qga} which are known to uplift to M-theory \cite{Corrado:2001nv,Bobev:2009ms}.

The structure of the paper is as follows. In the rest of the introduction we will give a brief  summary of the setup and of the results. In section \ref{sec:abjm}, we review the holographic dual of ABJM theory with flavor that in an appropriate limit is given by type IIA supergravity with D6-branes. Then in section \ref{sec:D6flat}, we consider flavor in flat space by studying probe D6-branes in flat slicing of AdS$_4$. In section \ref{sec:spherical}, we generalize the analysis to spherical slicing of AdS$_4$ to describe flavor on a 3-sphere. In section \ref{sec:freeen}, we derive the free energy of flavor degrees of freedom on $S^{3}$ from the holographically renormalized D6-brane on-shell action, which in section \ref{sec:Ffuncs}, is used to compute $F$-functions and study their monotonicity. We conclude with a discussion in section \ref{sec:conc}. Details of many of the computations are relegated in appendices.

\subsection{Summary of results}\label{subsec:sum}

On the field theory side, we consider flavored ABJM field theory living on $ S^{3}_{\alpha} $ of radius $ \alpha $, with action
\begin{equation}
	I_{\text{QFT}} = I_{\text{CFT}} + m \int_{S^{3}_{\alpha}}d^{3}\sigma\sqrt{\zeta}O(\sigma) \ ,
	\label{massivetheory}
\end{equation}
where $ I_{\text{CFT}} $ is the action of ABJM theory with $ N_f $ massless flavor fields, $O$ is the fermion bilinear of dimension $ \Delta = 2 $ and $m $ is the flavor mass. The action \eqref{massivetheory} defines a one-parameter family of theories, controlled by the dimensionless parameter $\alpha m$,  which we call  a {\em curvature RG flow}.

The quantity we are interested in is the renormalized free energy on the sphere as a function of $\alpha m$,
\begin{equation}
	F^{\text{ren}}(\alpha m)  = -\log{Z_{S^{3}_{\alpha}}^{\text{ren}}} \ .
	\label{spherefreesum}
\end{equation}
As $\alpha m$ varies between $0$ and $+\infty$, the curvature RG flow interpolates between two fixed point CFTs, for which \eqref{spherefreesum} is positive and scheme independent.

The first CFT$ _{3} $ at $ \alpha m = 0 $ is ABJM theory with massless flavor, and in the quenched approximation $ N_f\slash N \ll 1 $, its renormalized free energy is given by \cite{santamaria_unquenched_2011,conde_gravity_2011}:
\begin{equation}
	F_{\text{UV}} = \frac{\pi \sqrt{2}}{3}N^{3\slash 2}k^{1\slash 2} + \frac{\pi}{4}NN_f\sqrt{2 \lambda} +\mathcal{O}(N_f\slash N)^{2}\ .
	\label{fABJM}
\end{equation}
The IR CFT$ _{3} $ is  pure ABJM theory without flavor and its renormalized free energy is
\begin{equation}
	F_{\text{IR}} = \frac{\pi \sqrt{2}}{3}N^{3\slash 2}k^{1\slash 2}\ ,
	\label{pureABJM}
\end{equation}
{\emph{i.e.}}, the first term in (\ref{fABJM}). We recover this value as the $\alpha m \to +\infty$ limit of the curvature RG-flow free energy $F^{\text{ren}}(\alpha m)$ after a suitable subtraction of volume divergences.

The $ F $-functions, which interpolate between the fixed-point values (\ref{fABJM}) and (\ref{pureABJM}),   are constructed from the free energy for the massive theory along the curvature RG flow and we compute them holographically. Our computations are in the probe limit $ N_f\slash N \ll 1 $ where the background geometry is the ABJM solution $ \text{AdS}_{4}\times \mathbb{C}\mathbb{P}^{3} $ without any backreaction. Because the boundary field theory lives on $ S^{3}_{\alpha} $, we work in spherical slicing of Euclidean AdS$ _4 $ with the metric
\begin{equation}
	ds^{2} = \ell^{2}(du^{2} + \sinh^{2}{(u-c)}\,ds^{2}_{S^{3}}) \ ,
\end{equation}
where $ u $ is the holographic radial coordinate and $ c $ is a constant related to  the sphere radius $ \alpha $  by $c = -\log 2\alpha/\ell$ . The conformal boundary of $AdS_4$ is reached as $u \to +\infty$.

The D6-brane embeddings we consider extend along the directions of AdS$ _4 $ and are parametrized by the slipping mode $ \xi = \xi(u) $ which gives the location of the brane in the internal space $ \mathbb{C}\mathbb{P}^{3} $. In the probe limit, the embeddings are obtained from the variation of the D6-brane action which is a sum of DBI and WZ actions. After a careful analysis of regularity conditions for the WZ action, we show that the (unrenormalized) action takes the form\footnote{up to  boundary terms which are discussed in detail in the main body of the paper}
\begin{equation}
	I_{\text{D6}} = \frac{\pi}{4}N\sqrt{2 \lambda}\int du\,\sinh^{3}{(u-c)}\,\sin{\xi(u)}\,\left( \sqrt{1 + \xi'(u)^{2}} + \frac{1}{2}\sin{\xi(u)} \right) \ .
\end{equation}
Along the way, we also clarify some subtleties with regularity of the WZ term in this action.

The slipping mode $ \xi(u) $ is dual to the fermion bilinear in the flavor mass term. Close to the AdS$_4$ boundary, it  behaves as:
\begin{equation}\label{xi-asympt}
\xi(u) = {\pi\over 2} - y_- e^{-u}  +\mathcal{O}\left(e^{-2u}\right) \ , \quad u\to +\infty \ ,
\end{equation}
where the parameter $y_-$ is related by the holographic dictionary to the fermion mass,
\begin{equation} \label{mass-y-}
m =  {y_- \over \ell} \ .
\end{equation}

Along the curvature RG flow, we encounter two types of topologically distinct embeddings depending on whether the D6-branes extend all the way to the tip of the Euclidean AdS$ _4 $ cigar or whether they terminate at finite distance from the tip. There is a first-order transition between the two types of embeddings which is dual to a meson breaking quantum phase transition on the field theory side. As a result, the curvature RG flow crosses  a phase transition.

The holographic dictionary relates the bulk on-shell action to the free energy on the sphere \eqref{spherefreesum}. At leading order in $ N_f\slash N \ll 1 $, the contribution of massive flavor to the $ S^{3}_{\alpha} $ free energy comes from the on-shell action of the D6-branes. For massless flavor ($\alpha m = 0$), we find that the renormalized on-shell action of a single D6-brane
\begin{equation} \label{Free-ee}
	F^{\text{ren}}_{\text{D6}}\lvert_{m=0}\, = \frac{\pi}{4}N\sqrt{2 \lambda}
\end{equation}
reproduces the leading term in the expansion \eqref{fABJM}.  This corresponds to the UV end of the curvature RG flow (infinite curvature, or zero mass limit).

For massless flavor,   the result (\ref{Free-ee}) is scheme-independent and there are no possible finite boundary counterterms one can add to the action.  For non-zero mass instead there are two finite covariant counterterms. They produce shifts by $ \alpha m $ and $ (\alpha m)^{3} $ in the free energy, which is therefore scheme-dependent.   In a minimal subtraction scheme, we find that the free energy  behaves in the infinite volume limit ($\alpha m\to \infty)$ as
\begin{equation}\label{Free-e}
F^{\text{ren}}_{\text{D6}} = \frac{\pi}{4}N\sqrt{2 \lambda}\left[  \frac{3}{2}\,(\alpha m) - \frac{18\log{2}-11}{8}\,(\alpha m)^{-1} + \mathcal{O}(\alpha m)^{-3}\right], \quad \alpha m \rightarrow +\infty \ .
\end{equation}
The limit $\alpha m\to \infty$ corresponds to the IR limit of the curvature RG flow. The first term in (\ref{Free-e}) represents a  large-volume divergence. This term can be  eliminated  an appropriate change  of renormalization scheme. In such  a scheme, the D6-brane free energy  is now finite as  $\alpha m \to \infty$, and in fact it vanishes, as there is no ${\mathcal{O}}(1)$ term in (\ref{Free-e}). The IR limit of the total free energy (bulk + D6-branes) of the curvature RG flow  now reproduces the  value (\ref{pureABJM}) of the  sphere free energy of the IR CFT (pure ABJM without flavor).

The candidate $ F $-functions of \cite{ghosh_holographic_2019} are obtained by subtracting the infinite volume divergences appearing in (\ref{Free-e}) by either choosing an appropriate renormalization scheme or by acting with a suitable differential operator in $\alpha m$. In general, this produces four different functions $ \mathcal{F}_{1,2,3,4} $, and we show that they correctly interpolate between $ F_{\text{UV}} $ and $ F_{\text{IR}} $ as a function of $ \alpha m $.  However, they are \textit{not} monotonic. We also check that the Legendre transform of the free energy with respect to the mass, the quantum effective potential, \textit{does not} give a monotonic $ F $-function either.

We will discuss these results and their interpretation in the final section of the paper.

\section{Holographic dual of ABJM theory with quenched flavor}\label{sec:abjm}

ABJM theory is a 3-dimensional Chern--Simons--matter theory with gauge group $ U(N)_{k}\times U(N)_{-k} $ \cite{aharony_n6_2008}. It is an $ \mathcal{N} = 6 $ SCFT with two vector multiplets interacting via Chern--Simons terms that have opposite levels $ k,-k $. There are also four chiral multiplets (the matter part) transforming in bifundamental representations $ (N,\bar{N}) $ and $ (\bar{N},N) $ of $ U(N)\times U(N) $.

In the $ N \rightarrow \infty $ limit with $ k $ fixed, ABJM theory is conjectured to be holographically dual to M-theory on $ \text{AdS}_4\times S^{7}\slash \mathbb{Z}_k $ \cite{aharony_n6_2008}. Taking also $ k\rightarrow \infty $ such that
\begin{equation}
	\lambda \gg 1, \quad \frac{\lambda^{5\slash 2}}{N^{2}}\ll 1 \ ,
	\label{supgra}
\end{equation}
where $ \lambda = N\slash k $ is the 't Hooft coupling, the M-theory description is well-approximated by classical 10-dimensional type IIA supergravity on $ \text{AdS}_{4}\times \mathbb{C}\mathbb{P}^{3} $ \cite{aharony_n6_2008}.

The corresponding $ \text{AdS}_{4}\times \mathbb{C}\mathbb{P}^{3} $ supergravity solution in string frame is given by \cite{aharony_n6_2008,hikida_abjm_2009}
\begin{equation}
ds^{2}_{10} = \ell^{2}(ds^{2}_{\text{AdS}_{4}} + ds^{2}_{\mathbb{C}\mathbb{P}^{3}})
\label{ABJM}
\end{equation}
\begin{equation}
e^{\phi} = \frac{2\ell}{k}\quad F_{4} = \frac{3}{2}\,k\ell^{2}\,\vol{\text{AdS}_4}, \quad F_{2} = kJ \ ,
\end{equation}
where the curvature radius $\ell$ is given by the 't Hooft coupling
\begin{equation}
	\ell^{4} = 2\pi^{2}\lambda \ ,
\end{equation}
and we have set the string scale $ \ell_{s} = 1 $. Here $ ds_{\text{AdS}_{4}}^{2} $ in (\ref{ABJM}) denotes the metric of Euclidean AdS$ _{4} $ with unit radius and the metric of $ \mathbb{CP}^{3} $ is normalized such that its Ricci tensor is $ 8 $ times its metric. In addition, $ \vol{\text{AdS}_4} $ is the volume form of unit AdS$ _{4} $ and $ J $ is the Kähler form of $ \mathbb{C}\mathbb{P}^{3} $. The metric of $ \mathbb{C}\mathbb{P}^{3} $ is explicitly \cite{hikida_abjm_2009,zafrir_embedding_2012,conde_gravity_2011}
\begin{align}
ds^{2}_{\mathbb{C}\mathbb{P}^{3}} = d\xi^{2} +\sin^{2}{\xi}\,&\left( d\psi + \frac{\cos{\theta_{1}}}{2}\,d\phi_1 - \frac{\cos{\theta_{2}}}{2}\,d\phi_2 \right)^{2} + \cos^{2}{\Bigl(\frac{\xi}{2}\Bigr)}\,(d\theta_{1}^{2} + \sin^{2}{\theta_{1}}\,d\phi_{1}^{2})\nonumber\\
&\qquad\qquad\qquad\qquad\qquad\qquad+ \sin^{2}{\Bigl( \frac{\xi}{2}\Bigr) }\,(d\theta_{2}^{2} + \sin^{2}{\theta_{2}}\,d\phi_{2}^{2}) \ ,
\label{CP3}
\end{align}
and the coordinate ranges are
\begin{equation}
0 \leq \xi \leq \pi, \quad 0\leq \psi < 2\pi, \quad 0\leq \theta_{1,2} < \pi, \quad 0 \leq \phi_{1,2} < 2\pi \ .
\end{equation}
Our range for $ \xi $ differs from the literature where it is usually $ 0 \leq \xi \leq \pi\slash 2 $. Because of this, there is no factor of four in front of $ ds^{2}_{\mathbb{C}\mathbb{P}^{3}} $ in \eqref{ABJM}.

In these coordinates, the Kähler form is \cite{hikida_abjm_2009,zafrir_embedding_2012}
\begin{align}
J = -\frac{1}{4}\sin{\xi}\, d\xi \wedge (2d\psi + \cos{\theta_1}\,d\phi_1 - \cos{\theta_2}\,d\phi_2)& - \frac{1}{2}\cos^{2}{\Bigl( \frac{\xi}{2}\Bigr) }\,\sin{\theta_1}\,d\theta_1 \wedge d\phi_1\nonumber\\
&- \frac{1}{2}\sin^{2}{\Bigl( \frac{\xi}{2}\Bigr) }\,\sin{\theta_2}\,d\theta_2 \wedge d\phi_2 \ .
\end{align}
ABJM theory admits an extension with two sets of fermionic flavor fields $ (Q_i,\widetilde{Q}_i) $, with $ i=1,2 $, where $ Q_{1,2} $ transform in the fundamental representations $ (N,1) $, $ (1,N) $ respectively, while $ \widetilde{Q}_{1,2} $ transform in $ (\bar{N},1) $, $ (1,\bar{N}) $, respectively \cite{hohenegger_note_2009,hikida_abjm_2009,gaiotto_notes_2012,ammon_adding_2009}. The addition of flavor preserves $ \mathcal{N} = 3 $ supersymmetry.  Conformal symmetry is also preserved for massless flavor.

If we add a total number of $ N_f $ flavor fields of both kinds, then in the holographic dual, this corresponds to adding $ N_f $ D6-branes wrapping a 3-cycle $ M_{3}\subset \mathbb{C}\mathbb{P}^{3} $ such that near the conformal boundary $ M_3 = \mathbb{R}\mathbb{P}^{3} $ \cite{hohenegger_note_2009,hikida_abjm_2009,gaiotto_notes_2012,ammon_adding_2009}. The total Euclidean bulk action is
\begin{equation}
I_{\text{bulk}} = I_{\text{IIA}} + N_{f}I_{\text{D6}} \ ,
\label{fullaction}
\end{equation}
where $ I_{\text{IIA}} $ is the supergravity action and the action of a single D6-brane in the string frame is
\begin{equation} \label{D6action}
I_{\text{D6}} =  I_{\text{DBI}}+I_{\text{WZ}} =  T_{\text{D6}}\int e^{-\phi}\sqrt{\hat{g}_{7}}-T_{\text{D6}}\int \widehat{C}_{7} \ ,
\end{equation}
where we have turned off all world-volume gauge fields. Here $ \hat{g}_{7} $ is the 7-dimensional induced metric of the brane, $ \widehat{C}_7 $ is the pull-back of the background 7-form potential $ C_7 $ onto the brane and
$$ T_{\text{D6}} = {1\over  (2\pi)^{6}} $$ in our units. The background $ C_7 $ comes from the background 2-form flux via $ dC_7 = F_8 = *F_2 $.

In the regime \eqref{supgra}, the supergravity fields, that we denote collectively by  $ \Phi $, can be treated classically, and if we also take $ N_f\rightarrow \infty $, so can the embedding fields $ X $ of the D6-brane. In this limit, the classical Euclidean action (\ref{fullaction})  evaluated on-shell computes the free energy of the flavored ABJM model, which is the quantity we focus on in this work.

As an additional  approximation, we consider the probe-D6 limit, {\emph{i.e.}}, we take the large-$N$, large-$N_f$ limit such that $ \lambda\,(N_f\slash N) \ll 1 $ is fixed and small (see for instance \cite{nunez_unquenched_2010}). At fixed 't Hooft coupling, this is equivalent to the limit $ N_f\slash k \ll 1 $. Then,  the equations of motion for $ \{\Phi,X\} $ from \eqref{fullaction} can be solved perturbatively around the ABJM background \eqref{ABJM}. Let us expand
\begin{equation}
	\Phi = \Phi_{\text{ABJM}} + \delta \Phi, \quad X = X_{\text{ABJM}} + \delta X \ ,
\end{equation}
where $ X_{\text{ABJM}} $ is the brane embedding in $ \Phi_{\text{ABJM}} $. The backreaction $ \delta \Phi $ is sourced by the brane with a coefficient $ N_f $ so that $ \delta \Phi = \mathcal{O}(N_f \slash k) $ by the linearized supergravity equations. Since $ \Phi_{\text{ABJM}} $ is a solution, the leading correction to the on-shell expansion of $ I_{\text{IIA}} $ is of order $ \delta \Phi^{2} = \mathcal{O}(N_f \slash k)^{2} $. On the other hand, the perturbation $ \delta \Phi $ appears linearly in the equation for the embedding $ X $ implying that $ \delta X = \mathcal{O}(N_f \slash k) $. Since $ I_{\text{D6}} $ has the coefficient $ N_f $, $\delta X$-corrections coming from the brane on-shell action are also of order $ \mathcal{O}(N_f \slash k)^{2} $. This is the probe limit in which the leading $ (N_f\slash k) $-correction to the on-shell action of the ABJM solution comes only from the on-shell action of the D6-brane evaluated for the embedding in the ABJM background \eqref{ABJM}:
\begin{equation}
	I_{\text{bulk}}^{\text{on-shell}} = I_{\text{IIA}}^{\text{on-shell}} + N_{f}I_{\text{D6}}^{\text{on-shell}} + \mathcal{O}\left({N_f \over  k}\right)^{2} \ ,
\end{equation}
where
\begin{equation}
	I_{\text{IIA}}^{\text{on-shell}} \equiv I_{\text{IIA}}\lvert_{\Phi_{\text{ABJM}}}, \quad I_{\text{D6}}^{\text{on-shell}} \equiv I_{\text{D6}}\lvert_{\Phi_{\text{ABJM}},X_{\text{ABJM}}} \ .
	\label{onshellactions}
\end{equation}
In other words, the backreaction $ \delta \Phi $ of the brane on the background fields, and the corresponding response $ \delta X $ of the embedding, contribute only at subleading order $ \mathcal{O}(N_f \slash k)^{2} $. In the dual field theory, this is the quenched approximation.

\section{Holographic dynamics of flavor at zero curvature}\label{sec:D6flat}

We begin by reviewing the calculation of the probe D6-brane action in the flat slicing of Euclidean AdS$ _4$.  In this case,  the dual theory lives on $ \mathbb{R}^{3} $. Such a setup has been studied in  generality before, in \cite{zafrir_embedding_2012,jokela_thermodynamics_2013}, where the bulk geometry is a planar black hole and the dual theory lives on $ \mathbb{R}^{2}\times S^{1}_{\beta} $ with $ \mathbb{R}^{3} $ obtained in the zero temperature limit.

We shall derive a general form for the D6-brane action and show that the zero temperature actions in \cite{zafrir_embedding_2012,jokela_thermodynamics_2013} (see also \cite{jensen_more_2010}) are obtained by a specific gauge choice for the Wess--Zumino term. We also provide regularity conditions for the background 7-form $ C_7 $ that will be important when we generalize the derivation of the action to the spherical slicing of AdS$ _4 $.

\subsection{D6-brane embeddings in flat slicing}

For the massive flavor deformations studied in this paper, the dual D6-brane is embedded in the 8-dimensional submanifold
\begin{equation}
\text{AdS}_4\times M_4 \subset \text{AdS}_4\times \mathbb{C}\mathbb{P}^{3} \ ,
\end{equation}
where $ M_4\subset\mathbb{C}\mathbb{P}^{3} $ is defined by the conditions \cite{hikida_abjm_2009} in \eqref{CP3}
\begin{equation}
\theta_{1} = \theta_{2} \equiv \theta, \quad \phi_1 = -\phi_2 \equiv \phi \ .
\label{subspace}
\end{equation}
The condition $ \xi = \frac{\pi}{2} $ selects the 3-dimensional submanifold $ \mathbb{R}\mathbb{P}^{3} \subset M_4 $ which is wrapped by the massless brane \cite{hikida_abjm_2009}.
The metric on the submanifold \eqref{subspace} is
\begin{equation}
ds^{2}_{\text{AdS}_4\times M_4} = \ell^{2}(ds^{2}_{\text{AdS}_4} + ds^{2}_{M_4}) \ ,
\end{equation}
where the metric of $M_4$ is given by
\begin{equation}
ds^{2}_{M_4} = d\xi^{2} + \sin^{2}{\xi}\,(d\psi + \cos{\theta}\,d\phi)^{2} +d\theta^{2}+ \sin^{2}{\theta}\, d\phi^{2}\ ,
\end{equation}
and the metric of unit Euclidean AdS$ _4 $ in flat slicing (Poincar\'e coordinates) takes the form
\begin{equation}
	ds_{\text{AdS}_{4}}^{2} = \frac{dr^{2}}{r^{2}}+ r^{2}\left( (dx^{0})^{2}+ (dx^{1})^{2} + (dx^{2})^{2}\right) \ .
	\label{ads}
\end{equation}
The ranges are $ r \geq 0 $, $ (x^{0},x^{1},x^{2}) \in \mathbb{R}^{3} $ and the conformal boundary is located at $ r = \infty $. The relation to the Fefferham--Graham coordinate of appendix \ref{app:FG} is given by $ r = 1\slash z $.

The addition of massless flavor preserves conformal symmetry and the corresponding embedding $ \xi = \frac{\pi}{2} $ in $M_4$ preserves the isometries of AdS$_4$. One can turn-on the mass for the flavor (and break conformal symmetry) by considering more general brane embeddings, where the wrapping $ M_3 \subset \mathbb{C}\mathbb{P}^{3} $ depends on the AdS$ _4 $ radial coordinate:
\begin{equation}
\xi = \xi(r), \quad \lim_{r\rightarrow \infty}\xi(r) = \frac{\pi}{2}\ ,
\label{embedding}
\end{equation}
which defines the internal 3-cycle
\begin{equation}
	M_3 \equiv M_4\lvert_{\xi = \xi(r)}
\end{equation}
that the brane wraps. Asymptotically in $ r\rightarrow \infty $, it wraps an $ \mathbb{R}\mathbb{P}^{3} $ in the near-boundary limit.

The massless embedding $ \xi(r) = \frac{\pi}{2} $ reaches all the way to the Poincar\'e horizon $r=0$, preserving the isometries of AdS$ _{4} $. Massive embeddings have the brane ending smoothly at some finite $ r=r_0>0 $ before the Poincar\'e horizon  \cite{karch_adding_2002}. The brane can cap off smoothly, because the $ S^{1} $-fiber of $ \mathbb{R}\mathbb{P}^{3} $ shrinks to zero size at $ \sin{\xi(r_0)} = 0 $ where $ \mathbb{R}\mathbb{P}^{3} $ becomes an $ S^{2} $. Regularity of the brane world-volume metric requires that $ \xi'(r_0) = \infty $. Hence, embeddings dual to massive flavor satisfy\footnote{The $ S^{1} $ shrinks to zero also when $ \xi(r_0) = \pi $, but we shall see that these embeddings are related to embeddings $ \xi(r_0) = 0 $ by a parity symmetry of the equation of motion. Hence we  focus on embeddings \eqref{minkowskiembeddings}.}
\begin{equation}
\xi(r_0) = 0, \quad \xi'(r_0) = \infty, \quad r_0 > 0\ ,
\label{minkowskiembeddings}
\end{equation}
and they are called Minkowski embeddings \cite{mateos_holographic_2006}.  In section \ref{sec:spherical} however, we  refer to them as terminating embeddings. The parameter $ r_{0} $ is free and proportional to the flavor mass, as we shall see more explicitly below.

\subsection{Probe D6-brane action in flat slicing}

The  coordinates on the D6-brane world-volume are $ (r,x^{0},x^{1},x^{2},\psi,\theta,\phi) $ and its induced metric is
\begin{align}
ds^{2}_{\text{D6}} = &\ell^{2}\left[(1+r^{2}\xi'(r)^{2})\frac{dr^{2}}{r^{2}} +r^{2}\left( (dx^{0})^{2}+ (dx^{1})^{2} + (dx^{2})^{2}\right) \right.\nonumber\\
&\qquad\qquad\qquad\qquad\left.+ \sin^{2}{\xi(r)}\,(d\psi + \cos{\theta}\,d\phi)^{2}  + d\theta^{2} + \sin^{2}{\theta}\, d\phi^{2} \right] \ ,
\label{adsm3}
\end{align}
so that
\begin{equation}
\sqrt{\hat{g}_{7}} = \ell^{7} \sin{\theta}\,r^{2}\sin{\xi(r)}\,\sqrt{1 + r^{2}\xi'(r)^{2}} \ .
\end{equation}
The regularized DBI action (the first term in (\ref{D6action})) is
\begin{equation}
I^{\text{reg}}_{\text{DBI}} = \mathcal{N}\vol{\mathbb{R}^{3}} \int_{r_0}^{\Lambda} dr\,\ell^{3}r^{2}\sin{\xi(r)}\,\sqrt{1 + r^{2}\xi'(r)^{2}} \ ,
\label{ourdbi}
\end{equation}
where $ \Lambda $ is a geometrical cut-off near the boundary of AdS$ _4 $ and we performed the integral over $ M_3 $. The prefactor is given by
\begin{equation}
\mathcal{N}\ell^{3} = \frac{k\ell^{6}}{16\pi^{4}} =\frac{1}{8\pi}N\sqrt{2\lambda}\ .
\label{Ncoefficient}
\end{equation}
Next, we derive an expression for the WZ action. From $ F_{8} = *F_2 = *kJ $ we calculate \cite{hikida_abjm_2009,zafrir_embedding_2012}
\begin{equation}
F_8 = dC_{7} = k\ell^{6}\,J\wedge J \wedge \vol{\text{AdS}_4} \ .
\label{flatF8}
\end{equation}
On the subspace $ M_4 $ defined by \eqref{subspace}, we obtain
\begin{equation}
J\lvert_{M_4} = -\frac{1}{2}\sin{\xi}\, d\xi \wedge (d\psi + \cos{\theta}\,d\phi) - \frac{1}{2}\cos{\xi}\,\sin{\theta}\,d\theta \wedge d\phi
\end{equation}
so that
\begin{equation}
J\wedge J\lvert_{M_4} = \frac{1}{2} \sin{\xi}\,\cos{\xi}\,d\xi \wedge 8\vol{\mathbb{R}\mathbb{P}^{3}} \ .
\label{JJfirst}
\end{equation}
We have defined the volume form as
\begin{equation}
\vol{\mathbb{R}\mathbb{P}^{3}} = \frac{1}{8}\sin{\theta}\,d\psi\wedge d\theta\wedge d\phi \ .
\end{equation}
In the flat slicing (\ref{ads}),
\begin{equation}
\vol{\text{AdS}_4} = r^{2}\,dr\wedge \vol{\mathbb{R}^{3}}, \quad \vol{\mathbb{R}^{3}} = dx^{0}\wedge dx^{1}\wedge dx^{2}
\end{equation}
so that we obtain:
\begin{equation}
dC_7\lvert_{M_4} = \frac{k\ell^{6}}{2}\,r^{2}\sin{\xi}\,\cos{\xi}\,d\xi\wedge 8\vol{\mathbb{R}\mathbb{P}^{3}}\wedge dr\wedge \vol{\mathbb{R}^{3}} \ .
\label{flatdC}
\end{equation}
This equation determines $ C_7 $ up to an exact form $dU_6$. We consider 7-forms of the type
\begin{equation}
C_7\lvert_{M_4} = \frac{k\ell^{6}}{2}\left[ A(r,\xi)\,d\xi\wedge 8\vol{\mathbb{R}\mathbb{P}^{3}}\wedge \vol{\mathbb{R}^{3}} + B(r,\xi)\, dr\wedge 8\vol{\mathbb{R}\mathbb{P}^{3}}\wedge \vol{\mathbb{R}^{3}}\right]
\label{genC7flat}
\end{equation}
and we impose regularity conditions at the Poincar\'e horizon $ r = 0 $:
\begin{equation} \label{conditions}
A(0,\xi) = B(r,0) = B(r,\pi) = 0 \ .
\end{equation}
These regularity conditions ensure that $ dU_6 $ integrates to a boundary term on the cut-off surface $ r=\Lambda $ which is the only true boundary of spacetime. The same conditions are proven in Appendix~\ref{app:reg} to ensure regularity in the spherical slicing of AdS$ _4 $.

We still have the freedom of adding an exact form $ dU_6 $ to \eqref{genC7flat} without changing the corresponding $ F_8 $, provided we do not violate the conditions (\ref{conditions}).  To preserve the structure \eqref{genC7flat}, we must have
\begin{equation}
U_6 = \frac{k\ell^{6}}{2}\,U(r,\xi)\,8\vol{\mathbb{R}\mathbb{P}^{3}}\wedge \vol{\mathbb{R}^{3}} \ ,
\end{equation}
where $ U(r,\xi) $ obeys the regularity conditions
\begin{equation}
U(0,\xi) = U(r,0)  = U(r,\pi) = 0 \ .
\label{Uregflat}
\end{equation}
For \eqref{genC7flat}, the pull-back onto the brane is given by
\begin{equation}
\widehat{C}_7 = \frac{k\ell^{6}}{2}\left[ A(r,\xi(r))\,\xi'(r) + B(r,\xi(r))\right] dr\wedge 8\vol{\mathbb{R}\mathbb{P}^{3}}\wedge \vol{\mathbb{R}^{3}}
\end{equation}
and the corresponding WZ action is
\begin{equation}
	I_{\text{WZ}}^{\text{reg}} = \mathcal{N}\vol{\mathbb{R}^{3}} \int_{r_0}^{\Lambda}dr\,\ell^{3}\left[ A(r,\xi(r))\,\xi'(r) + B(r,\xi(r))\right] \ .
	\label{genWZflat}
\end{equation}
Now from equation \eqref{flatdC} it follows that
\begin{equation}
\partial_r A-\partial_\xi B = r^{2}\sin{\xi}\,\cos{\xi}
\label{ABeqflat}
\end{equation}
which we want to solve for $ A $ and $ B $. Performing a gauge transformation (which corresponds to the freedom of adding a $dU_6$)
\begin{equation}
	A\rightarrow A + \partial_\xi U, \quad B \rightarrow B + \partial_r U
\end{equation}
we fix $ U $ such that $ A + \partial_\xi U = 0 $. We obtain the equation
\begin{equation}
\partial_\xi (B + \partial_r U) = -r^{2}\sin{\xi}\,\cos{\xi}
\end{equation}
which can be integrated to give
\begin{equation}
B(r,\xi) = -\frac{1}{2}\,r^{2}\sin^{2}{\xi} - \partial_r U(r,\xi) + f(r) \ ,
\end{equation}
where $ f(r) $ is an arbitrary function of $ r $ such that $ f(0) = 0 $. We can absorb it into $ U $ by redefining
\begin{equation}
U \rightarrow U + \int_0^{r} dr'\,f(r')
\end{equation}
which does not affect $ A + \partial_\xi U = 0 $. Hence the general regular solution is
\begin{equation}
A(r,\xi) = -\partial_\xi U(r,\xi), \quad B(r,\xi) = -\frac{1}{2}\,r^{2}\sin^{2}{\xi} - \partial_r U(r,\xi) \ .
\label{gensolflat}
\end{equation}
Substituting into the WZ action \eqref{genWZflat}, we obtain:
\begin{equation}
I_{\text{WZ}}^{\text{reg}} = -\frac{\mathcal{N}}{2}\vol{\mathbb{R}^{3}} \int_{r_0}^{\Lambda}dr\,\ell^{3}r^{2}\sin^{2}{\xi(r)} - \mathcal{N}\vol{\mathbb{R}^{3}}\,U(r,\xi(r))\lvert_{r=\Lambda} \ ,
\label{WZflat}
\end{equation}
where the contribution $ U\lvert_{r=r_0} $ from the tip of the brane vanishes by regularity \eqref{Uregflat} and the function $ U $ integrates to a boundary term on the cut-off surface.\footnote{This follows from $ \frac{dU}{dr} = (\partial_{\xi}U)\,\xi'(r) + \partial_{r}U $.} Hence by regularity, the WZ action is unique up to a boundary term $ U $. We observe below that after holographic renormalization, the renormalized on-shell action is unique, because there is no scheme dependence.

Combining \eqref{ourdbi} and \eqref{WZflat}, the regularized D6-brane action \eqref{D6action} is then
\begin{equation}
I_{\text{D6}}^{\text{reg}} = \mathcal{N}\vol{\mathbb{R}^{3}}\int_{r_0}^{\Lambda}dr\,\ell^{3}r^{2}\sin{\xi(r)}\left( \sqrt{1 + r^{2}\xi'(r)^{2}} + \frac{1}{2}\sin{\xi(r)} \right)+ \mathcal{N}\vol{\mathbb{R}^{3}}\,U\lvert_{r=\Lambda} \ .
\label{fullactionflat}
\end{equation}
It is expressed in a general gauge $ U $ and includes various gauge fixed actions that have appeared in the literature as special cases \cite{zafrir_embedding_2012,jokela_thermodynamics_2013}. Setting
\begin{equation}
U(r,\xi) = -\frac{1}{6}\,r^{3}\sin^{2}{\xi}%+V(r,\xi).
\end{equation}
and using equations \eqref{genWZflat} and \eqref{gensolflat}, the action becomes
\begin{equation}
I_{\text{D6}}^{\text{reg}} = \mathcal{N}\vol{\mathbb{R}^{3}} \int_{r_0}^{\Lambda} dr\,\ell^{3}r^{3}\sin{\xi(r)}\left( \sqrt{r^{-2} + \xi'(r)^{2}}-\frac{1}{3}\cos{\xi(r)}\,\xi'(r)\right)%+ \mathcal{N}\vol{\mathbb{R}^{3}}\,V\lvert_{r=\Lambda}
\label{zafrirbrane}
\end{equation}
as in \cite{zafrir_embedding_2012} at zero temperature.\footnote{Our $ r $-coordinate is related to the radial coordinate in \cite{zafrir_embedding_2012} as $ r_{\text{here}} = r^2_{\text{there}}\slash 4 $.} Another gauge choice is given by
\begin{equation}
U(r,\xi) = -\frac{1}{2}\,r^{3}\sin^{2}{\xi}% + V(r,\xi).
\end{equation}
which gives the zero-temperature action in \cite{jokela_thermodynamics_2013}:
\begin{equation}
I_{\text{D6}}^{\text{reg}} = \mathcal{N}\vol{\mathbb{R}^{3}} \int_{r_0}^{\Lambda} dr\,\ell^{3}r^{2}\sin{\xi(r)}\left( \sqrt{1 + r^{2}\xi'(r)^{2}}-r\cos{\xi(r)}\,\xi'(r)  - \sin{\xi(r)}\right) \ .%+ \mathcal{N}\vol{\mathbb{R}^{3}}\,V\lvert_{r=\Lambda}.
\label{jokelabrane}
\end{equation}

\subsection{Renormalized on-shell action in flat slicing}

The equation of motion derived from the general action \eqref{fullactionflat} is given by
\begin{equation}
r^{2}\cos{\xi(r)}\,\Bigl(\sqrt{1 + r^{2}\xi'(r)^{2}}+\sin{\xi(r)}\Bigr)-\frac{d}{dr}\biggl(\frac{r^{4}\sin{\xi(r)}\xi'(r)}{\sqrt{1 + r^{2}\xi'(r)^{2}}}\biggr) = 0 \ .
\end{equation}
Imposing regularity conditions \eqref{minkowskiembeddings}, the equation of motion has the solution near the AdS boundary
\begin{equation}
\xi(r) = \arccos{\frac{r_0}{r}} = \frac{\pi}{2}-\frac{r_0}{r} + \mathcal{O}(r^{-3}) \ .
\label{massiveflat}
\end{equation}
where $ r_0 > 0 $ is a constant and the solution dual to massless flavor is the special case $ r_0 = 0 $. A general solution has the asymptotics
\begin{equation}
\xi(r) = \frac{\pi}{2}+\frac{\xi_-}{r} + \frac{\xi_+}{r^{2}} + \mathcal{O}(r^{-3})\ , \quad r\rightarrow \infty\ ,
\end{equation}
so that the solution \eqref{massiveflat} has $ \xi_+ = 0 $. In terms of the Fefferham--Graham coordinate $ z = 1\slash r $, this expansion is the same as for a bulk scalar field dual to an operator of dimension $ \Delta = 2 $:
\begin{equation}
\xi-\frac{\pi}{2} = \xi_{-}z^{3-\Delta} + \xi_{+}z^{\Delta} + \ldots, \quad z\rightarrow 0 \ .
\end{equation}
The dual operator is the fermion bilinear $O$ that appears in the flavor mass term. The holographic dictionary then states that the non-normalizable mode,  $ \xi_- $,  is the flavor mass and the normalizable mode $ \xi_+ $ is related to the vev $ \langle O\rangle $ of the fermion bilinear. We prove this dictionary in detail in Appendix \ref{sec:vevproof} by varying the D6-brane on-shell action with respect to $ \xi_- $. Hence the mass dual to the solution \eqref{massiveflat} is given by $ r_0 $ and the vev vanishes. The vanishing vev is a result of the $ \mathcal{N} = 3 $ residual supersymmetry and one can check that the solution \eqref{massiveflat} is supersymmetric from the kappa symmetry condition \cite{conde_gravity_2011}.

For the exact solution \eqref{massiveflat}, the action \eqref{fullactionflat} is UV divergent. In general, there are divergences coming from both the integral part and from the boundary part $ U\lvert_{r = \Lambda} $ of the action. However, the gauge fixed action \eqref{jokelabrane} is completely finite on-shell and actually vanishes identically as observed in \cite{jokela_thermodynamics_2013}. In a general gauge, one also has a boundary term $ U\lvert_{r = \Lambda} $ that now contains all of the UV divergences. The general divergence structure in flat slicing is of the form
\begin{equation}
	U(r,\xi(r))\lvert_{r = \Lambda}\,= a_1\,\Lambda^{3}+  a_2\,\xi_-^{2}\, \Lambda + 2a_2\,\xi_- \xi_+ + \mathcal{O}(\Lambda^{-1}) \ .
	\label{flatdiv}
\end{equation}
The divergences are removable by the two divergent covariant counterterms (after integrating out the internal space $ M_{3} $ of the brane)
\begin{equation}
	\mathcal{N}\int_{r = \Lambda} \sqrt{\gamma_{3}}, \quad \mathcal{N}\int_{r = \Lambda} \sqrt{\gamma_{3}}\,\sin^{2}{\xi(r)} \ .
	\label{flatcts}
\end{equation}
and $ \gamma_{3} $ is the induced metric on the cut-off surface $ r = \Lambda $.
There is a possible finite boundary term one can add to the action,
\begin{equation} \label{finitect-flat}
 \int_{r=\Lambda} \sqrt{\gamma_3} \sin \xi(r) \cos^3 \xi(r) \simeq  \vol{\mathbb{R}^{3}}\, \xi_-^3 \ .
\end{equation}
However, adding this boundary term  breaks supersymmetry. If we use  a supersymmetric scheme, we have to set its coefficient to zero, and
\begin{equation}
I^{\text{ren}}_{\text{D6,on-shell}} = 0 \ .\label{vanishingfree}
\end{equation}
The dual statement is that the flavor contribution to the renormalized free energy of the supersymmetric  massive  theory on $ \mathbb{R}^{3} $ vanishes in the quenched approximation. In the massless limit, the boundary term (\ref{finitect-flat}) vanishes and the result (\ref{vanishingfree})  is scheme-independent.

Notice that the finite term proportional to $ \xi_-\xi_+ $ in \eqref{flatdiv} is always canceled by the counterterm canceling the $ a_{2} $ divergence. This happens, because $ U\lvert_{r = \Lambda} $ is independent of the derivative $ \xi'(\Lambda) $ and depends on the embedding only through $ \xi(\Lambda) $. This is different compared to a massive scalar field in AdS space where the regularized on-shell action also involves the $ r $-derivative of the scalar field at $ r = \Lambda $. In that case, the finite term is not canceled and the renormalized on-shell action is proportional to $ \xi_-\xi_+ $ \cite{klebanov_adscft_1999}.

\section{Holographic dynamics of flavor at finite curvature}\label{sec:spherical}

In this section, we generalize the previous result for the D6-brane action in flat slicing to spherical slicing of Euclidean AdS$ _{4} $. This corresponds to computing the flavor contribution to the free energy of flavored ABJM on $S^3$, in the quenched-flavor limit. We consider brane embeddings which depend on the new radial coordinate which appears when we write AdS$ _{4} $ in a spherical foliation. The resulting embeddings are topologically different from the flat slicing embeddings studied in the previous section. The spherical foliation we consider is not to be confused with more commonly appearing global coordinates of AdS$ _4 $ that correspond to an $ S^{2}\times S^{1} $-slicing in Euclidean signature.

\subsection{Spherical slicing of Euclidean AdS$ _4 $}

Unit Euclidean AdS$ _{4} $ foliated by spherical $ S^{3} $-slices in Fefferman--Graham gauge is\footnote{See Appendix \ref{app:FG} for a review of the Fefferman--Graham coordinate system.}
\begin{equation}
	ds^{2}_{\text{AdS}_{4}} = \frac{1}{z^{2}}\biggl[ dz^{2} +\left( 1 - \frac{z^{2}}{4\alpha^{2}}\right)^{2} \zeta_{ij}d\sigma^{i}d\sigma^{j}\biggr] \ ,
		\label{zspherslice}
\end{equation}
where $ \sigma^{i} $ denote the three boundary coordinates, $ \zeta_{ij} $ is the background metric of the dual theory (it is scaled by a Weyl transformation when performing a bulk coordinate transformation that preserves the FG gauge) which we take to be a 3-sphere of radius $ \alpha $:\footnote{Analytic continuation to de Sitter space $ dS^{3} $ is obtained by setting $ \tau = -it $ with $ t\in \mathbb{R} $.}
\begin{equation}
	\zeta_{ij}d\sigma^{i}d\sigma^{j} = \alpha^{2}( d\tilde{\theta}^{2} + \cos^{2}{\tilde{\theta}}\,d\tau^{2} + \sin^{2}{\tilde{\theta}}\,d\tilde{\phi}^{2}) \equiv \alpha^{2}ds_{S^{3}}^{2}
\end{equation}
with ranges
\begin{equation}
	0\leq z \leq 2\alpha, \quad 0\leq\tilde{\theta}< \frac{\pi}{2} \ , \quad 0\leq \tau,\tilde{\phi} < 2\pi \ .
\end{equation}
The conformal boundary is located at $ z=0 $ and the tip of the Euclidean AdS$ _4 $ cigar is at $ z=2\alpha $. In our computations, we use a different radial coordinate $ u $ in which the metric takes the form \cite{ghosh_holographic_2018,ghosh_holographic_2019}
\begin{equation}
	ds_{\text{AdS}_{4}}^{2} = du^{2} + \sinh^{2}{(u-c)}\,ds_{S^{3}}^{2}, \qquad c < u < +\infty
	\label{sphericalslice}
\end{equation}
which is obtained from (\ref{zspherslice}) by setting:
\begin{equation}
	z =  e^{-u}, \quad \alpha = \frac{1}{2}\,e^{-c} \ .
	\label{zalpha}
\end{equation}
Now the tip of Euclidean AdS$ _{4} $ is at $ u=c $ and the conformal boundary at $ u=\infty $.

Notice that in our conventions, the AdS-metric \eqref{sphericalslice} is dimensionless: the physical dimensionful metric is the ABJM metric \eqref{ABJM} which contains the factor of $ \ell^{2} $. Hence the coordinates and the radius $ \alpha $ appearing in \eqref{sphericalslice} are dimensionless. The physical radius of the field theory sphere (with dimensions of length) is $ \alpha\ell $. (This convention is different from the introduction section \ref{subsec:sum} where we denoted the physical dimensionful radius by $ \alpha $ to make the units more transparent.)

The coordinates $ (u,\tilde{\theta},\tau,\tilde{\phi}) $ are related to the flat slicing coordinates $ (r,x^{0},x^{1},x^{2}) $ in the infinite volume limit. Setting $ z = 1\slash \rho $ and taking $ \alpha \rho \rightarrow \infty $ (large volume or conformal boundary limit), the metric \eqref{zspherslice} becomes
\begin{equation}
	ds_{\text{AdS}_{4}}^{2} = \frac{d\rho^{2}}{\rho^{2}}+ \rho^{2}\alpha^{2}ds_{S^{3}}^{2} + \mathcal{O}(\alpha \rho)^{3}, \quad \alpha \rho \rightarrow \infty \ ,
	\label{limitofads}
\end{equation}
which is approximately AdS$ _4 $ in flat slicing \eqref{ads} if we identify $ \rho = r $. Thus up to corrections
\begin{equation}
	r = e^{u} + \ldots, \quad \alpha \rightarrow \infty \ .
	\label{rident}
\end{equation}

\subsection{D6-brane embeddings in spherical slicing}\label{sec:twotypes}

\begin{figure}[t]
	\begin{subfigure}[t]{0.5\textwidth}
		\centering
		\begin{tikzpicture}
			\node (img1)  {\includegraphics[scale=0.8]{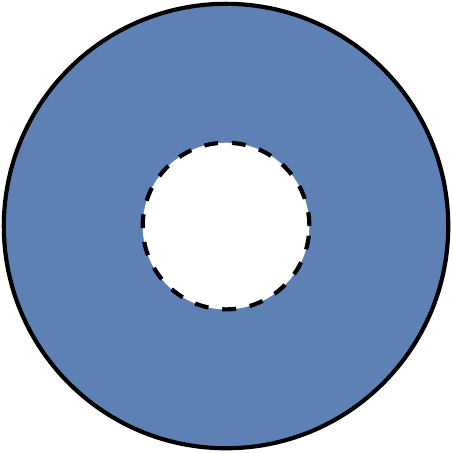}};
		\end{tikzpicture}
		\subcaption{}
		\label{terminatingemb}
	\end{subfigure}
	\begin{subfigure}[t]{0.5\textwidth}
		\centering
		\begin{tikzpicture}
			\node (img1)  {\includegraphics[scale=0.8]{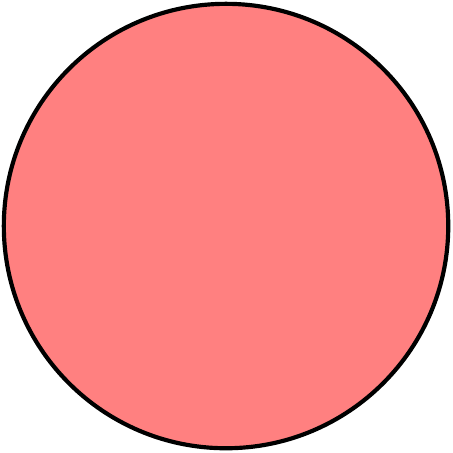}};
		\end{tikzpicture}
		\subcaption{}
		\label{Space-fillingemb}
	\end{subfigure}
	\caption{Visualization of the two types of topologically distinct D6-brane embeddings in spherically sliced AdS$ _{4} $.}
	\label{twoembeddings}
\end{figure}

As in flat slicing, we consider brane embeddings that wrap the 3-cycle $ M_3 \subset M_4 \subset \mathbb{C}\mathbb{P}^{3} $ inside the internal space with the remaining four world-volume directions extending along AdS$ _{4} $. However, now the angle $ \xi $ is taken to be a function of $ u $ instead of $ r $:
\begin{equation}
\xi = \xi(u), \quad \lim_{u\rightarrow \infty}\xi(u) = \frac{\pi}{2} \ .
\end{equation}
The induced metric on the world-volume of the brane is
\begin{align}
	ds^{2}_{\text{D6}} = &\ell^{2}\left[(1+\xi'(u)^{2})du^{2}+\sinh^{2}{(u-c)}\,( d\tilde{\theta}^{2} + \cos^{2}{\tilde{\theta}}\,d\tau^{2} + \sin^{2}{\tilde{\theta}}\,d\tilde{\phi}^{2}) \right.\nonumber\\
	&\qquad\qquad\qquad\qquad\left.+ \sin^{2}{\xi(u)}\,(d\psi + \cos{\theta}\,d\phi)^{2}  + d\theta^{2} + \sin^{2}{\theta}\, d\phi^{2} \right] \ .
	\label{D6indspher}
\end{align}
As in flat slicing, the leading on-shell asymptotic corrections to $ \xi(u) $ as $ u\rightarrow \infty $ are controlled by $ \xi_{\pm} $ that encode the flavor mass and the vev of the fermion bilinear. This will be shown below in section \ref{subsec:D6dynamics}. But in contrast to flat slicing, the interior behavior is modified and there are now two types of topologically distinct embeddings:
\begin{enumerate}
\item Terminating embeddings
\item Space-filling embeddings.
\end{enumerate}
Space-filling embeddings reach all the way to the tip $ u=c $ and fill the whole Euclidean AdS$ _{4} $. Terminating embeddings, on the other hand, cap off at radial distance $ u = u_0 > c $ where an $ S^{1} $-cycle of $ M_3 $ shrinks to zero size ($ \sin{\xi(u_0)} = 0 $). They are analogous to the Minkowski embeddings in flat slicing. The topological difference comes from the space that shrinks to zero size: for space-filling embeddings, it is $ S^{3} \subset \text{AdS}_{4} $, while for terminating embeddings, it is $ S^{1} \subset M_3 $. The embeddings are visualized in figure \ref{twoembeddings}. We shall also see later that terminating embeddings are connected to space-filling embeddings by a singular \textit{critical embedding} for which $ S^{3} \subset \text{AdS}_{4} $ and $ S^{1} \subset M_3 $ shrink at the same time.

Regularity of the brane world-volume geometry imposes conditions on $ \xi(u) $ at the tip $ u=u_{0} $ of the brane. As in flat slicing, regularity of terminating embeddings imposes:
\begin{equation}
\text{terminating:}\quad\xi(u_{0}) = 0, \quad \xi'(u_{0}) = \infty \ .
\label{mink}
\end{equation}
Hence there is again a one-parameter family of terminating embeddings parametrized by $ u_0 > c $.

For space-filling embeddings, the tip of the brane is at $ u=c $ and the brane induced metric \eqref{D6indspher} as a function of $ u = c + \delta u $ with $ \delta u \rightarrow 0 $ is
\begin{equation}
ds^{2}_{\text{D6}} = \left( 1 + \xi'(c)^{2}\right) d(\delta u)^{2} + \delta u^{2}\,ds_{S^{3}}^{2} + \ldots \ .
\end{equation}
Therefore, there is a conical singularity at the tip unless $ \xi'(c) = 0 $. The value  $ \xi(c) $ of the angle at the tip is left as a free parameter leading to a one-parameter family of space-filling embeddings:
\begin{equation}
\text{space-filling:}\quad \xi(c) = \xi_0, \quad \xi'(c) = 0 \ .
\label{Space-filling}
\end{equation}
Space-filling embeddings are analogous to black hole embeddings, that appear when studying theories at finite temperature \cite{mateos_holographic_2006,mateos_thermodynamics_2007,zafrir_embedding_2012,jokela_thermodynamics_2013}. In those cases, the embeddings reach  the horizon of the black hole where the Euclidean time circle $ S^{1}_{\beta} $ shrinks to zero size.

\begin{figure}[t]
	\centering
	\begin{tikzpicture}
	\node (img1)  {\includegraphics[scale=0.9]{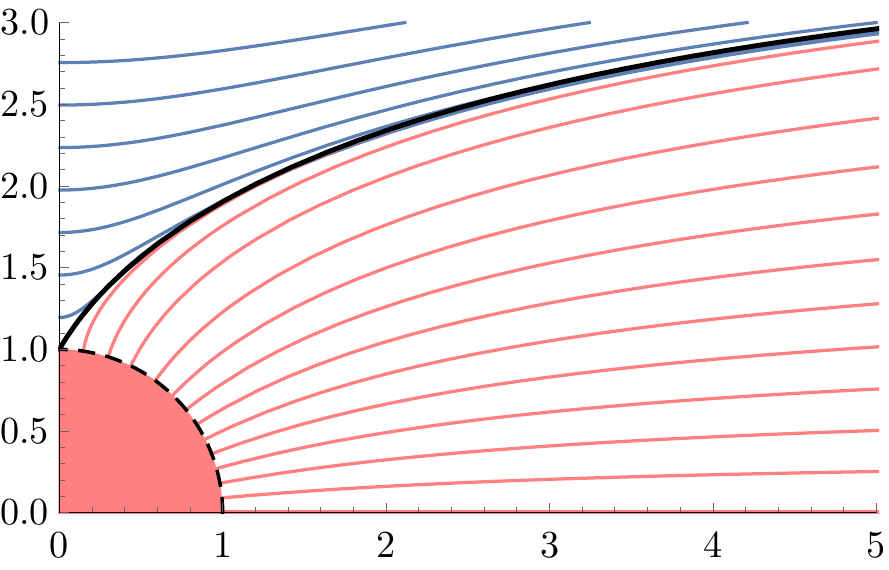}};
	\node[] at (7,-0.4) {Type of embedding};
	\draw[line width=1.5pt, color=mathematica1] (5.5,-1) -- (7,-1) node[right, color=black]  {terminating};
	\draw[line width=1.5pt, color=pink] (5.5,-1.5) -- (7,-1.5) node[right, color=black] {space-filling};
	\draw[line width=1.5pt, color=black] (5.5,-2) -- (7,-2) node[right, color=black] {critical};
	\node[right=of img1, node distance=0cm, yshift=-2.15cm, xshift=-1cm] {$ x $};
	\node[above=of img1, node distance=0cm, anchor=center,yshift=-0.8cm, xshift=-3.65cm] {$ y(x) $};
	\end{tikzpicture}
	\caption{Numerical solutions of the equation of motion \eqref{yeomtext} for the embedding $ y(x) $ in the Cartesian coordinates for $ \alpha = 1\slash 2 $ or $ R_0 = 1 $. Terminating embeddings are in blue (top curves) while space-filling embeddings in pink (bottom curves) and there is a critical embedding in black separating the two. The massless solution $ y(x) = 0 $ is also drawn at the bottom of the plot. The pink shaded region $ x^{2}+y^{2} < R_0^{2} $ is not covered by the coordinates and the circle $ x^{2}+y^{2}=R_0^{2} $ corresponds to the tip $ u = c $ of Euclidean AdS$_{4} $.}
	\label{minkowskicart}
\end{figure}
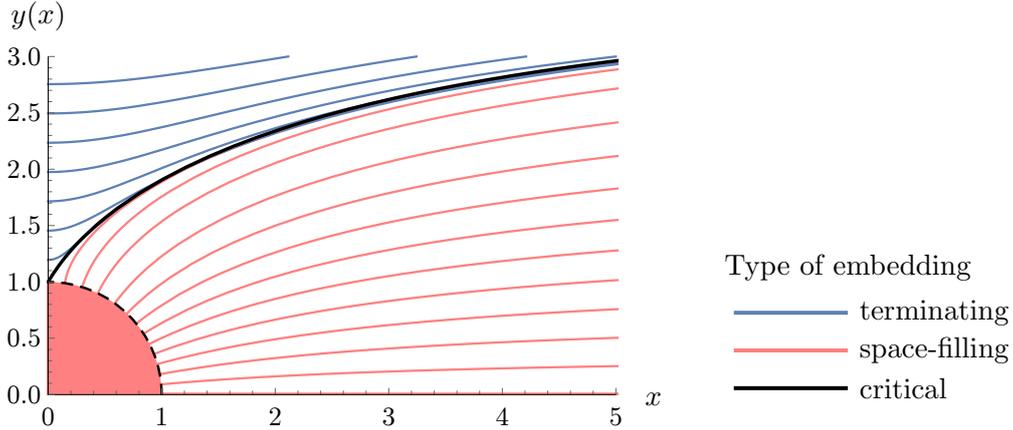

\paragraph{Cartesian parametrization} It will be useful to express the embedding $ \xi(u) $ in a different parametrization. Note  that in the spherical slicing of AdS$ _4 $, the metric of $ \text{AdS}_{4}\times M_4 $ contains\footnote{Here $ du^{2} $ comes from the AdS$ _{4} $ part of the metric while $ d\xi^{2} $ from $ M_4 $.}
\begin{equation}
ds^{2}_{\text{AdS}_{4}\times M_4} \supset \ell^{2}(du^{2} + d\xi^{2}) = \frac{\ell^{2}}{\rho^{2}}(d\rho^{2} + \rho^{2}d\xi^{2}) \ ,
\end{equation}
where $ \rho = e^{u} $ was defined above \eqref{limitofads} and it has the range
$$ \rho\geq e^{c} = 1\slash (2\alpha)\;. $$
 The polar coordinates $ (\rho,\xi) $ together cover the quadrant $ \mathbb{R}^{2}_+ $ with a disk of radius $ R_0 = {1\over  (2\alpha)} $ removed. It is then natural to define Cartesian coordinates\footnote{The $ x $-coordinate here should not to be confused with the flat coordinates $ (x^{0},x^{1},x^{2}) $ of previous sections.}
\begin{equation}
y = e^{u}\cos{\xi}, \quad x = e^{u}\sin{\xi}
\label{cartesian}
\end{equation}
with inverse
\begin{equation}
\xi = \arctan{\frac{x}{y}}, \quad  u = \frac{1}{2} \log{\left( x^{2} + y^{2}\right)} \ .
\end{equation}
In these coordinates, the conformal boundary is located at $ x^{2} + y^{2} \rightarrow \infty $ while the tip $ u=c $ of AdS$ _4 $ is the circle $ x^{2} + y^{2} = R_0^{2} $. Now we can parametrize the D6-brane embedding as
\begin{equation}
y=y(x)
\end{equation}
which is related to the parametrization $ \xi = \xi(u) $ via
\begin{equation}
\xi(u) = \arctan{\frac{x}{y(x)}}, \quad \xi'(u) = \frac{y(x)-x \dot{y}(x)}{x + y(x)\dot{y}(x)} \ ,
\label{oldemb}
\end{equation}
where the dot denotes derivative with respect to $ x $. In this parametrization, the regularity conditions \eqref{mink} for terminating embeddings are
\begin{equation}
\text{terminating:}\quad y(0) = y_0, \quad \dot{y}(0) = 0 \ ,
\label{regularity}
\end{equation}
where the initial condition $ y_0 = e^{u_0} $ takes values in the range $ [R_0,\infty) $. For space-filling embeddings \eqref{Space-filling}, we obtain
\begin{equation}
\text{space-filling:}\quad y(x_0) = y_0, \quad \dot{y}(x_0) = \frac{y_0}{x_0}, \quad x_0 = \sqrt{R_0^{2} - y_0^{2}} \ ,
\label{regularityspace}
\end{equation}
where now $ y_0 = R_0\cos{\xi_0} $ is in the range $ [0,R_0] $. Together, the two types of embeddings cover the range $ y_0 \in [0,\infty) $ of the initial condition (see figure \ref{minkowskicart} for the corresponding embeddings).

\subsection{Probe D6-brane dynamics in spherical slicing}\label{subsec:D6dynamics}

For the embedding $ \xi = \xi(u) $, the DBI action in spherical slicing becomes
\begin{equation}
I_{\text{DBI}}^{\text{reg}} =\mathcal{N}\vol{S^{3}}\int_{u_0}^{\Lambda} du \sinh^{3}{(u-c)}\,\sin{\xi(u)}\,\sqrt{1 + \xi'(u)^{2}} \ ,
\label{DBIspher}
\end{equation}
where we integrated over $ S^{3}, M_3 $ and the constant $ \mathcal{N} $ is the same as in flat slicing \eqref{Ncoefficient}. The derivation of the Wess--Zumino action can be done similarly as in flat slicing. In spherical slicing,
\begin{equation}
\vol{\text{AdS}_4} = \sinh^{3}{(u-c)}\,du\wedge \vol{S^{3}}, \quad \vol{S^{3}} = \sin{\tilde{\theta}}\cos{\tilde{\theta}}\,d\tilde{\theta}\wedge d\tau \wedge d\tilde{\phi}
\end{equation}
so that the 8-form flux \ref{flatF8} on the subspace $ M_4 $ is
\begin{equation}
	dC_7\lvert_{M_4} = \frac{k\ell^{6}}{2}\sinh^{3}{(u-c)}\,\sin{\xi}\,\cos{\xi}\,
	d\xi\wedge 8\vol{\mathbb{R}\mathbb{P}^{3}}\wedge du\wedge \vol{S^{3}} \ .
	\label{spherdC}
\end{equation}
Now the 7-form potential adapted to spherical slicing is of the type
\begin{equation}
C_7\lvert_{M_4} = \frac{k\ell^{6}}{2}\left[ A(u,\xi)\,d\xi\wedge 8\vol{\mathbb{R}\mathbb{P}^{3}}\wedge \vol{S^{3}} + B(u,\xi)\, du\wedge 8\vol{\mathbb{R}\mathbb{P}^{3}}\wedge \vol{S^{3}}\right]
\label{genC7}
\end{equation}
and we impose regularity conditions
\begin{equation}
A(c,\xi) = B(u,0) = B(u,\pi) = 0
\label{regAB}
\end{equation}
that are proved to ensure regularity in Appendix~\ref{app:reg}. The Wess--Zumino action becomes
\begin{equation}
I_{\text{WZ}}^{\text{reg}} = \mathcal{N}\vol{S^{3}} \int_{u_0}^{\Lambda}du\,\ell^{3}\left[ A(u,\xi(u))\,\xi'(u) + B(u,\xi(u))\right]
\label{spherWZ}
\end{equation}
which has essentially the same form as in flat slicing. By \eqref{spherdC} the components satisfy
\begin{equation}
\partial_u A-\partial_\xi B = \sinh^{3}{(u-c)}\,\sin{\xi}\,\cos{\xi}
\label{ABeq}
\end{equation}
and the general solution is
\begin{equation}
A(u,\xi) = -\partial_\xi U(u,\xi), \quad B(u,\xi) = -\frac{1}{2}\sinh^{3}{(u-c)}\,\sin^{2}{\xi} - \partial_u U(u,\xi) \ ,
\label{gensol}
\end{equation}
where $ U(u,\xi) $ must also obey the regularity conditions
\begin{equation}
U(c,\xi) = U(u,0)  = U(u,\pi) = 0 \ .
\label{ureg}
\end{equation}
Substituting to the WZ action \eqref{spherWZ}, we obtain
\begin{equation}
I_{\text{WZ}}^{\text{reg}} = -\frac{\mathcal{N}}{2}\vol{S^{3}} \int_{u_0}^{\Lambda}du\,\ell^{3}\sinh^{3}{(u-c)}\,\sin^{2}{\xi(u)} - \mathcal{N}\vol{S^{3}}\,U\lvert_{u=\Lambda} \ .
\label{WZspher}
\end{equation}
Finally the D6-brane action \eqref{D6action} becomes
\begin{equation}
I_{\text{D6}}^{\text{reg}} = \mathcal{N}\vol{S^{3}}\int_{u_0}^{\Lambda}du\,\ell^{3}\sinh^{3}{(u-c)}\,\sin{\xi(u)}\,\left( \sqrt{1 + \xi'(u)^{2}} + \frac{1}{2}\sin{\xi(u)} \right)+ \mathcal{N}\vol{S^{3}}\,U\lvert_{u=\Lambda} \ .
\label{fullactionspher}
\end{equation}
In the infinite volume limit $ \alpha \rightarrow \infty $, the coefficient of the leading $ \alpha^{3} $-term in this  action gives the D6-brane action in flat slicing \eqref{fullactionflat} based on the relation between the radial coordinates \eqref{rident}.

The variational principle for the action \eqref{fullactionspher} is well-defined due to regularity conditions imposed on the brane world-volume metric and on the 7-form potential $ C_7 $, as proven in Appendix~\ref{app:actvar}. If regularity was not imposed, the variation would pick up boundary terms from the interior that are non-vanishing.\footnote{See \cite{jensen_more_2010} for a discussion on interior contributions.} The resulting equation of motion is explicitly
\begin{equation}
\sinh^{3}{(u-c)}\,\cos{\xi(u)}\,\left( \sqrt{1 + \xi'(u)^{2}} + \sin{\xi(u)}\right)  - \frac{d}{du}\left(\frac{\sinh^{3}{(u-c)}\,\sin{\xi(u)}\,\xi'(u)}{\sqrt{1 + \xi'(u)^{2}}} \right) = 0
\label{xieom}
\end{equation}
which in Cartesian coordinates for the embedding $ y = y(x) $ is
\begin{equation}
\frac{y(x^{2} + y^{2} - R_0^{2})^{3}}{(x^{2} + y^{2})^{5\slash 2}}\left(x + \frac{(x^{2}+y^{2})\sqrt{1+\dot{y}^{2}}}{x+y\dot{y}} \right) -\frac{x^{2}+y^{2}}{x+y\dot{y}}\frac{d}{dx}\left(\frac{x( x^{2} + y^{2} - R_0^{2})^{3}(y-x \dot{y})}{(x^{2} + y^{2})^{5\slash 2}\sqrt{1+\dot{y}^{2}}} \right)  = 0 \ ,
\label{yeomtext}
\end{equation}
where $ R_0 = 1\slash (2\alpha) $. We can immediately see that the equation of motion has the massless solution
\begin{equation}
\xi(u) = \frac{\pi}{2} \quad \Leftrightarrow \quad y(x) = 0
\label{massless}
\end{equation}
and that it is invariant under the replacement
\begin{equation}
\xi(u) \rightarrow \pi - \xi(u) \quad \Leftrightarrow \quad  y(x) \rightarrow -y(x)
\label{symmetry}
\end{equation}
which maps the massless solution onto itself. Focusing on solutions that do not break this symmetry, we restrict ourselves to the sector $ \xi(u) \in [0,\frac{\pi}{2}] $ or equivalently to  $ y(x) \geq 0 $.

Asymptotically the branes wrap an $ \mathbb{R}\mathbb{P}^{3} $ which means $ \xi(u) - \frac{\pi}{2}\rightarrow 0 $ as $ u\rightarrow \infty $. This is equivalent to
\begin{equation}
y(x) \rightarrow y_-, \quad x\rightarrow \infty \ ,
\end{equation}
where $ y_- $ is a positive constant. The subleading behavior is obtained by studying the equation of motion. Expanding \eqref{yeomtext} in $ x \rightarrow \infty $ and small $ \dot{y}(x),\ddot{y}(x) $, we obtain at leading order
\begin{equation}
x\ddot{y}(x)+2\dot{y}(x) = 0
\end{equation}
which determines the asymptotic behavior to be
\begin{equation}
y(x) = y_- - \frac{y_+}{x} + \mathcal{O}(x)^{-2} \ ,
\end{equation}
where $ y_{+} $ is another constant. This translates to
\begin{equation}
\xi(u) = \frac{\pi}{2} - \frac{y_-}{x} + \frac{y_+}{x^{2}} + \mathcal{O}(x)^{-3}, \quad x \rightarrow \infty \ ,
\end{equation}
and since asymptotically $ x = e^{u} + \ldots $, this is equivalent to
\begin{equation}
\xi(u) = \frac{\pi}{2}+\xi_-e^{-u} +  \xi_+ e^{-2u} + \mathcal{O}(e^{-3u}), \quad u \rightarrow \infty
\label{spherxiasymp}
\end{equation}
with the identification
\begin{equation}
y_- = -\xi_-, \quad y_+ = \xi_+ \ .
\label{ypm}
\end{equation}
The holographic dictionary identifies $ y_- $ as the flavor mass and $ y_+ $ as the vev of the dual fermion bilinear on $ S^{3}_{\alpha} $. This identification is proven in Appendix~\ref{sec:vevproof}. Since there is a one-parameter family of (regular) embeddings, $ y_{+} $ is not an independent parameter, but a function of the flavor mass $ y_{-} $.

Now let $ y(x) $ be a solution of the equation of motion \eqref{yeomtext} for sphere radius $ \alpha $ and flavor mass $ y_{-} $. Then $ \tilde{y}(x) = \lambda^{-1} y(\lambda x) $ is a solution of the equation with radius $ \tilde{\alpha} = \lambda \alpha $. From the asymptotics of the new embedding, we obtain
\begin{equation}
	\tilde{y}_{-} = \lambda^{-1}y_{-}, \quad \tilde{y}_{+} = \lambda^{-2}y_{+}
\end{equation}
so that the two embeddings correspond to the same value of the dimensionless parameter $ \tilde{\alpha}\tilde{y}_{-} = \alpha y_{-} $. One can also check that the embedding $ y $ with radius $ \alpha $ has the same renormalized on-shell action $ I_{\text{D6,on-shell}}^{\text{ren}} $ \eqref{D6onshellren} as the embedding $ \tilde{y} $ with radius $ \tilde{\alpha} $, which implies that
\begin{equation}
	I_{\text{D6,on-shell}}^{\text{ren}}(\alpha,y_-) = I_{\text{D6,on-shell}}^{\text{ren}}(\alpha y_-) \ .
\end{equation}

\subsection{Embedding solutions and the quantum phase transition}\label{sec:massive}

\begin{figure}[t]
	\begin{subfigure}[t]{0.5\textwidth}
		\centering
		\begin{tikzpicture}
		\node (img1)  {\includegraphics[scale=1]{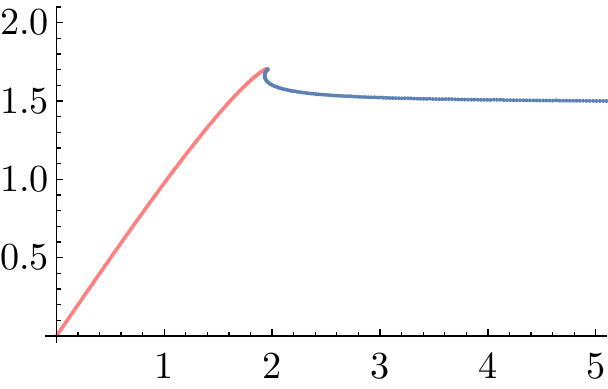}};
		\node[right=of img1, node distance=0cm, yshift=-1.5cm, xshift=-1cm] {$  \alpha y_- $};
		\node[above=of img1, node distance=0cm, anchor=center,yshift=-0.8cm, xshift=-2.6cm] {$ \alpha^{2}y_+ $};
		\draw[line width=1.5pt, color=mathematica1] (-0.5,-0.5) -- (1,-0.5) node[right, color=black] {terminating};
		\draw[line width=1.5pt, color=pink] (-0.5,-1) -- (1,-1) node[right, color=black] {space-filling};
		\end{tikzpicture}
		\subcaption{}
	\end{subfigure}
	\begin{subfigure}[t]{0.5\textwidth}
		\centering
		\begin{tikzpicture}
		\node (img1)  {\includegraphics[scale=1]{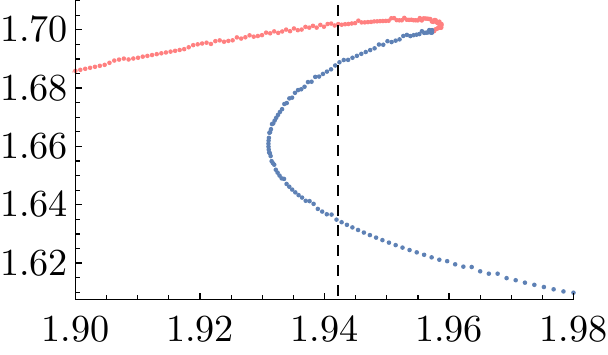}};
		\node[right=of img1, node distance=0cm, yshift=-1.3cm, xshift=-1.2cm] {$ \alpha y_- $};
		\node[above=of img1, node distance=0cm, anchor=center,yshift=-0.8cm, xshift=-2.4cm] {$ \alpha^{2}y_+ $};
		\end{tikzpicture}
		\subcaption{}
	\end{subfigure}
	\caption{The normalizable mode $ y_+ $, which is proportional to the vev, as a function of the non-normalizable mode $ y_- $, which is proportional to the source. The black vertical line shows the location of the quantum phase transition as determined from minimization of the free energy.}
	\label{vevsource}
\end{figure}
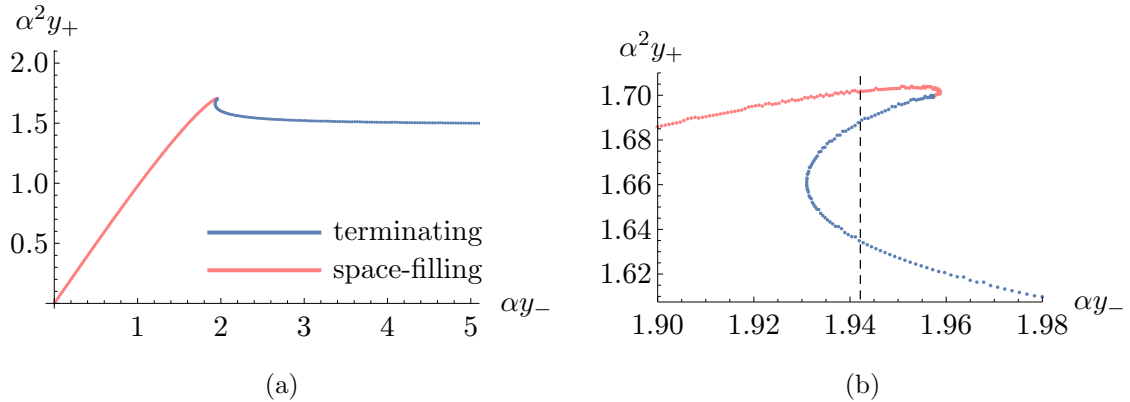

The equation of motion \eqref{xieom} for the D6-brane embedding $ \xi(u) $ is highly non-linear and has to be solved numerically in general. We have done this in Cartesian coordinates for the embedding $ y(x) $ with the equation of motion \eqref{yeomtext}. The solutions are obtained by imposing regularity conditions corresponding to either a space-filling or a terminating embedding and integrating the equation to the conformal boundary. From the asymptotics of the resulting solutions, we can then determine the source $ y_- $ and the vev $ y_{+} $ as a function of each other. The numerical solutions $ y(x) $ are plotted in figure \ref{minkowskicart} and the corresponding $ y_{\pm} $ are plotted in figure \ref{vevsource}.

We find that terminating embeddings exist only in a semi-infinite range of sources $ y_- \in [y_{\text{t}},\infty) $ while space-filling embeddings exist in the range $ y_- \in [0,y_{\text{s}}] $ where $ y_{\text{s}}> y_{\text{t}} $. In the overlap region, $ y_- \in [y_{\text{t}},y_{\text{s}}] $, the terminating and space-filling embeddings are separated by a critical embedding that exhibits self-similarity. These types of critical solutions were originally discovered in brane--black hole systems in \cite{christensen_soap_1998,frolov_domain_1999,frolov_merger_2006}. These systems were later generalized to a holographic D3\slash D7 system in \cite{mateos_holographic_2006,mateos_thermodynamics_2007} and to D6-branes in the ABJM background in \cite{jokela_thermodynamics_2013,Bea:2016fcj}.

In our case, there is no black hole in the bulk, but the collapsing $ S^{3} $-cycle in spherical slicing of AdS$_4 $ plays an analogous role (see \cite{hirayama_holographic_2006,vaganov_holographic_2016} for studies in spherical slicing). The fact that neither space-filling or terminating embeddings exist for all masses and the existence of the critical embedding suggests the presence of a quantum phase transition driven by curvature (since the system is at zero temperature). The nature of the transition is encoded in certain scaling exponents (not to be confused with critical exponents) that control the behavior of embeddings near the critical one \cite{karch_critical_2009}: for complex-valued exponents one expects a first-order transition in the dual field theory, while for real-valued exponents, the transition is expected to be continuous.\footnote{The argument is roughly as follows. The imaginary part of the exponent produces spiraling behavior in the vev--source plot near the phase transition. As a result, a first-order phase transition becomes possible since the solution can jump from one spiral arm to another leading to discontinuity in the derivative of the free energy (i.e. the vev).} In our setup, the corresponding scaling exponents of the critical embedding are computed in Appendix~\ref{par:crit} and they are given by
\begin{equation}
\beta_{\pm} = -\frac{3}{2} \pm i\frac{\sqrt{7}}{2} \ .
\label{exponents}
\end{equation}
Hence we expect a first-order phase transition which will indeed be confirmed by the computation of the free energy in section \ref{subsec:freen} (see figure \ref{freesource}). The exponents \eqref{exponents} are the same as in \cite{karch_critical_2009} for a D5-brane on $ \text{AdS}_{5}\times S^{5} $ in $ S^{3}\times S^{1} $-slicing of $ \text{AdS}_{5} $ (global coordinates) which is a different system compared to our case. We also observe that the vev--source plot in figure \ref{vevsource} has the same qualitative shape as in \cite{karch_critical_2009,erdmenger_mesons_2011} which is most likely due to universality of the transition.

%\noindent\JK{Modified the last sentence above in the way Elias said it: we observe similar vev--source plots $ \Rightarrow $ this is probably due to universality.}

The exponents \eqref{exponents} also have the same imaginary part as the exponents $ \beta_{\pm} = -\frac{1}{2} \pm i\frac{\sqrt{7}}{2} $ computed in \cite{jokela_thermodynamics_2013} for D6-branes in a backreacted planar black hole background which is dual to flavored ABJM theory at finite temperature. The different real part is explained by dimensions of the shrinking submanifolds of the D6-brane: in our case, either an $ S^{3}\subset \text{AdS}_{4} $ or an $ S^{1}\subset M_3 $ shrink, while in \cite{jokela_thermodynamics_2013} it is either an $ S^{1} \subset \text{BH}_{4} $ or $ S^{1}\subset M_3 $ that shrink.\footnote{The general formula for the critical exponents involving shrinking manifolds of dimensions $ n $ and $ m $ is \cite{karch_critical_2009} $\beta_{\pm} = \bigl[(1-a)\pm \sqrt{a^{2}-6a+1}\,\bigr]/2$, where $ a = n+m $.}

In addition to numerical solutions, we can solve the equation of motion analytically in the infinite volume  $ \alpha y_{-} \rightarrow \infty $ and the massless $ \alpha y_{-} \rightarrow 0 $ limits. As shown in \cite{ghosh_holographic_2018,ghosh_holographic_2019} the $ \alpha y_{-} \rightarrow \infty $ limit is an IR limit while the $ \alpha y_{-} \rightarrow 0 $ limit is a UV limit and both are approximated by the appropriate UV and IR CFTs. The detailed calculations are given in Appendix~\ref{app:asympts} and here we simply cite the results. In the infinite volume limit $ \alpha y_- \rightarrow \infty $, we find the expansion
\begin{equation}
\alpha^{2}y_+ =  \frac{3}{2}+\frac{18\log{2}-11}{8}\,(\alpha y_-)^{-2}+\mathcal{O}(\alpha y_-)^{-4}, \quad \alpha y_- \rightarrow \infty
\label{vevIRtext}
\end{equation}
which provides a good fit to the numerically computed result in figure~\ref{vevsource} in the terminating phase. In the opposite massless limit $ \alpha y_- \rightarrow 0 $, we get
\begin{equation}
\alpha^{2}y_+ =  \alpha y_- + \ldots, \quad \alpha y_- \rightarrow 0 \ ,
\label{sourcevevUVtext}
\end{equation}
where the ellipsis denote higher-order corrections in $ \alpha y_- $. This reproduces the linear slope observed numerically in figure~\ref{vevsource}. These analytical expansions will play an important role in upcoming sections.

\section{Free energy of flavor at finite curvature}\label{sec:freeen}

In this section, we compute renormalized free energy of the flavor fields on the 3-sphere in the quenched approximation. The free energy is holographically computed by the renormalized on-shell action of the D6-brane in spherical slicing and it is a function of the dimensionless parameter $ \alpha y_{-} $. In contrast to flat slicing, the renormalized free energy is not unique, but fixed up to two parameters that parametrize the freedom in the choice of a renormalization scheme. In general, the renormalized free energy has infinite volume divergences when $ \alpha y_{-} \rightarrow \infty $, but they can be canceled by a shift of renormalization scheme as will be done in section \ref{sec:Ffuncs}. In the opposite $ \alpha y_{-} \rightarrow 0 $ massless limit, the free energy matches with that of massless flavor computed previously in the literature by other methods \cite{conde_gravity_2011,herzog_multi-matrix_2011}.

\subsection{Holographic renormalization in spherical slicing}\label{subsec:holren}

The on-shell regularized action \eqref{fullactionspher} is divergent when the near conformal boundary cut-off is taken to infinity $ \Lambda \rightarrow \infty $. There are divergences coming from both the $ u $-integral and the boundary term $ U\lvert_{u=\Lambda} $. The divergences of the regularized on-shell action are computed in Appendix \ref{app:holren} and they are given by
\begin{equation}
	I^{\text{reg}}_{\text{D6,on-shell}} = (\alpha \ell)^{3}\vol{S^{3}}\,\biggl[a_1e^{3\Lambda} +  \bigg( \frac{a_2}{\alpha^{2}}+ a_3\,\xi_-^{2}\bigg) e^{\Lambda} + 2a_3\,\xi_- \xi_+ + \mathcal{O}(e^{-\Lambda})\biggr], \quad \Lambda \rightarrow \infty
	\label{divtext}
\end{equation}
with three free coefficients $ a_{1,2,3} $ corresponding to the freedom in $ U $ (there also is more freedom in finite terms taken care of below) and the factor of $ \mathcal{N} $ has been absorbed to the coefficients. The divergences in \eqref{divtext} correspond to the three possible (divergent) counterterms, given by (after integrating out the internal space $ M_{3} $ of the brane):
\begin{equation}
	\mathcal{N}\int_{u=\Lambda} \sqrt{\gamma_3}, \quad \mathcal{N}\int_{u=\Lambda} \sqrt{\gamma_3}\,\sin^{2}{\xi(u)}, \quad \mathcal{N}\int_{u=\Lambda} \sqrt{\gamma_3}\,\ell^{2}R_{3} \ ,
	\label{invariantstext}
\end{equation}
where $ \gamma_{3} $ is the metric of the $ u=\Lambda $ slice of AdS$ _4 $ of radius $ \ell $ and $ R_{3} $ is its Ricci scalar. Compared to flat slicing counterterms \eqref{flatcts}, there is a third counterterm involving $ R_{3} $ which takes care of the divergence proportional $ a_2 $ in \eqref{divtext}. As in flat slicing, the finite term proportional to $ \xi_-\xi_+ $ in \eqref{divtext} is always canceled by the counterterm canceling the $ a_{3} $ divergence.

In addition to the three divergent counterterms \eqref{invariantstext}, there are two finite counterterms given by
\begin{equation}
\mathcal{N}\int_{u=\Lambda} \sqrt{\gamma_{3}}\,\sin{\xi(u)}\cos^{3}{\xi(u)}, \quad  \mathcal{N}\int_{u=\Lambda} \sqrt{\gamma_{3}}\;\ell^{2}R_{3}\,\sin{\xi(u)}\cos{\xi(u)} \ .
\label{finctstext}
\end{equation}
These counterterms do not contain divergences, but contribute a finite piece: the first counterterm contributes a term $ \propto(\alpha y_{-})^{3} $ and the second one a term $ \propto(\alpha y_-) $ after identifying $ y_{-} = -\xi_- $. In general, $ U $ can generate such finite terms, so we shall keep the coefficients of the finite counterterms free and they parametrize the renormalization scheme.  After subtraction, the renormalized on-shell action is unique up to these two coefficients. The resulting renormalized on-shell action is (see equation \eqref{D6onshellren} in appendix \ref{app:holren})
\begin{equation}
I_{\text{D6,on-shell}}^{\text{ren}} = (I_{\text{fin}}-\,I_{\text{B}}\lvert_{u=u_0})\lvert_{\text{on-shell}}\,+B_{3}\,(\alpha y_{-})^{3}+B_{1}\,(\alpha y_{-}) \ ,
\label{D6onshellrentext}
\end{equation}
where the constants $ (B_3,B_1) $ are related to the coefficients of the finite counterterms \eqref{finctstext}, the first term is
\begin{align}
	I_{\text{fin}} = \mathcal{N}\vol{S^{3}}\int_{u_0}^{\infty}du\,\ell^{3}&\sinh^{3}{(u-c)}\,\sin{\xi(u)}\,\left( \sqrt{1 + \xi'(u)^{2}} -\sin{\xi(u)}\right.\nonumber\\
	&\;\;\left.-\coth{(u-c)}\,\left[ 1 - \frac{2}{\sinh^{2}{(u-c)}}\right] \,\cos{\xi(u)}\,\xi'(u) \right)
	\label{Ifintext}
\end{align}
and the second term is
\begin{equation}
	I_{\text{B}}\lvert_{u=u_0} = -\mathcal{N}\ell^{3}\vol{S^{3}}
	\begin{cases}
		0, \quad &\text{terminating}\\
		\sin^{2}{\xi(c)}, \quad &\text{space-filling} \ .
	\end{cases}
	\label{intconttext}
\end{equation}
The renormalized action can also be written in Cartesian coordinates for the embedding $ y = y(x) $ and the explicit expression can be found in Appendix~\ref{app:holren}.

\subsection{The free energy of the probe D6-brane}\label{subsec:freen}

In the probe limit, the total renormalized free energy $F^{\text{ren}}(\alpha y_-)$ of flavored-ABJM theory is given by
\begin{equation}
	F^{\text{ren}}(\alpha y_-) = F^{\text{ren}}_{\text{IIA}} + N_{f}F^{\text{ren}}_{\text{D6}}(\alpha y_-) + \mathcal{O}(N_{f}\slash N)^{2} \ ,
\end{equation}
where by the AdS\slash CFT duality
\begin{equation}
	F^{\text{ren}}_{\text{IIA}} = I_{\text{IIA,on-shell}}^{\text{ren}}, \quad F^{\text{ren}}_{\text{D6}} = I^{\text{ren}}_{\text{D6,on-shell}}
	\label{FrenIren}
\end{equation}
with on-shell actions as defined in \eqref{onshellactions}. We  focus on the flavor contribution $ F^{\text{ren}}_{\text{D6}} $ to the free energy given by the on-shell action \eqref{D6onshellren}. To simplify notation, we denote
\begin{equation}
\widetilde{F}^{\text{ren}}_{\text{D6}} \equiv F^{\text{ren}}_{\text{D6}}\lvert_{B_{3} = B_{1} = 0}
\label{D6freeen}
\end{equation}
for the flavor free energy in a holographic minimal subtraction scheme where no finite counterterms are added in holographic renormalization. In Appendix~\ref{sec:vevproof}, we show that
\begin{equation}
\frac{\partial \widetilde{F}^{\text{ren}}_{\text{D6}}}{\partial(\alpha y_-)} = F_{\text{D6,UV}}\;\alpha^{2}y_+ \ ,
\label{Fwiggleder}
\end{equation}
where
\begin{equation}
F_{\text{D6,UV}} = \mathcal{N}\ell^{3}\vol{S^{3}} = \frac{\pi}{4}N \sqrt{2\lambda}
\label{probeFUV}
\end{equation}
and $ y_+ = y_+(\alpha y_-) $ is known numerically from the asymptotics of the brane embeddings. Integrating this equation gives
\begin{equation}
\widetilde{F}^{\text{ren}}_{\text{D6}}(\alpha y_-) = \widetilde{F}^{\text{ren}}_{\text{D6}}(0) + F_{\text{D6,UV}} \int_{0}^{\alpha y_-}ds\,\alpha^{2}y_+(s) \ .
\label{freeintegral}
\end{equation}
where we understand that the function $ y_{+}(s) $ is made single-valued by defining it piecewise with a jump at the location of the quantum phase transition.
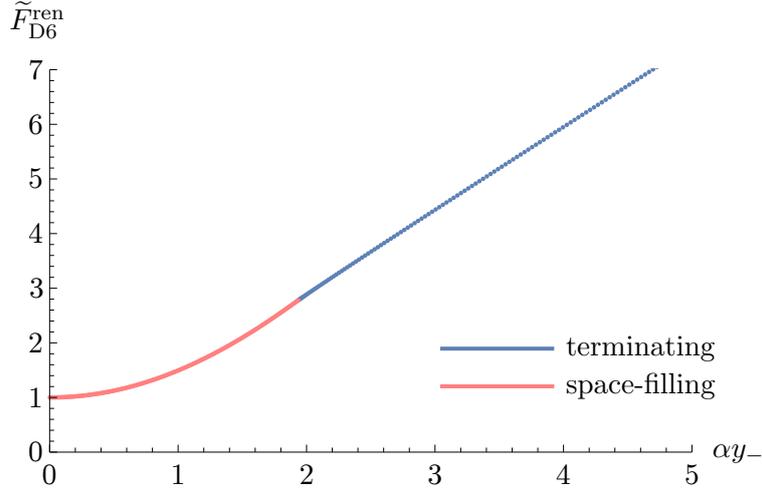
\begin{figure}[t]
	\centering
	\begin{tikzpicture}
		\node (img1)  {\includegraphics{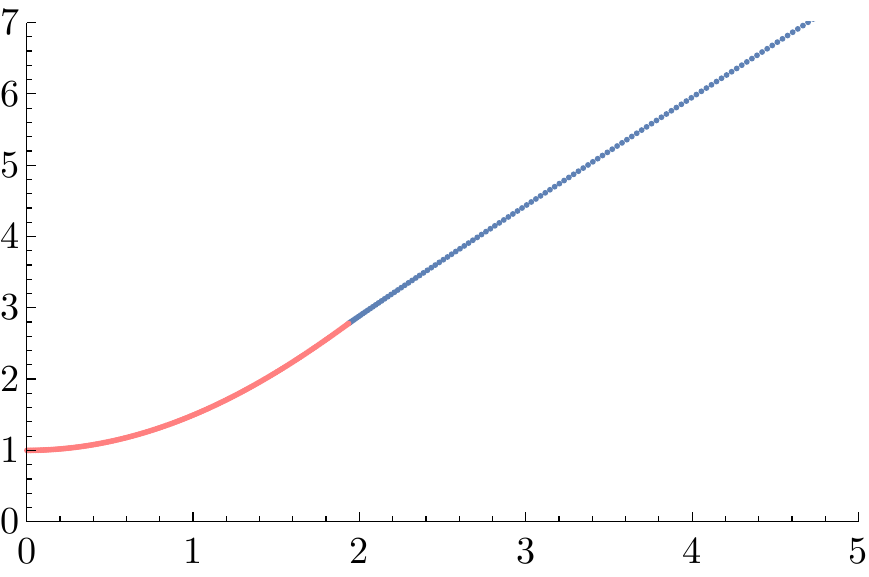}};
		\node[right=of img1, node distance=0cm, yshift=-2.4cm, xshift=-1.1cm] {$ \alpha y_- $};
		\node[above=of img1, node distance=0cm, anchor=center,yshift=-0.7cm,xshift=-4.3cm] {$ \widetilde{F}^{\text{ren}}_{\text{D6}} $};
		\draw[line width=1.5pt, color=mathematica1] (1,-1) -- (2.5,-1) node[right, color=black] {terminating};
		\draw[line width=1.5pt, color=pink] (1,-1.5) -- (2.5,-1.5) node[right, color=black] {space-filling};
	\end{tikzpicture}
	\caption{Renormalized on-shell action of the D6-brane \eqref{D6onshellren} in the minimal subtraction scheme \eqref{D6freeen} as a function of the source in units with $ F_{\text{D6,UV}} \equiv 1 $. It diverges linearly as $ \alpha y_-\rightarrow \infty $.}
	\label{freesource}
\end{figure}

The fixed point theory $ \alpha y_- = 0 $ is ABJM theory with massless flavor on $ S^{3}_{\alpha} $. This theory has conformal symmetry and its free energy $F^{\text{ren}}(0)$ is a positive constant and independent of $ \alpha $. Holographically, the dual embedding is the trivial embedding $ \xi(u) = \frac{\pi}{2} $ whose on-shell action simply computes the renormalized volume $ V^{\text{ren}}_{\text{AdS}_{4}} $ of AdS$ _{4} $:
\begin{equation}
	\widetilde{F}^{\text{ren}}_{\text{D6}}(0) = \frac{3}{2}\,\mathcal{N}\ell^{3}\,V^{\text{ren}}_{\text{AdS}_{4}} \ .
	\label{renads4}
\end{equation}
Renormalized volume is defined as the finite term in the expansion of the regularized volume
\begin{equation}
V^{\text{reg}}_{\text{AdS}_{d+1}} = \vol{S^{d}}\int_{c}^{\Lambda}\sinh^{d}{(u-c)}
\end{equation}
as $ \Lambda \rightarrow \infty $. For odd $ d = 2n+1 $, the renormalized volume is given by \cite{graham_volume_1999}
\begin{equation}
V^{\text{ren}}_{\text{AdS}_{d+1}} = \frac{(-1)^{n}\sqrt{\pi}\,n!}{2\Gamma\bigl(n+\frac{3}{2}\bigr)}\,\vol{S^{2n+1}} = \frac{(-1)^{n+1}\pi^{n+\frac{3}{2}}}{\Gamma\bigl(n+\frac{3}{2}\bigr)} \ ,
\end{equation}
which for $ d=3 $ gives $ V^{\text{ren}}_{\text{AdS}_{4}} = (2\slash 3)\vol{S^{3}} $. Substituting to \eqref{renads4} gives
\begin{equation}
	F^{\text{ren}}_{\text{D6}}(0) = \widetilde{F}^{\text{ren}}_{\text{D6}}(0)  = F_{\text{D6,UV}}
	\label{uvflavor}
\end{equation}
with the right hand side given by \eqref{probeFUV}.\footnote{To obtain the correct $ F_{\text{D6,UV}} $, regularity of $ \widehat{C}_7 $ in the interior is necessary. Non-regular $ \widehat{C}_7 $ would lead to an interior $ u=c $ contribution to the renormalized Wess--Zumino action which would shift the value of $ F_{\text{UV}} $.} This result is scheme independent, because both of the finite counterterms vanish when $ \alpha y_- = 0 $. Substituting \eqref{uvflavor} to \eqref{freeintegral} and performing the integral then gives us the free energy. We can also obtain the free energy directly by numerically integrating the D6-brane on-shell action \eqref{D6onshellren}. The results agree and the free energy is plotted in figure \ref{freesource}.

The total free energy $ F^{\text{ren}} $ for massless flavor has been computed by various methods in the literature. On the field theory side, it can be computed using matrix model localization \cite{santamaria_unquenched_2011} (see also \cite{marino_lectures_2012,herzog_multi-matrix_2011}), while on the gravity side, one can use the M-theory uplift \cite{santamaria_unquenched_2011} or include backreaction of the D6-branes by smearing \cite{conde_gravity_2011} (the smearing method agrees with the exact M-theory result only up to order $ N_f\slash N $). The result to any order in $ N_f\slash N $ is \cite{santamaria_unquenched_2011}
\begin{equation}
F^{\text{ren}}_{\text{IIA}} = \frac{\pi\sqrt{2}}{3}N^{3\slash 2}k^{1\slash 2}f(N_f\slash k), \quad f(a) =\frac{1+a}{\sqrt{1+a\slash 2}} \ .
\label{abjmtot}
\end{equation}
An expansion in the probe limit $ N_f\slash N \ll 1 $ is the same as an expansion in $ N_f \slash k \ll 1 $ at fixed 't Hooft coupling $ \lambda = N\slash k $. Therefore,  expanding the function in \eqref{abjmtot} we obtain
\begin{equation}
	F^{\text{ren}}(0) = \frac{\pi \sqrt{2}}{3}N^{3\slash 2}k^{1\slash 2} + \frac{\pi}{4}NN_f\sqrt{2 \lambda} +\mathcal{O}(N_f\slash N)^{2}  \ . 
	\label{Mtheory}
\end{equation}
We observe that the leading flavor correction matches with the on-shell action $ N_f F_{\text{D6,UV}} $ of $ N_f $ probe D6-branes with $ F_{\text{D6,UV}} $ given by \eqref{probeFUV}. The first term corresponds to the on-shell supergravity action.

\paragraph{Large curvature expansion}

Using the expression \eqref{freeintegral} we can also compute mass corrections to the UV fixed point flavor free energy $ F_{\text{D6,UV}} $. Using $ \alpha^{2}y_+(s) \approx s $ \eqref{sourcevevUVtext} in the space-filling phase, we obtain:
\begin{equation}
\widetilde{F}^{\text{ren}}_{\text{D6}}(\alpha y_-) = F_{\text{D6,UV}}\left[ 1 + \frac{1}{2}\, (\alpha y_-)^{2} + \mathcal{O}(\alpha y_-)^{4}\right] , \quad \alpha y_- \rightarrow 0 \ .
\end{equation}
The leading correction $ (\alpha y_-)^{2} $ is what one expects from conformal perturbation theory for a fermionic mass deformation with an operator $O$ of dimension $ \Delta = 2 $. In that case, the mass term is proportional to $ y_- $ and the leading correction comes from the 2-point function of $O$ at order $ y_{-}^{2} $ since the 1-point function of $O$ on $ S^{3}_{\alpha} $ vanishes at zero source by conformal flatness. It would be interesting to understand whether this correction could be obtained directly from the M-theory up-lift or from field theory localization calculations.

\paragraph{Small curvature (flat) expansion} In the opposite $ \alpha y_{-}\rightarrow \infty $ limit in the terminating phase, the behavior of $ \alpha^{2}y_{+} $ as a function of $ \alpha y_- $ is given by \eqref{vevIRtext}. Using \eqref{freeintegral}, we obtain
\begin{equation}
	\widetilde{F}^{\text{ren}}_{\text{D6}}(\alpha y_-) = F_{\text{D6,UV}}\,\left[  \frac{3}{2}\,(\alpha y_-) - \frac{18\log{2}-11}{8}\,(\alpha y_-)^{-1} + \mathcal{O}(\alpha y_-)^{-3}\right], \quad \alpha y_-\rightarrow \infty \ .
	\label{lineardiv}
\end{equation}
We observe that the free energy is IR divergent as $ \sim \alpha y_- $ which corresponds to a subleading volume divergence proportional to the sphere radius. The absence of the cubic divergence $ \sim (\alpha y_{-})^{3} $ is expected, because  its coefficient computes the free energy of massive flavor on flat space which we have shown to vanish in the  minimal subtraction scheme which defines $ \widetilde{F}^{\text{ren}}_{\text{D6}} $, see equation \eqref{vanishingfree}. However, in a general scheme, the cubic divergence does not vanish. We also observe that the finite term in the expansion \eqref{lineardiv} vanishes. This  is a scheme independent statement, because in our case, finite counterterms only modify coefficients of the terms which  diverge in the infinite-volume limit.

\section{On the $ F $-theorem and possible $ F $-functions}\label{sec:Ffuncs}

The renormalized free energy of a CFT on a sphere is an unambiguous scheme independent quantity that provides a good measure of the degrees of freedom of the theory. As we  discuss below, this also true for the UV and IR limit (namely the massless and infinite mass limit) of the probe-D6 free energy we have computed in the previous section, after a suitable subtraction of infinite-volume divergence in the IR case. This motivates one to search for a suitable interpolating function, constructed from the free energy, which plays the role of an $ F $-function and matches the UV and IR free energies.

The first proposal for such an $ F $-function was an appropriately renormalized entanglement entropy of a disk in the theory on flat space \cite{liu_refinement_2013} and it was proven to be monotonic in \cite{casini_rg_2012}. Going beyond entanglement entropy, alternative holographic $ F $-functions constructed from the renormalized on-shell bulk action were proposed in \cite{ghosh_holographic_2019}. These $ F $-functions were shown to be monotonic in the case of RG-flows driven by a canonical scalar field coupled to Einstein gravity \cite{ghosh_holographic_2019}, and we want to test it here in the case of RG-flows constructed in a top-down holographic model, flavored-ABJM theory.

Motivated by these considerations, we study different proposals for an $ F $-function whose purpose is to count the number of degrees of freedom carried by the massive flavor and to match the CFT $F$-value at the fixed points.  Holographically, the quantity of interest is the on-shell action of the D6-brane which equals the free energy of the flavor. Based on the proposals of \cite{ghosh_holographic_2019}, we attempt to construct $ F $-functions from the renormalized free energy and also from its Legendre transform, the quantum effective potential.

\subsection{Proposals for holographic $ F $-functions}

Consider a 3-dimensional Euclidean CFT on $ S^{3}_{\alpha} $ deformed by a relevant operator of dimension $ 1\slash 2 < \Delta < 3 $ (the lower limit is the unitary bound) with bare coupling $ \lambda $. The generating functional of the theory is a function of the dimensionless coupling
\begin{equation}
j = \alpha \lvert\lambda\lvert^{1\slash (3-\Delta)}
\end{equation}
that defines a one-parameter family of different theories. The flow in the dimensionless coupling defines what is called a \textit{curvature RG flow} \cite{ghosh_holographic_2018,ghosh_holographic_2019}. This should not be confused with the standard RG flow of a single theory, in which the parameter along the flow is energy scale. Nevertheless, curvature RG flows also describe an interpolation between UV and IR features of the theory.

Instead of the dimensionless coupling, one can parametrize the flow by any single-valued parameter $ \mathcal{P} = \mathcal{P}(j) $. Then an $ F $-function $ \mathcal{F} = \mathcal{F}(\mathcal{P}) $ is a function that monotonically interpolates between UV and IR theories, such that at the end-points $ \mathcal{P}(0) $ and $ \mathcal{P}(\infty) $ of the flow, it coincides with the $ S^{3}_{\alpha} $ free energy of the corresponding theory. In other words, the $ F $-function is required to satisfy
\begin{equation}
\mathcal{F}(\mathcal{P}) =
\begin{cases}
F_{\text{UV}}, \quad &j \rightarrow 0\\
F_{\text{IR}}, \quad &j \rightarrow \infty
\end{cases}, \qquad
\frac{d\mathcal{F}}{dj}\leq 0 \ ,
\label{Frequirements}
\end{equation}
where $ F_{\text{UV}} $ and $ F_{\text{IR}} $ are renormalized free energies of two fixed point theories. The free energy $ F_{\text{UV}} $ is computed in the $ \lambda = 0 $ CFT while $ F_{\text{IR}}$ is computed in the CFT that is left after the degrees of freedom involved in the operator deformation are gapped out.

For holographic RG flows in Einstein--Scalar theory, four candidate $ F $-functions, that were constructed from the renormalized on-shell bulk action, were shown to satisfy the monotonicity properties \eqref{Frequirements} in \cite{ghosh_holographic_2019} for regular holographic flows. The scalar field $ \varphi $ dual to the operator deformation behaves as
\begin{equation}
	\varphi(u) = e^{-\Delta_- u}\,[\varphi_{-} + \ldots] + e^{-\Delta_+ u}\,[\varphi_{+} + \ldots] \ , \quad u \rightarrow \infty \ ,
\end{equation}
where $ \Delta_- = 3-\Delta_+ $. The curvature RG flow was parametrized by the dimensionless curvature
\begin{equation}
\mathcal{R} = \frac{6}{\alpha^{2} \lvert\varphi_{-}\lvert^{2\slash \Delta_{-}}}
\end{equation}
and the renormalized Einstein--Scalar on-shell action $ I^{\text{ren}}_{\text{bulk}} = I^{\text{ren}}_{\text{bulk}}(\mathcal{R}) $ is a function of this parameter. The $ F $-functions differ by the four different ways of removing the two IR (infinite volume) divergences $ \sim \mathcal{R}^{-3\slash 2} $ and $ \sim \mathcal{R}^{-1\slash 2} $ that appear in a general renormalization scheme. More precisely, the renormalized action diverges as
\begin{equation}
	I^{\text{ren}}_{\text{bulk}}(\mathcal{R}) = b_3\mathcal{R}^{-3\slash 2}+b_1\mathcal{R}^{-1\slash 2} + \mathcal{O}(1), \quad \mathcal{R}\rightarrow 0 \ ,
\end{equation}
with coefficients $ b_{1,3} $ parametrizing the renormalization scheme. The volume divergences can be removed by an appropriate choice of scheme, by acting with suitable differential operators or by a combination of the two. Define the differential operators
\begin{equation}
\mathcal{D}_{3} = \frac{2}{3}\mathcal{R}\frac{\partial}{\partial \mathcal{R}} + 1, \quad \mathcal{D}_{1} = 2\mathcal{R}\frac{\partial}{\partial \mathcal{R}} + 1\ ,
\label{diffops}
\end{equation}
which annihilate the two divergences respectively
\begin{equation}
\mathcal{D}_{3}\,\mathcal{R}^{-3\slash 2} = 0, \quad \mathcal{D}_{1}\,\mathcal{R}^{-1\slash 2} = 0 \ .
\end{equation}
The four candidate $ F $-functions are then defined as \cite{ghosh_holographic_2019}
\begin{alignat}{3}
\mathcal{F}_1(\mathcal{R}) &= \mathcal{D}_{1}&\mathcal{D}_{3}&\,I^{\text{ren}}_{\text{bulk}}(\mathcal{R}) \nonumber\\
\mathcal{F}_2(\mathcal{R}) &= \mathcal{D}_{1}& &\,I^{\text{ren}}_{\text{bulk}}(\mathcal{R})\lvert_{b_3 = 0} \nonumber\\
\mathcal{F}_3(\mathcal{R}) &= &\mathcal{D}_{3}&\,I^{\text{ren}}_{\text{bulk}}(\mathcal{R})\lvert_{b_1 = 0} \nonumber\\
\mathcal{F}_4(\mathcal{R}) &=  & \, &\,I^{\text{ren}}_{\text{bulk}}(\mathcal{R})\lvert_{b_{3} = b_1 = 0} \ .
\label{holFs}
\end{alignat}
For a theory deformed by an operator of dimension $ \Delta > 3\slash 2 $, the on-shell action computes the free energy of the dual field theory and the non-normalizable mode of the bulk scalar field is dual to the source of the operator deformation:
\begin{equation}
I^{\text{ren}}_{\text{bulk}}(\mathcal{R}) = F^{\text{ren}}(\mathcal{R}), \quad \varphi_{-} = \lambda, \quad \Delta_+ = \Delta
\end{equation}
so that $ \mathcal{R} = 6\slash j^{2} $ is curvature in units of the coupling.

For $ \Delta < 3\slash 2 $, it is instead the Legendre transform of the free energy with respect to the relevant coupling $ \lambda $, the quantum effective potential, that appears on the field theory side. In addition, the non-normalizable mode is identified with the vev of the dual operator:
\begin{equation}
I^{\text{ren}}_{\text{bulk}}(\mathcal{R}) = \Gamma^{\text{ren}}(\mathcal{R}), \quad \varphi_{-} \propto \langle O\rangle, \quad \Delta_- = \Delta
\end{equation}
so that $ \mathcal{R} \propto 6\slash (\alpha^{2}\langle  O\rangle^{2\slash \Delta}) $ is curvature in units of the vev. As a result, holography suggests that for $ \Delta < 3\slash 2 $ it is the quantum effective potential instead of the free energy which is monotonic. This was proven to be the case in the holographic models of \cite{ghosh_holographic_2019} and also in particular for the  free massive boson theory, without reference to holography \cite{ghosh_holographic_2019}.

In our setup of ABJM theory with massive flavor, the dimension $ \Delta = 2> 3\slash 2 $ of the fermion bilinear falls to the first range and the $ F $-functions \eqref{holFs} have dual interpretation as being constructed from free energy. Nevertheless, we also compute the quantum effective potential in our case and try to construct an $ F $-function that way. Both constructions fail to produce monotonic $ F $-functions as we show next. In the discussion section \ref{sec:conc}, we surmise that the non-monotonicity is caused by the first order phase transition.

\subsection{Candidate $ F $-functions from the free energy}

We now compute the $ F $-functions \eqref{holFs} of \cite{ghosh_holographic_2019} in ABJM theory with massive flavor. In our case, the coupling is the non-normalizable mode $ \lambda =  y_- = -\xi_- $ of the D6-brane embedding and it sources an operator of dimension $ \Delta = 2 > 3\slash 2 $. The dimensionless parameter is then $ j = \alpha y_{-} $ and the D6-brane on-shell action is identified with flavor free energy as in \eqref{FrenIren}. The UV fixed point CFT $ y_{-} = 0 $ is ABJM theory with massless flavor on $ S^{3}_{\alpha} $, and in the probe limit, it has the free energy \eqref{Mtheory}:
\begin{equation}
F_{\text{UV}} = \frac{\pi \sqrt{2}}{3}N^{3\slash 2}k^{1\slash 2} + \frac{\pi}{4}NN_f\sqrt{2 \lambda} +\mathcal{O}(N_f\slash N)^{2} \ .
\end{equation}
The IR fixed point CFT is simply pure ABJM theory without flavor, because the flavor sector completely decouples from the theory when the flavor mass is taken to infinity. The free energy at the IR fixed point is then:
\begin{equation}
F_{\text{IR}} = \frac{\pi \sqrt{2}}{3}N^{3\slash 2}k^{1\slash 2} \ .
\end{equation}
We see that the fixed point free energies satisfy the $ F $-theorem $ F_{\text{UV}} > F_{\text{IR}} $ and the question then is whether the candidate $ F $-functions \eqref{holFs} interpolate monotonically between these two values.
\begin{figure}
	\centering
	\begin{tikzpicture}
		\node (img1)  {\includegraphics{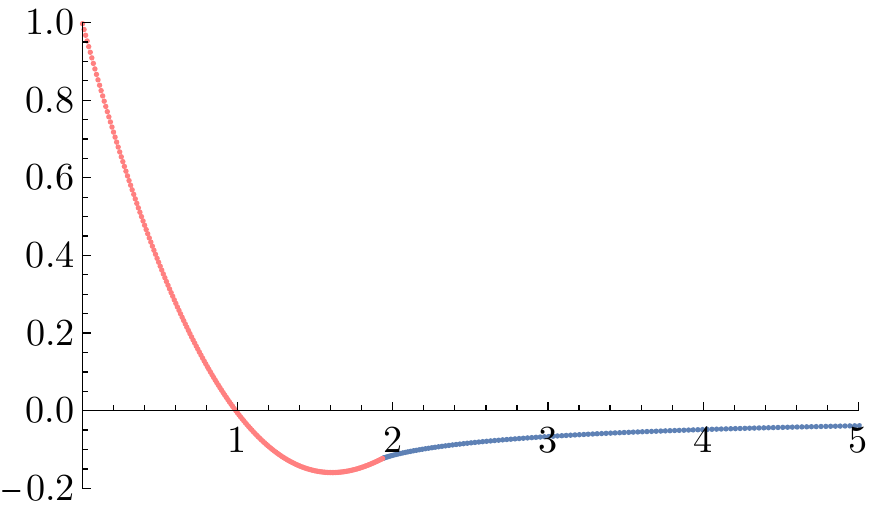}};
		\node[right=of img1, node distance=0cm, yshift=-1.5cm, xshift=-1cm] {$ \alpha y_- $};
		\node[above=of img1, node distance=0cm, anchor=center,yshift=-0.8cm, xshift=-3.5cm] {$ \mathcal{F}_4 $};
		\draw[line width=1.5pt, color=mathematica1] (1,-0.25) -- (2.5,-0.25) node[right, color=black] {terminating};
		\draw[line width=1.5pt, color=pink] (1,-0.75) -- (2.5,-0.75) node[right, color=black] {space-filling};
	\end{tikzpicture}
	\caption{Free energy in the IR finite scheme $ (b_3,b_1) = (0,0) $.}
	\label{freeeneergyfin}
\end{figure}
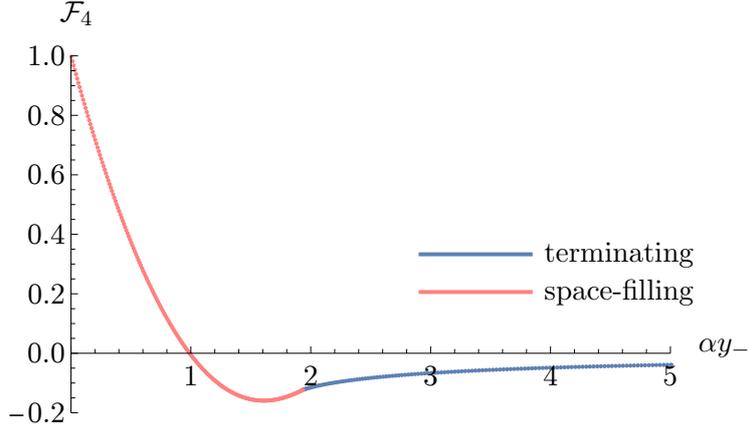

To study monotonicity, it is enough to focus on the free energy of a single D6-brane since the gravity contribution is independent of $ \alpha y_{-} $. The D6-brane free energy $ \widetilde{F}^{\text{ren}}_{\text{D6}} $ in the holographic minimal subtraction scheme is IR divergent and we define an IR finite free energy $ \mathcal{F}_4 $ by a shift of renormalization scheme:
\begin{equation}
\mathcal{F}_4(\alpha y_-) \equiv \widetilde{F}^{\text{ren}}_{\text{D6}}(\alpha y_-) - \frac{3}{2}\,(\alpha y_-) \ .
\label{IRfinitescheme}
\end{equation}
This defines what we call the IR finite scheme and the corresponding free energy is plotted in figure \ref{freeeneergyfin}. The free energy $F^{\text{ren}}_{\text{D6}}$ in a general scheme parametrized by $ (b_3,b_1) $ is defined relative to the IR finite scheme $ (0,0) $ as
\begin{equation}
F^{\text{ren}}_{\text{D6}} = \mathcal{F}_4 + b_{3}\,(\alpha y_-)^{3}+ b_{1}\,(\alpha y_-) = \widetilde{F}^{\text{ren}}_{\text{D6}} + b_{3}\,(\alpha y_-)^{3}+ \biggl(b_{1}- \frac{3}{2}\biggr)\,(\alpha y_-)
\label{generalscheme}
\end{equation}
with IR divergences controlled by the two parameters $ (b_3,b_1) $:
\begin{equation}\label{generalschemeIR}
F^{\text{ren}}_{\text{D6}}(\alpha y_-) = b_{3}\,(\alpha y_-)^{3}+ b_{1}\,(\alpha y_-)-\frac{18\log{2}-11}{8}\,(\alpha y_-)^{-1} + \ldots, \quad \alpha y_- \rightarrow \infty \ .
\end{equation}
The relations to the coefficients $ (B_3,B_1) $ of the finite counterterms \eqref{finctstext} appearing in \eqref{D6onshellrentext} are $ B_3 = b_3 $ and $ B_1 = b_1-3\slash 2 $.

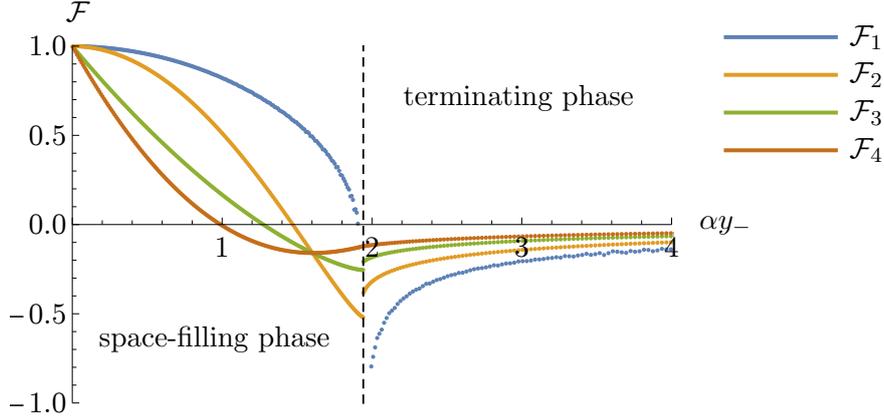
\begin{figure}[t]
	\centering
	\begin{tikzpicture}
		\node (img1)  {\includegraphics[scale=1]{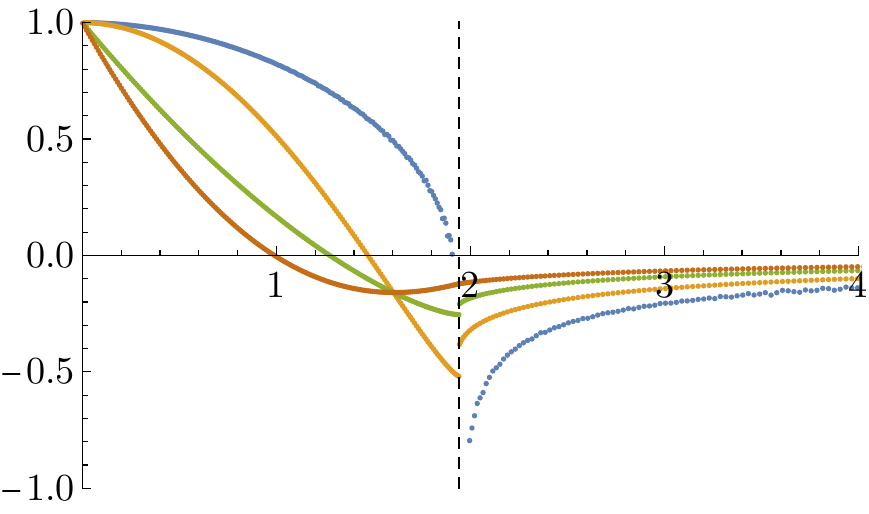}};
		\node[right=of img1, node distance=0cm, yshift=0cm, xshift=-1cm] {$ \alpha y_- $};
		\node[above=of img1, node distance=0cm, anchor=center,yshift=-0.9cm, xshift=-3.5cm] {$ \mathcal{F} $};
		\node[node distance=0cm, anchor=center,yshift=-1.5cm, xshift=-1.7cm] {space-filling phase};
		\node[node distance=0cm, anchor=center,yshift=1.7cm, xshift=2.3cm] {terminating phase};
		\draw[line width=1.5pt, color=mathematica1] (5,2.5) -- (6.5,2.5) node[right, color=black] {$ \mathcal{F}_1 $};
		\draw[line width=1.5pt, color=mathematica2] (5,2) -- (6.5,2) node[right, color=black] {$ \mathcal{F}_2 $};
		\draw[line width=1.5pt, color=mathematica3] (5,1.5) -- (6.5,1.5) node[right, color=black] {$ \mathcal{F}_3 $};
		\draw[line width=1.5pt, color=mathematica6] (5,1) -- (6.5,1) node[right, color=black] {$ \mathcal{F}_4 $};
	\end{tikzpicture}
	\caption{Plots of the $F$-functions \eqref{Ffunctions} constructed from the D6-brane free energy in units with $ F_{\text{D6,UV}} = 1 $. The location of the quantum phase transition is indicated by the dashed vertical line.}
	\label{F234}
\end{figure}

In our case, the differential operators \eqref{diffops} are
\begin{equation}
\mathcal{D}_{3} = -\frac{1}{3}\,j\frac{\partial}{\partial j} + 1, \quad \mathcal{D}_{1} = -j\frac{\partial}{\partial j} + 1\ ,
\label{diffopsj}
\end{equation}
where $ j  = \alpha y_- $ and they annihilate the cubic $ \mathcal{D}_3\,j^{3} = 0 $ and linear $ \mathcal{D}_1\,j = 0 $ divergences in \eqref{generalscheme}. The candidate $ F $-functions \eqref{holFs} are
\begin{alignat}{3}
\mathcal{F}_1(\alpha y_-) &= \mathcal{D}_{1}&\mathcal{D}_{3}&\,F^{\text{ren}}_{\text{D6}}(\alpha y_-) \nonumber\\
\mathcal{F}_2(\alpha y_-) &= \mathcal{D}_{1}& &\,F^{\text{ren}}_{\text{D6}}(\alpha y_-)\lvert_{b_3 = 0} \nonumber\\
\mathcal{F}_3(\alpha y_-) &= &\mathcal{D}_{3}&\,F^{\text{ren}}_{\text{D6}}(\alpha y_-)\lvert_{b_1 = 0} \nonumber\\
\mathcal{F}_4(\alpha y_-) &=  & \, &\,F^{\text{ren}}_{\text{D6}}(\alpha y_-)\lvert_{b_{3} = b_1 = 0} \ ,
\label{Ffunctions}
\end{alignat}
where $ \mathcal{F}_4 $ is just the D6-brane free energy in the IR finite scheme \eqref{IRfinitescheme}. By construction, all of the candidate $ F $-functions satisfy
\begin{equation}
\lim_{\alpha y_-\rightarrow 0}\mathcal{F}_{1,2,3,4} = F_{\text{D6,UV}} = \frac{\pi}{4}N\sqrt{2\lambda}, \quad \lim_{\alpha y_- \rightarrow \infty}\mathcal{F}_{1,2,3,4} = 0
\end{equation}
as required. However, as can be seen from figure \ref{F234}, these values are not interpolated monotonically. We can see this analytically from the small curvature expansion of the free energy \eqref{lineardiv} in the IR finite scheme
\begin{equation}
\mathcal{F}_4 = -\frac{18\log{2}-11}{8}\,(\alpha y_{-})^{-1} + \ldots, \quad \alpha y_{-} \rightarrow \infty \ ,
\label{leadcoeffneg}
\end{equation}
where the coefficient of the leading term is negative. The sign of this coefficient is not changed by the action of the differential operators \eqref{diffopsj} so that all of the candidate $ F $-functions are also non-monotonic. One can also see from figure \ref{F234} that all candidate $F$-functions except ${\mathcal F}_4$ are discontinuous across the phase transition and  are monotonic in the UV phase.

The fact that the leading coefficient in \eqref{leadcoeffneg} is negative can already be inferred from the plot of the vev as a function of mass: the spiral spins downwards. If the vev in the terminating phase had a larger value than in the space-filling phase, the coefficient in \eqref{leadcoeffneg} would be positive.

\subsection{Candidate $ F $-functions from the quantum effective potential}

In \cite{ghosh_holographic_2019}, it was proposed that for deformations with an operator of dimension $ \Delta < 3\slash 2 $, monotonic $ F $-functions can be constructed from the quantum effective potential instead of the free energy. In our case, we are not in this range of scaling dimensions since $ \Delta = 2 $, but we can nevertheless test whether the quantum effective potential would provide a good $ F $-function.

We define the (dimensionless) conjugate variable $ p $ of the source $ y_- $ in a general scheme via
\begin{equation}
\frac{\partial F^{\text{ren}}_{\text{D6}}}{\partial (\alpha y_-)} = p  \ ,
\end{equation}
where the free energy is given by \eqref{generalscheme}. Using
\begin{equation}
\frac{\partial F^{\text{ren}}_{\text{D6}}}{\partial (\alpha y_-)} = \frac{\partial \widetilde{F}^{\text{ren}}_{\text{D6}}}{\partial (\alpha y_-)}+ 3\,b_{3}\,(\alpha y_-)^{2}+ \Bigl(b_{1}- \frac{3}{2}\Bigr)
\end{equation}
and \eqref{Fwiggleder}, it follows that
\begin{equation}
p = \alpha^{2}y_+ + 3\,b_{3}\,(\alpha y_-)^{2} + \Bigl(b_{1}-\frac{3}{2}\Bigr) \ ,
\end{equation}
where we have set $ F_{\text{D6,UV}} \equiv 1 $ which we assume below as well. Here $ y_+ $ as a function of $ \alpha y_- $ is known numerically from the asymptotic behavior of the embeddings.

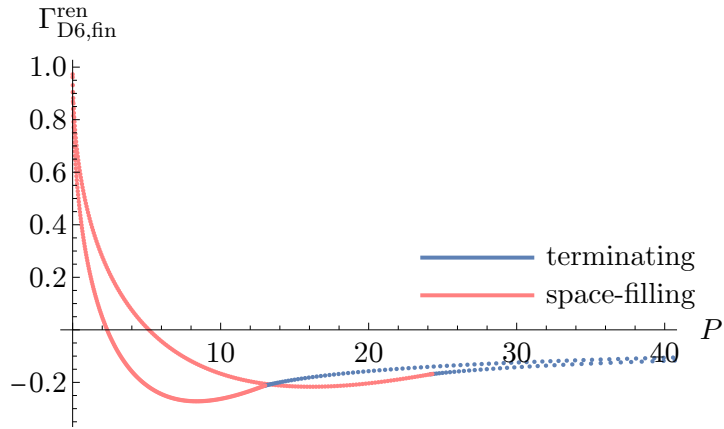
\begin{figure}
	\centering
	\begin{tikzpicture}
		\node (img1)  {\includegraphics{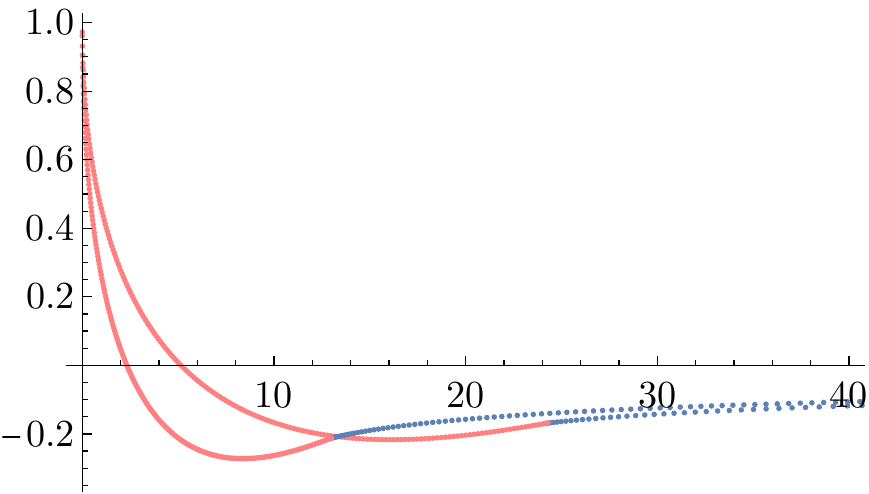}};
		\node[right=of img1, node distance=0cm, yshift=-1.2cm, xshift=-1cm] {$ P $};
		\node[above=of img1, node distance=0cm, anchor=center,yshift=-0.8cm, xshift=-3.5cm] {$ \Gamma^{\text{ren}}_{\text{D6,fin}} $};
		\draw[line width=1.5pt, color=mathematica1] (1,-0.25) -- (2.5,-0.25) node[right, color=black] {terminating};
		\draw[line width=1.5pt, color=pink] (1,-0.75) -- (2.5,-0.75) node[right, color=black] {space-filling};
	\end{tikzpicture}
	\caption{Quantum effective potential with IR divergences subtracted \eqref{gammafin} (not the $ F $-function) in two schemes $ b_3 = 1,2 $ where the vev is invertible. Because $ \Gamma^{\text{ren}}_{\text{D6,fin}}(P) $ is not monotonic, neither is $ \mathcal{F}(P) $.}
	\label{legendreplot}
\end{figure}

The quantum effective potential is defined as the Legendre transform
\begin{equation}
\Gamma^{\text{ren}}_{\text{D6}}(p) = F^{\text{ren}}_{\text{D6}}(\alpha y_-)-(\alpha y_-)\,\frac{\partial F^{\text{ren}}_{\text{D6}}}{\partial (\alpha y_-)} \ ,
\end{equation}
where $ y_- = y_-(p) $ is understood. It is well defined only in renormalization schemes $ (b_3,b_1) $ where $ p $ is invertible as a function of $ y_- $. In Appendix \ref{app:invvev}, we prove that $ p $ is invertible only in schemes where $ b_3 > 0 $ without constraints on $ b_1 $. In such schemes, the end-points of the flow correspond to
\begin{equation}
p(0) = b_1-\frac{3}{2}, \quad p(\infty) = \infty \ .
\end{equation}
Notice that the value of the quantum effective potential does not depend on $ b_1 $:
\begin{equation}
\Gamma^{\text{ren}}_{\text{D6}}(p) = \biggr[\widetilde{F}^{\text{ren}}_{\text{D6}}(\alpha y_-)-(\alpha y_-)(\alpha^{2}y_+)\biggl] - 2\,b_{3}\,(\alpha y_-)^{3} \ .
\label{gammatilde}
\end{equation}
Thus without loss of generality it is useful to introduce the shifted vev
\begin{equation}
	P \equiv p -\Bigl(b_{1}-\frac{3}{2}\Bigr) \ ,
\end{equation}
which has the range $ P \in [0,\infty) $. After inverting $ P $ as a function of $ \alpha y_- $ order by order, we find the following expansion for \eqref{gammatilde} (see Appendix \ref{app:invvev} for more details)
\begin{equation}
\Gamma^{\text{ren}}_{\text{D6}}(p) = -\frac{2}{\sqrt{27\,b_3}}\,P^{3\slash 2} +\frac{3}{\sqrt{12\,b_3}}\,P^{1\slash 2} +\frac{1}{\sqrt{48\,b_3}}\,\biggl(-\frac{9}{4}+12\,b_3\,b_{-1}\biggr)\,P^{-1\slash 2} + \ldots , \quad P \rightarrow \infty \ ,
\label{gammaIR}
\end{equation}
where
\begin{equation}
b_{-1} = -\frac{18\log{2}-11}{8}
\end{equation}
and the expansion is well defined since we assume $ b_3 >0 $. In the opposite limit $ P\rightarrow 0 $, we obtain
\begin{equation}
\Gamma^{\text{ren}}_{\text{D6}}(p) = 1 - \frac{1}{2}\,P^{2} + \ldots, \quad P \rightarrow 0 \ .
\label{gammaUV}
\end{equation}
We can now attempt to construct an $ F $-function from the quantum effective potential with the flow parametrized by $ P \in [0,\infty) $. In contrast to above, we no longer have the freedom of shifting the renormalization scheme arbitrarily: we have to stay in schemes where the vev is invertible. But in such schemes, we are never able to remove the IR divergences \eqref{gammaIR}. Thus the only possible $ F $-function is the one where the IR divergences are removed by the action of differential operators:\footnote{One can attempt to construct the $ F $-function by replacing $ P $ with $ p $ in the differential operator. This does remove the IR divergences, but the resulting $ F $-function does not have the correct UV limit $ \mathcal{F}(0) = 1 $.}
\begin{equation}
\mathcal{F}(P) = \biggl(\frac{2}{3}\,P\frac{d}{dP}-1\biggr)\biggl(2\,P\frac{d}{dP}-1\biggr)\,\Gamma^{\text{ren}}_{\text{D6}}(p) \ .
\label{Ffunctionlegendre}
\end{equation}
Based on the asymptotics \eqref{gammaIR}, we obtain
\begin{equation}
\mathcal{F}(P) = -\frac{8}{\sqrt{432\,b_3}}\,\biggl(\frac{9}{4}+12\,b_3\,\frac{18\log{2}-11}{8}\biggr)\,P^{-1\slash 2} + \ldots, \quad P \rightarrow \infty \ .
\label{FIR}
\end{equation}
Similarly from \eqref{gammaUV} we obtain
\begin{equation}
\mathcal{F}(P) = 1  -\frac{1}{2}\, P^{2} + \ldots, \quad P\rightarrow 0 \ .
\label{FUV}
\end{equation}
Clearly then $ \mathcal{F}(0) = F_{\text{D6,UV}} \equiv 1 $ and $ \mathcal{F}(\infty) = 0 $ as required. From \eqref{FIR} it follows that a necessary condition for monotonicity is
\begin{equation}
\frac{3}{2}+(18\log{2}-11)\,b_3\leq 0
\end{equation}
or equivalently
\begin{equation}
b_3 \leq -\frac{3\slash 2}{18\log{2}-11} \approx -1 \ .
\label{monotonicity}
\end{equation}
This violates the convexity condition $ b_3 > 0 $. Hence in renormalization schemes where the vev is invertible and the Legendre transform exists, the $ F $-function $ \mathcal{F}(P) $ is never monotonic. This is already seen from the IR finite part of the quantum effective potential
\begin{equation}
\Gamma^{\text{ren}}_{\text{D6,fin}}(p) \equiv \Gamma^{\text{ren}}_{\text{D6}}(p)-\biggl( -\frac{2}{\sqrt{27\,b_3}}\,P^{3\slash 2} +\frac{3}{\sqrt{12\,b_3}}\,P^{1\slash 2}\biggr)
\label{gammafin}
\end{equation}
which is plotted in two schemes in figure \ref{legendreplot}. The differential operators cannot change the monotonicity of $ \Gamma^{\text{ren}}_{\text{D6,fin}} $ in $ P\rightarrow \infty $ so that $ \mathcal{F}(P) $ is also non-monotonic.

\section{Conclusions and discussion}\label{sec:conc}

In this work, we studied strongly coupled ABJM theory with massive flavor on a three-sphere using the AdS\slash CFT duality. We obtained results in the quenched flavor approximation, which correspond to the probe flavor-brane approximation in the gravity dual. In the conformal limit of massless flavors, our results reproduce the leading flavor contribution found in the backreacted case. For non-zero flavor mass, the theory  flows in the IR to the pure ABJM CFT.

As in other instances involving probe flavor D-brane embeddings in curved foliations of Euclidean AdS space, we found two types of topologically distinct embeddings. One of them reaches all the way down the end of space in the bulk radial direction, it dominates the (semiclassical) partition function at large curvature and connects to the massless flavor limit;  the other terminates before reaching the end of radial coordinate, it  dominates at small curvature, and connects to the flat-space limit. At the critical value of the curvature (for fixed flavor mass) we observe a first-order phase transition, which in the dual field theory corresponds to a meson-melting transition like the ones previously observed in other holographic models \cite{mateos_holographic_2006,mateos_thermodynamics_2007,karch_chiral_2006,karch_precision_2015,karch_supersymmetric_2015}. 

The main focus of the paper is on the holographic calculation of the flavor contribution to the sphere free energy for an arbitrary value of the  fermion mass  $m$ (as defined in the UV CFT), as a function of the dimensionless parameter $\alpha m$,  where $\alpha$ is the sphere radius. We verified analytically that, as can be expected on general grounds, the renormalized free energy behaves, at the extreme limits of parameter space,  as:
\begin{equation} \label{disc1}
	F^{\text{ren}}(\alpha m)  = 
	\begin{cases}
		F_{\text{UV}} + \mathcal{O}(\alpha m), \quad & \alpha m \to 0\\
		b_3 (\alpha m)^3 + b_1 (\alpha m) + F_* + \mathcal{O}\left((\alpha m)^{-1}\right), \quad &\alpha m \to +\infty \ .
	\end{cases}
\end{equation}
In the above expression, $F_{\text{UV}}$ is the free energy of the massless flavor CFT on $S^3$, $b_1$ and $b_3$ are renormalization scheme-dependent parameters, and  $F_*$ is a scheme-independent constant. At face value, the free energy diverges as $\alpha m \to \infty$ because space volume becomes infinite in this limit.  

Starting from the computation of the free energy, we  tested a recent conjecture \cite{ghosh_holographic_2019}, according to which  one can construct,  out of $F^{\text{ren}}(\alpha m)$,  monotonic $F$-functions  which interpolate between the universal ``$ F $-values'' $F_{\text{UV}}$ and $F_{\text{IR}}$, {\emph{i.e.}}, the sphere free energies  of the UV and IR CFTs,  reached for $\alpha m \to 0$ and $\alpha m \to +\infty$, respectively. In our case, these CFTs  are: in the UV,  the massless flavored ABJM; in the IR, pure ABJM with no flavor.

Let us discuss this conjecture, and its fate,  in more detail. Concretely, we  have four  candidate $ F $-functions ${\cal F}_i(\alpha m)$ which are constructed by removing the infinite volume divergences in (\ref{disc1})  by either a suitable choice of scheme, or by acting with appropriate differential operators. By construction, all of these  $ F $-functions satisfy the following properties:
\begin{equation} \label{disc2}
\lim_{j\to 0} {\cal F}_i(j) = F_{\text{UV}}, \qquad \lim_{j\to +\infty} {\cal F}(j) = F_{*} \qquad j \equiv \alpha m \ .
\end{equation}
One can break  up the conjecture into three different statements (from the weakest to the strongest), which we discuss one by one in light of  our results:
\begin{enumerate}
	\item  The $\alpha m \to +\infty$ limit  of the $F$-functions coincide with the renormalized free energy of   IR CFTs on $ S^{3}_{\alpha} $. This translates into the statement about the constant $F_*$ appearing in (\ref{disc1}):
	\begin{equation} \label{conj1}
	F_* = F_{\text{IR}} \ ,
	\end{equation}
	independently of the details of the RG flow solutions which provide the functions ${\cal F}(j)$.
	\item The UV value is larger than the  IR value,
	\begin{equation} \label{conj2}
	F_{\text{UV}} > F_{\text{IR}} \ .
	\end{equation}
	This is actually guaranteed because of the  monotonicity of the Liu-Mezei $F$-function, which was proved in \cite{Casini:2012ei}.
	\item the $F$-functions are monotonically decreasing  over the range $j\in (0, +\infty)$.
\end{enumerate}
Concerning the first point, it is not at all a priori obvious  that (\ref{conj1}) should hold: the value $F_*$ enters as a   {\em subleading} contribution (at large $\alpha m$) to the renormalized free energy, as one can see from equation (\ref{disc1}). While the true free energy diverges,  this contribution remains finite in the infinite-volume limit.\footnote{It is therefore not the same thing as the finite free energy {\em density} per unit volume of the flat-space theory.}

It is therefore a non-trivial statement that this finite term $F_*$ in (\ref{disc1}) matches the (finite) free energy of the IR CFT on the sphere.   In the model we analyzed here, we found that this  is indeed the case: our computation shows that, for the flavor free energy in the probe limit,  $F_*^{(\text{flavor})}=F_{\text{IR}}^{(\text{flavor})}=0$, see equation (\ref{generalschemeIR}). Therefore, the total contribution  to the finite term in (\ref{disc1})  is $F_* = F_{\text{ABJM}} = F_{\text{IR}}$, since the IR theory is the pure ABJM CFT, and  in the probe limit the ABJM contribution is a constant independent of $\alpha m$. Therefore, point 1 of the conjecture holds.

The first point of the conjecture was already shown  to hold in minimal Einstein-scalar theories \cite{ghosh_holographic_2019}, where the IR CFT is realized  holographically as a minimum of the scalar potential: in that case one can  show in full generality, by expanding Einstein's equation around the IR extremum, that the correct  IR limit is obtained in (\ref{conj1}). Here, we found that this result extends to the present setup, which  cannot be reduced to a minimally coupled scalar field flowing between two extrema of a potential.  This  suggests that this result holds in general.  It may be possible to prove it directly in the field theory  by studying conformal perturbation theory around the IR CFT. We leave this open question for future work.

The second point (\ref{conj2}) also holds as expected, we have:
\begin{equation}
F_{\text{UV}} = F_{\text{ABJM}} + F^{(\text{flavor})}_{\text{UV}} > F_{\text{IR}} = F_{\text{ABJM}} \ .
\end{equation}
%%%%%%%%%%%%%%%%%%%%%
On the contrary,  we found that the third point does not hold: the  interpolation is  non-monotonic. Therefore, the strongest version of the conjecture fails. As a result, the free energy does not provide  monotonic $ F $-functions in  strongly coupled, flavored ABJM.

An important remark is that the phase transition is crucial in order to obtain the correct IR value for $F_*$: as we can see from figure \ref{F234}, all the F-functions overshoot the IR value ({\emph{i.e.}}, zero) in the large-curvature (UV) phase. The correct value $F_*=0$ is instead a property of the low-curvature (IR) phase free energy.

It is also tempting to think  that the presence of the  phase transition plays a significant  role in violating monotonicity. By looking at figure \ref{F234},  we see that three out of the four $F$-functions are monotonic in the large-curvature phase, all the way down to the phase transition (the fourth one, denoted ${\cal F}_4$, fails to be monotonic before the phase transition). This may lead one to think that monotonicity of the $F$-functions is a property of the phase that connects to the UV, but this property does not extend across the phase transition\footnote{In contrast, the Liu-Mezei $F$-function is unaffected by phase transitions of this kind, because it is always computed in the flat-space vacuum.}

Notice that the wrong monotonicity seems to be a property of the {\em whole} small-curvature phase, all the way to the IR limit. In fact,  ultimately it is the sign of the coefficient of the  first subleading correction in $(\alpha m)^{-1}$ in equation (\ref{disc1}) which determines the sign of ${\cal F}_i'(j)$ as $j \to \infty$. We have computed this quantity from the RG-flow solutions (see  equation (\ref{generalschemeIR})), but  it may be possible in general to relate it directly to a property of the IR CFT. In that case, one may formulate a necessary condition for monotonicity in terms of the IR CFT data alone.

One may notice that the behavior of the IR subtracted free energy (6.13) in our model is very similar to the behavior of the same quantity in a free conformally coupled massive boson on $ S^{3}_{\alpha} $ \cite{ghosh_holographic_2019}. However in that case it was possible to obtain monotonic $F$-functions from the Legendre transform of the free energy, {\emph{i.e.}}, the quantum effective potential \cite{ghosh_holographic_2019}. With this in mind,  we have also constructed similar interpolating functions  using the quantum effective potential, by taking the Legendre transform of the D6-brane free energy. When the Legendre transform  exists, by construction these new functions also interpolate between $F_{\text{UV}}$ and $F_{\text{IR}}$. However, we have found that  none of them is monotonic.

This work leaves several open questions. On the one hand, we think it will be worth investigating these issues further in top-down models which do not have phase transitions, like the scalar-field RG-flows which uplift to M-theory which can be found in the literature. On the other hand, one would like to study the role of phase transitions  in a controlled field theoretical setting, where one may have a perturbative description. Finally, it would be interesting to gain a general understanding of the matching between the sphere free energy of the IR CFT and the  finite term of the curvature RG-flow free energy in the large-volume limit.

%%%%%%%%%%%%%%%%%%%%%%%%%%%%%%%%%
\section*{Acknowledgements}\label{ACKNOWL}
\addcontentsline{toc}{section}{Acknowledgements}

We thank I.~Bena, G.~Cuomo, A.~V.~Ramallo, and N.~Warner for useful discussions. N.~J. has been supported in part by the Academy of Finland grant no.~1322307. This work was supported in part by CNRS contract IAE 199430.
J.~K. was supported by the Osk. Huttunen Foundation.

\appendix
\renewcommand{\theequation}{\thesection.\arabic{equation}}
\addcontentsline{toc}{section}{Appendix\label{app}}
\section*{Appendix}

\section{Conformally flat foliations of Euclidean AdS$ _{d+1} $}\label{app:FG}
	
Any asymptotically locally AdS$ _{d+1} $ metric $ G_{ab} $ can be written in a Fefferham--Graham expansion of the form\footnote{The indices $ a,b,\ldots $ run over the $ d+1 $ bulk coordinates while the indices $ i,j, \ldots $ run over the $ d $ boundary coordinates.}
\begin{equation}
G_{ab}dX^{a}dX^{b} = \frac{\ell^{2}}{z^{2}}(dz^{2} + \gamma_{ij}(z,\sigma)\,d\sigma^{i}d\sigma^{j}) \ ,
\label{FGgauge}
\end{equation}
where $ \gamma_{ij} $ has the expansion
\begin{equation}
\gamma(z,\sigma) = g_{(0)}(\sigma) + g_{(2)}(\sigma)\,z^{2} +g_{(4)}(\sigma)\,z^{4}+ \ldots
\end{equation}
and ellipsis denote subleading terms in the $ z\rightarrow 0 $ expansion.\footnote{Note that for even $ d $, there is also a logarithmic term characterizing the Weyl anomaly. Another common radial coordinate used in the literature is given by $ \sqrt{z} $.} The metric $ g_{(0)} $ is determined up to a Weyl transformation: a diffeomorphism preserving the FG gauge \eqref{FGgauge} induces a Weyl scaling on $ g_{(0)} $ \cite{imbimbo_diffeomorphisms_2000}. Hence $ g_{(0)} $ is identified as the boundary metric on which the dual field theory lives.
	
Let $ d+1\geq 3 $ and assume that $ G_{ab} $ solves the vacuum Einstein equations with a negative cosmological constant in the bulk (it is an Einstein metric):
\begin{equation}
	\mathcal{R}_{ab} = -\frac{d}{\ell^{2}}G_{ab} \ .
\end{equation}
In \cite{skenderis_quantum_2000} it was proven that if $ G $ is also conformally flat everywhere in the bulk, the Fefferham--Graham expansion truncates:
\begin{align}
	\gamma_{ij}(z,\sigma) &= g_{ij(0)}(\sigma) - S_{ij}(\sigma)\,z^{2} + \frac{1}{4}(S^{2})_{ij}(\sigma)\,z^{4}\\
	&= \left( \delta^{k}_{i}- \frac{1}{2}S_{i}^{k}(\sigma)\,z^{2} \right)\left( \delta^{l}_{j}- \frac{1}{2}S_{j}^{l}(\sigma)\,z^{2} \right)g_{kl(0)}(\sigma) \ ,
	\label{truncates}
\end{align}
where $ S_{ij} $ is the Schouten tensor of the boundary metric $ g_{(0)} $
\begin{equation}
	S_{ij} =  \frac{1}{d-2}\left(R_{ij} - \frac{1}{2(d-1)}\,R\,g_{ij(0)} \right)
	\label{g0S}
\end{equation}
whose indices are raised and lowered with $ g_{ij(0)} $. Here $ R_{ij} $ denotes the Ricci tensor and $ R $ the Ricci scalar of the boundary metric $ g_{(0)} $.

In fact, an asymptotically locally AdS$ _{d+1} $ metric, which is also Einstein and conformally flat, is actually locally AdS$ _{d+1} $ everywhere:\footnote{It is not necessarily globally AdS$ _{d+1} $, for example, it could be a quotient of AdS$ _{d+1} $.}
\begin{equation}
	\mathcal{R}^{ab}_{cd} = -\frac{1}{\ell^{2}}\,(\delta^{a}_{c}\delta^{b}_{d} - \delta^{a}_{d}\delta^{b}_{c}) \ ,
	\label{riem}
\end{equation}
where $ \mathcal{R}^{ab}_{cd} $ is the Riemann tensor of $ G $. For $ d+1 > 3 $, this follows from the expression
\begin{equation}
	0=W^{ab}_{cd} = \mathcal{R}^{ab}_{cd} - 4 \delta^{[a}_{[c}\mathcal{S}^{b]}_{d]} \ ,
\end{equation}
where $ W_{abcd} $ is the Weyl tensor of $ G $ and $ \mathcal{S}^{a}_{b} $ is the Schouten tensor of $ G $:
\begin{equation}
	\mathcal{S}^{a}_{b} = \frac{1}{d-1}\left(\mathcal{R}^{a}_{b} - \frac{1}{2d}\,\mathcal{R}\,\delta^{a}_{b} \right) = -\frac{1}{2\ell^{2}}\,\delta^{a}_{b} \ . %\frac{1}{2d(d+1)}\mathcal{R}\,\delta^{a}_{b}
\end{equation}
For $ d+1=3 $, any Einstein metric is automatically locally AdS$ _{3} $ (for the same reason why three-dimensional Einstein gravity is topological).

Thus a locally AdS$ _{d+1} $ manifold has a truncated FG expansion of the type \eqref{truncates}. In addition, \cite{skenderis_quantum_2000} showed that $ g_{(0)} $ cannot be anything, but must be locally conformally flat:
\begin{equation}
	\text{locally AdS$ _{d+1} $ bulk} \quad \Rightarrow \quad
	\begin{cases}
		C_{ijkl} &= 0, \quad d>3\\
		C_{ij} &= 0 , \quad d=3 \ ,
	\end{cases}
	\label{implication}
\end{equation}
where $ C_{ijkl} $ is the Weyl tensor of $ g_{(0)} $ and $ C_{ij} $ is the Cotton tensor of $ g_{(0)} $.\footnote{When $ d=2 $, the metric is automatically conformally flat.} As a result, only locally conformally flat metrics can appear on the conformal boundary of AdS$ _{d+1} $, or in other words, AdS$ _{d+1} $ admits foliations only by conformally flat slices. For example, assuming further that the boundary metric $ g_{(0)} $ is locally a direct product space, conformal flatness restricts the geometry of the factors \cite{brozos-vazquez_complete_2006}: a global direct product $ S^2\times S^2 $ is not locally conformally flat and cannot appear on the conformal boundary of AdS$ _{4} $, but $ S^2 \times \mathbb{H}^{2} $ is (see for instance \cite{nishioka_free_2021}).

\paragraph{Maximally symmetric foliations of EAdS$ _{d+1} $}

The boundary metric $ g_{(0)} $ of EAdS$ _{d+1} $ has to be locally conformally flat which includes maximally symmetric spaces as a special case:
\begin{equation}
	R^{ij}_{kl} = \pm\frac{1}{\alpha^{2}}\,(\delta^{i}_{k}\delta^{j}_{l}-\delta^{i}_{l}\delta^{j}_{k}), \quad R = \pm\frac{d(d-1)}{\alpha^{2}} \ ,
	\label{maxsym}
\end{equation}
where $ \alpha $ is the curvature radius of $ g_{(0)} $. To write the truncated FG expansion in this case, we obtain from \eqref{g0S}:
\begin{equation}
	S^{i}_{j} = \frac{1}{2d(d-1)}R\,\delta^{i}_{j} = \pm \frac{1}{2\alpha^{2}}\,\delta^{i}_{j}\ ,
	\label{confschouten}
\end{equation}
which when substituted to \eqref{truncates} gives, in the case of the sphere:
\begin{equation}
\ell^{2}ds^{2}_{\text{AdS}_{d+1}} = \frac{\ell^{2}}{z^{2}}\biggl[ dz^{2} +\biggl( 1 - \frac{z^{2}}{4\alpha^{2}}\biggr)^{2} \alpha^{2}ds^{2}_{S^{d}}\biggr]
\end{equation}
which is used in the main text.

\section{Variation of the D6-brane action in spherical slicing}\label{app:var}

In this Appendix, we show how regularity conditions at the tip of the brane ensure that the variational principle for the D6-brane action is well defined. The same regularity conditions also ensure that the derivative of the on-shell action with respect to the non-normalizable mode $ \xi_- $ is proportional to the normalizable mode $ \xi_+ $ as expected from the standard holographic dictionary of a scalar field.

\subsection{Regularity conditions for the 7-form}\label{app:reg}

The 7-form lives on $ \text{AdS}_{4}\times M_4 $ where $ M_4 $ is defined by the conditions \eqref{subspace}. Its boundary is topologically $ \partial \text{AdS}_4\times M_4 = S^{3}\times M_{4} $ where $ \partial \text{AdS}_4 $ is the cut-off surface $ u=\Lambda $. If a 7-form on $ \text{AdS}_{4}\times M_4 $ is regular, then it satisfies Stokes' theorem
\begin{equation}
\int_{\text{AdS}_4\times M_4}dC_7 = \int_{\partial \text{AdS}_4\times M_4}C_7 \ ,
\label{stokes}
\end{equation}
in other words, there are contributions from the physical boundary only (the cut-off surface).\footnote{One can see this as a definition of regularity. If there were singularities, one would have to introduce boundaries around them which would require extra boundary conditions that are not fixed by the dual theory.} In spherical slicing, the integral on the left hand side involving the 7-form \eqref{genC7} is proportional to
\begin{equation}
	\int_{c}^{\Lambda} du \int_{0}^{\pi} d\xi\, (\partial_u A - \partial_\xi B) = \int_{0}^{\pi} d\xi\, \left[ A(\Lambda,\xi)-A(c,\xi)\right]  - \int_{c}^{\Lambda} du\,\left[ B(u,\pi)-B(u,0)\right] \ .
\end{equation}
Imposing the conditions
\begin{equation}
A(c,\xi) = 0, \quad B(u,0) = B(u,\pi) = 0 \ ,
\label{appreg}
\end{equation}
we get
\begin{equation}
	\int_{c}^{\Lambda} du \int_{0}^{\pi} d\xi\, (\partial_u A - \partial_\xi B) =\int_{0}^{\pi} d\xi\, A(\Lambda,\xi)
\end{equation}
and Stokes' theorem \eqref{stokes} is satisfied. The conditions \eqref{appreg} are the regularity conditions for the 7-form. $B(u,\xi)$ has to vanish separately at $ \xi = 0$ and $ \xi = \pi $ since they correspond to different points on $M_4$.

\subsection{Variation of the action}\label{app:actvar}

In this section, we prove that the regularity conditions \eqref{appreg} for $ C_7 $ ensure a well defined variation principle for the D6-brane action. The D6-brane embeddings we consider are independent of $ M_3, S^{3} $ angles and reach to the conformal boundary where they wrap an $ \mathbb{R}\mathbb{P}^{3} $ at $ \xi = \frac{\pi}{2} $. Hence we can parametrize the embeddings by $ (u,\xi) = (u(s),\xi(s)) $ where $ s \in [0,1] $ runs over a finite interval and
\begin{equation}
u(1) = \infty, \quad \xi(1) = \frac{\pi}{2}
\label{dir}
\end{equation}
are Dirichlet boundary conditions at the conformal boundary. In the interior, the boundary conditions depend on the brane topology and are required by regularity of the brane world-volume metric to be (see section \ref{sec:twotypes})
\begin{equation}
u(0) = c, \quad \text{space-filling}; \quad \xi(0) = 0, \quad \text{terminating};
\label{branereg}
\end{equation}
with the other coordinate left free in both cases. In this parametrization, the pull-back of \eqref{genC7} to the brane is
\begin{equation}
\widehat{C}_7 = \frac{k\ell^{6}}{2}\left[ A(u(s),\xi(s))\,\xi'(s) + B(u(s),\xi(s))\,u'(s)\right] ds\wedge 8\vol{\mathbb{R}\mathbb{P}^{3}}\wedge \vol{S^{3}}
\end{equation}
and the WZ action becomes
\begin{equation}
I_{\text{WZ}} = \mathcal{N}\vol{S^{3}}\int_{0}^{1} ds\,\ell^{3}\left[ A(u(s),\xi(s))\,\xi'(s) + B(u(s),\xi(s))\,u'(s)\right] \ .
\label{wzab}
\end{equation}
This is an action for two scalar fields $ u(s) $ and $ \xi(s) $ and the variation is given by
\begin{equation}
\delta I_{\text{WZ}} = \mathcal{N}\vol{S^{3}} \int_0^{1} ds\,\ell^{3}\, (\partial_u A-\partial_\xi B )\left[ u'(s)\,\delta\xi - \xi'(s)\,\delta u\right] +\mathcal{N}\vol{S^{3}}\,\ell^{3} (A\,\delta\xi + B\,\delta u)\lvert_{s = 0}^{s=1} \ .
\label{wzvar}
\end{equation}
The boundary terms in the UV $ s = 1 $ vanish by the Dirichlet boundary conditions \eqref{dir}. For space-filling embeddings, $ u(s) $ enjoys Dirichlet boundary conditions in the interior so that
\begin{equation}
(A\,\delta\xi + B\,\delta u)\lvert_{s = 0} = A(c,\xi(c))\,\delta\xi\lvert_{s = 0} = 0 \ ,
\end{equation}
where the last equality follows from regularity $ A(c,\xi) = 0 $ of $ C_7 $ at the tip of Euclidean AdS$ _4 $. For terminating embeddings, it is $ \xi(s) $ that enjoys Dirichlet boundary conditions so that
\begin{equation}
(A\,\delta\xi + B\,\delta u)\lvert_{s = 0} = B(u,0)\,\delta u\lvert_{s = 0} = 0
\end{equation}
by regularity $ B(u,0) = 0 $ of $ C_7 $ at the tip of $ M_4 $. Since all of the boundary contributions vanish, the variational principle is well defined. Regularity of the brane embedding \eqref{branereg} and of $ C_7 $ \eqref{appreg} in the interior is what ensures this.

Note that DBI action does not require as much care, because all the contributions from the interior vanish automatically and boundary terms in the UV vanish by \eqref{dir}.

\section{Analytical solutions of the equations of motion}

In this Appendix, we solve the (spherical slicing) equations of motion of the D6-brane embedding analytically in various limits. First we consider the limits $ \alpha y_- \rightarrow 0, \infty $ and then we solve for the critical embedding near the point of the quantum phase transition.

\subsection{Asymptotics of solutions}\label{app:asympts}

We first solve the equation of motion \eqref{yeomtext} for $ y(x) $ perturbatively in the limit $ \alpha y_- \rightarrow \infty $ where $ y_- $ is the flavor mass. The analysis is formally similar to the finite temperature analysis in \cite{mateos_thermodynamics_2007,jokela_thermodynamics_2013} where the CFT lives on $ S^{1}_{\beta}\times\mathbb{R}^{2} $. In that case, the equation is expanded in the small temperature limit.

We start by writing the ansatz
\begin{equation}
y(x) = y^{(0)}(x) + \frac{1}{\alpha^{2}}\,y^{(1)}(x) +  \frac{1}{\alpha^{4}}\,y^{(2)}(x) + \mathcal{O}(\alpha^{-6}) \ ,
\end{equation}
where the corrections to $ y^{(0)} $ are taken to be small. At the end of the computation, this expansion translates to an expansion in inverse powers of the parameter $ \alpha y_- $ as expected by the scaling argument at the end of section \ref{subsec:D6dynamics}. In the infinite volume limit, the solution will be in the terminating phase so regularity requires
\begin{equation}
y^{(0)}(0) = y_0, \quad \dot{y}^{(0)}(0) = 0, \quad y^{(n)}(0) = \dot{y}^{(n)}(0) = 0, \quad n \geq 1 \ .
\label{pertin}
\end{equation}
We substitute this ansatz to the equation of motion \eqref{yeomtext} and expand in powers of $ 1\slash \alpha^{2} $. At leading order, we find  the solution obeying the initial conditions is
\begin{equation}
y^{(0)}(x) = y_0 = \text{constant} \ .
\label{y0sol}
\end{equation}
In the limit $ \alpha \rightarrow \infty $, this is just the supersymmetric Minkowski embedding \eqref{massiveflat} seen in flat slicing. This follows from the definition \eqref{cartesian} of the Cartesian coordinate as $ \xi = \arccos{\frac{y(x)}{e^{u}}} \sim \arccos{\frac{y_{0}}{r}} $ where we used the identification \eqref{rident} between the flat slicing and the spherical slicing radial coordinates in the $ \alpha \rightarrow \infty $ limit.

At second order, the general solution has two free parameters whose behavior near $ x = 0 $ is of the form $ y^{(1)}(x) = C_1\log{x} + C_2 + \mathcal{O}(x) $. Regularity imposes $ C_{1,2} = 0 $ which fixes the two free parameters and the resulting solution is
\begin{equation}
y^{(1)}(x) = \frac{3}{2 y_0}\log{\left( \frac{2\sqrt{x^{2}+y_0^{2}}}{y_0 + \sqrt{x^{2}+y_0^{2}}}\right)} \ .
\end{equation}
By substituting $y^{(0)}$, $y^{(1)}$ into the equation at next order we can solve for $ y^{(2)}(x) $ in terms of dilogarithms. There are again two free parameters that are fixed by requiring that the coefficients of the logarithmic divergence and the finite term vanish at $ x = 0 $. We shall not write down the explicit expression, but track its contribution to the source and vev $ y_{\pm} $. Therefore, we find that massive embeddings in the flat limit are given by
\begin{equation}
y(x) = y_0\left[ 1 + \frac{3}{2}\,(\alpha y_0)^{-2}\log{\left( \frac{2\sqrt{x^{2}+y_0^{2}}}{y_0 + \sqrt{x^{2}+y_0^{2}}}\right)} +\mathcal{O}(\alpha y_0)^{-4}\right] \ .
\label{minkowskiIR}
\end{equation}
We observe that the expansion is in the dimensionless variable $ \alpha y_0 $. From the asymptotic $ x \rightarrow \infty $ behavior of the solution, we obtain
\begin{align}
y_- &= y_0\left[  1 + \frac{3\log{2}}{2}\,(\alpha y_0)^{-2} + \mathcal{O}(\alpha y_0)^{-6}\right]\nonumber\\
y_+ &= \frac{1}{\alpha^{2}}\left[ \frac{3}{2} + \frac{18\log{2}-11}{8}\,(\alpha y_0)^{-2} + \mathcal{O}(\alpha y_0)^{-4}\right] \ .
\label{sourcevevIR}
\end{align}
From the first equation we obtain an expansion of $ \alpha y_0 $ in powers of $ \alpha y_- $ which when substituted to the second equation gives
\begin{equation}
\alpha^{2}y_+ =  \frac{3}{2}+\frac{18\log{2}-11}{8}\,(\alpha y_-)^{-2}+\mathcal{O}(\alpha y_-)^{-4}, \quad \alpha y_- \rightarrow \infty \ .
\label{vevIR}
\end{equation}
In the large curvature limit $ \alpha y_{-}\rightarrow 0 $, or equivalently in the massless limit, we expect the embedding $ y(x) $ to approach the massless embedding $ y(x) = 0 $. Thus, expanding the equation of motion \eqref{yeomtext} in small $ y(x) $ and $ \dot{y}(x) $ gives at leading order
\begin{equation}
x^3 \left(4 \alpha ^2 x^2-1\right)
\ddot{y}(x)+x^2 \left(8 \alpha ^2 x^2+4\right) \dot{y}(x)-6 x y(x) = 0 \ .
\end{equation}
The general solution is
\begin{equation}
y(x) = C_1\frac{x^{2}}{(x+R_0)^{2}} + C_2\frac{x^{3}}{(x^{2}-R_0^{2})^{2}} \ ,
\end{equation}
where $ R_0 = 1\slash (2\alpha) $. Regularity conditions of space-filling embeddings \eqref{regularityspace} imposes
\begin{equation}
y\Bigl(\sqrt{R_0^{2} - y_{0}^{2}}\,\Bigr) = y_{0}, \quad \dot{y}\Bigl(\sqrt{R_0^{2} - y_{0}^{2}}\,\Bigr) = \frac{y_{0}}{\sqrt{R_0^{2} - y_{0}^{2}}} \ ,
\label{regconds}
\end{equation}
where $ y_{0} $ parametrizes the embeddings and has to be small for our approximation to be valid. Expanding the first condition in small $ y_{0} $, we obtain the equation
\begin{equation}
\frac{C_{2}}{y_{0}}\biggl(\frac{R_{0}^{3}}{y_{0}^{3}}-\frac{3}{2}\frac{R_{0}}{y_{0}}\biggr)+\biggl(\frac{C_1}{4}+\frac{3}{8}\frac{C_{2}}{R_{0}}\biggr)+\mathcal{O}(y_{0}^{2}) = y_{0}, \quad y_{0}\rightarrow 0
\end{equation}
which sets $ C_2 = 0 $ and $ C_{1} = 4y_{0} $. Thus the solution near the massless solution $ y(x) = 0 $ is given by
\begin{equation}
y(x) = y_{0}\,\frac{4x^{2}}{(x+R_0)^{2}}
\label{nearmassless}
\end{equation}
which also satisfies the second derivative condition in \eqref{regconds} when $ y_{0}\rightarrow 0 $. Taking $ x\rightarrow \infty $, we obtain
\begin{equation}
\alpha y_- = \frac{4y_{0}}{\alpha}, \quad \alpha^{2}y_+ = \frac{4y_{0}}{\alpha} 
\end{equation}
so that
\begin{equation}
\alpha^{2}y_+ =  \alpha y_- + \ldots, \quad \alpha y_- \rightarrow 0 \ ,
\label{sourcevevUV}
\end{equation}
where the ellipsis denote higher-order corrections in $ \alpha y_- $.

\subsection{The critical solution}\label{par:crit}

To find the critical solution, we write $ u-c \rightarrow \epsilon\,(u-c) $ and $ \xi(u)\rightarrow \epsilon\,\xi(u) $ and expand the D6-brane action in $ \epsilon \rightarrow 0 $ so that we are in the overlap region of the two types of embeddings. The WZ action expands as
\begin{equation}
I_{\text{WZ}}^{\text{reg}} = -\epsilon^{5}\,\mathcal{N}\vol{S^{3}} \int du\,\xi(u)^{2} + \mathcal{O}(\epsilon)^{7}
\end{equation}
and the DBI action as
\begin{equation}
I^{\text{reg}}_{\text{DBI}}  =  \epsilon^{4}\,\mathcal{N}\vol{S^{3}} \int du\,\frac{(u-c)^{3}}{8}\,\xi(u)\sqrt{1+\xi'(u)^{2}} + \mathcal{O}(\epsilon)^{6} \ .
\label{expdbi}
\end{equation}
The DBI action dominates over the WZ action in this limit.\footnote{This is ultimately a result of the regularity conditions imposed on $ C_{7} $.} The leading order equation of motion obtained from \eqref{expdbi} is given by
\begin{equation}
(u-c)\xi\xi''+[3\xi \xi'-(u-c) ] (1+\xi'^{2}) = 0, \quad u-c \ll 1 \ .
\label{selfsimeq}
\end{equation}
It has the solution
\begin{equation}
\xi_{*}(u) = \frac{u-c}{\sqrt{3}}, \quad u-c \ll 1
\end{equation}
which is the critical solution in the deep interior plotted in figure \ref{minkowskicart}. Small perturbations of the critical solution take the form
\begin{equation}
\xi(u) = \xi_{*}(u) + \delta \xi(u) \ ,
\end{equation}
where the perturbation satisfies the equation
\begin{equation}
(u-c)^{2}\,\delta \xi''-4\,(u-c)\,\delta \xi'+4\,\delta \xi=0 \ .
\end{equation}
The solution is given by
\begin{equation}
\delta \xi(u) = \delta b_{+}\,(u-c)^{\beta_+} + \delta  b_-\,(u-c)^{\beta_-}\ , \quad u-c \ll 1 \ ,
\end{equation}
where the exponents are given by
\begin{equation}
\beta_{\pm} = -\frac{3}{2} \pm i\frac{\sqrt{7}}{2} \ .
\label{exponentsapp}
\end{equation}

\section{Renormalized action and vev in spherical slicing}\label{app:renormalization}

In this Appendix, we holographically renormalize the D6-brane on-shell action in spherical slicing. Then we show that the variation of the renormalized action with respect to the non-normalizable mode $ y_- $ of the embedding gives the normalizable mode $ y_+ $ as expected from the relation between a source and a vev.

\subsection{Holographic renormalization}\label{app:holren}

To renormalize the D6-brane action \eqref{fullactionspher}, we use the following integration by parts trick that allows to separate the divergences from the $ u $-integral and encode them in a boundary term. We start by writing the action as
\begin{align}
I^{\text{reg}}_{\text{D6}} = \mathcal{N}\vol{S^{3}}\int_{u_0}^{\Lambda}du\,\ell^{3}\sinh^{3}{(u-c)}\,\sin{\xi(u)}\,&\left( \sqrt{1 + \xi'(u)^{2}} -\sin{\xi(u)} + \frac{3}{2}\sin{\xi(u)} \right)\nonumber\\
&\qquad\qquad\qquad\qquad + \mathcal{N}\vol{S^{3}}\,U\lvert_{u=\Lambda} \ .
\label{inint}
\end{align}
Then integrating by parts the last term in round brackets, we can write the regularized action as
\begin{equation}
I_{\text{D6}}^{\text{reg}} = (I_{\text{fin}}-I_{\text{B}}\lvert_{u=u_0}) + (I_{\text{B}}+ U)\lvert_{u=\Lambda} \ ,
\label{D6regspher}
\end{equation}
where the first term is
\begin{align}
I_{\text{fin}} = \mathcal{N}\vol{S^{3}}\int_{u_0}^{\Lambda}du\,\ell^{3}&\sinh^{3}{(u-c)}\,\sin{\xi(u)}\,\left( \sqrt{1 + \xi'(u)^{2}} -\sin{\xi(u)}\right.\nonumber\\
&\;\;\left.-\coth{(u-c)}\,\left[ 1 - \frac{2}{\sinh^{2}{(u-c)}}\right] \,\cos{\xi(u)}\,\xi'(u) \right)
\label{Ifin}
\end{align}
and the remaining terms are given in terms of
\begin{equation}
I_{\text{B}} = \frac{\mathcal{N}}{2}\vol{S^{3}}\,\ell^{3}\cosh{(u-c)}\,[\sinh^{2}{(u-c)} - 2]\,\sin^{2}{\xi(u)} \ .
\label{Idiv}
\end{equation}
The contribution from $ u=u_0 $ depends on the type of embedding in question and is explicitly
\begin{equation}
I_{\text{B}}\lvert_{u=u_0} = -\mathcal{N}\ell^{3}\vol{S^{3}}
\begin{cases}
0, \quad &\text{terminating}\\
\sin^{2}{\xi(c)}, \quad &\text{space-filling}\ .
\end{cases}
\label{intcont}
\end{equation}
The contribution from $ u=\Lambda $, on the other hand, is UV divergent on-shell:
\begin{equation}
I_{\text{B}}\lvert_{u=\Lambda} = \frac{\mathcal{N}}{2}\,(\alpha \ell^{3})\vol{S^{3}}\,\biggl[ e^{3\Lambda} - \biggl(\frac{9}{4\alpha^{2}} + \xi_-^{2} \biggr)  e^{\Lambda} -2\,\xi_- \xi_+  + \mathcal{O}(e^{-\Lambda})\biggr] \ .
\end{equation}
Because $ U $ is in general arbitrary, we find the on-shell divergence structure
\begin{equation}
(I_{\text{B}}+U)\lvert_{u=\Lambda} = (\alpha \ell)^{3}\vol{S^{3}}\,\biggl[a_1e^{3\Lambda} +  \bigg( \frac{a_2}{\alpha^{2}}+ a_3\,\xi_-^{2}\bigg) e^{\Lambda} + 2a_3\,\xi_- \xi_+ + \mathcal{O}(e^{-\Lambda})\biggr], \quad \Lambda \rightarrow \infty
\label{div}
\end{equation}
with three free coefficients $ a_{1,2,3} $ corresponding to the freedom in $ U $ (there is more freedom in finite counterterms below). On-shell, this includes all of the divergences of the regularized action \eqref{D6regspher}, because the integral \eqref{Ifin} is convergent on-shell when $ \Lambda \rightarrow \infty $ and $ I_{\text{B}}\lvert_{u=u_0} $ is finite.

There are three covariant counterterms that are divergent:
\begin{equation}
d_1\int_{u=\Lambda} \sqrt{\gamma_3}, \quad d_2\int_{u=\Lambda} \sqrt{\gamma_3}\,\sin^{2}{\xi(u)}, \quad  d_3\int_{u=\Lambda} \sqrt{\gamma_3}\,\ell^{2}R_{3} \ ,
\label{invariants}
\end{equation}
where $ \gamma_{3} $ is the metric of the $ u=\Lambda $ slice of AdS$ _4 $ of radius $ \ell $ and $ R_{3} $ is its Ricci scalar. Their sum produces the same on-shell divergences as in \eqref{div} with the identification
\begin{equation}
a_1 = d_1+d_2, \quad a_2 = -\frac{3}{4}(d_1+d_2-8d_3), \quad a_3 = -d_2  \ .
\end{equation}
By appropriately tuning the coefficients $ d_{1,2,3} $, the divergences \eqref{div} of the regularized on-shell action can be canceled. In addition, the finite term $ 2a_3\xi_-\xi_+ $ in \eqref{div} is always canceled by the $ d_2 $ counterterm. On top of the divergent counterterms \eqref{invariants}, there are two covariant finite counterterms
\begin{equation}
	d_4\int_{u=\Lambda} \sqrt{\gamma_{3}}\,\sin{\xi(u)}\cos^{3}{\xi(u)}, \quad  d_5\int_{u=\Lambda} \sqrt{\gamma_{3}}\,\ell^{2}R_{3}\,\sin{\xi(u)}\cos{\xi(u)} \ ,
\end{equation}
whose sum goes as (using $ y_- = -\xi_- $)
\begin{equation}
\ell^{3}\vol{S^{3}}\,\biggl[d_4\,(\alpha y_-)^{3} + 6\,d_5\,(\alpha y_-) + \mathcal{O}(e^{-\Lambda})\biggr], \quad \Lambda \rightarrow \infty \ ,
\end{equation}
with no divergent contributions. These counterterms parametrize the renormalization scheme and in general they can be generated by $ U\lvert_{u=\Lambda} $ so we keep the coefficients $ (d_4,d_4) $ free.

After adding all covariant counterterms to the regularized action \eqref{D6regspher} such that divergences are canceled, the renormalized on-shell D6-brane action becomes
\begin{equation}
I_{\text{D6,on-shell}}^{\text{ren}} = (I_{\text{fin}}-\,I_{\text{B}}\lvert_{u=u_0})\lvert_{\text{on-shell}}\,+B_{3}\,(\alpha y_{-})^{3}+B_{1}\,(\alpha y_{-}) \ ,
\label{D6onshellren}
\end{equation}
where we have defined
\begin{equation}
B_3 \equiv \ell^{3}\vol{S^{3}}\,d_4, \quad B_1 \equiv 6\,\ell^{3}\vol{S^{3}}\,d_5
\end{equation}
and the first two terms are given in \eqref{Ifin} and \eqref{intcont}. Thus, the renormalized on-shell action in spherical slicing is unique up to two free parameters $ (B_3,B_1) $. In Cartesian coordinates for the embedding $ y = y(x) $, the renormalized action is explicitly (setting $ B_3 = B_1 = 0 $ for simplicity)
\begin{equation}
I_{\text{fin}} = \mathcal{N}\ell^{3}\vol{S^{3}}\int_{x_0}^{\Lambda}dx\,\mathcal{L}(x,y,\dot{y}), \quad \mathcal{L}(x,y,\dot{y}) = \alpha^{3}\sum_{k=0}^{3}\alpha^{-2k}\mathcal{L}_{k}(x,y,\dot{y}) \ ,
\label{conv}
\end{equation}
where
\begin{align}
\mathcal{L}_{0}(x,y,\dot{y}) &= x\sqrt{x^{2} + y^{2}}\left(\sqrt{1+\dot{y}^{2}} - 1 \right)\nonumber\\
\mathcal{L}_1(x,y,\dot{y}) &= -\frac{3x}{4\sqrt{x^{2} + y^{2}}}\left[\sqrt{1+\dot{y}^{2}} - \frac{2y(y-x\dot{y})}{x^{2} + y^{2}} -1\right]\nonumber\\
\mathcal{L}_2(x,y,\dot{y}) &= \frac{3x}{16(x^{2} + y^{2})^{3\slash 2}}\left[\sqrt{1+\dot{y}^{2}} + \frac{4y(y-x\dot{y})}{x^{2} + y^{2}} -1\right]\nonumber\\
\mathcal{L}_3(x,y,\dot{y}) &= -\frac{x}{64(x^{2} + y^{2})^{5\slash 2}}\left[\sqrt{1+\dot{y}^{2}} + \frac{2y(y-x\dot{y})}{x^{2} + y^{2}} -1\right] \ .
\end{align}
and
\begin{equation}
I_{\text{B}}\lvert_{u = u_0} =
\begin{cases}
0, \quad &\text{terminating}\\
\mathcal{N}\ell^{3}\vol{S^{3}}\left(1 - 4\alpha^{2}y_0^{2} \right), \quad &\text{space-filling} \ .
\end{cases}
\label{add}
\end{equation}
The parameter $ y_{0} $ is defined in \eqref{regularityspace}.

\subsection{Holographic dual of the vev}\label{sec:vevproof}

In the holographic minimal subtraction scheme with $ B_{3} = B_{1} = 0 $, the renormalized on-shell D6-brane action \eqref{D6onshellren} is given by
\begin{equation}
	I_{\text{D6,on-shell}}^{\text{ren}} = (I_{\text{fin}} -I_{\text{B}}\lvert_{u=u_0})\lvert_{\text{on-shell}} \ ,
	\label{appD6ren}
\end{equation}
where we write \eqref{Ifin} as
\begin{equation}
	I_{\text{fin}} = \int_{u_0}^{\Lambda}du\, \mathcal{L}_{\text{fin}}(u,\xi(u),\xi'(u)) \ .
\end{equation}
The on-shell variation of the renormalized action picks up contributions only from the UV boundary $ u = \Lambda $ which is ensured by the $ C_7 $ regularity conditions as proven in Appendix~\ref{app:actvar}. However, let us see how it works explicitly. The on-shell variation of the first term in \eqref{appD6ren} with respect to $ \xi_- $ gives
\begin{equation}
	\frac{\partial I_{\text{fin}}}{\partial \xi_-}\bigg\lvert_{\text{on-shell}} = \frac{\partial \mathcal{L}_{\text{fin}}}{\partial \xi'(u)}\frac{\partial \xi(u)}{\partial \xi_-}\bigg\lvert_{u=\Lambda} - \frac{\partial \mathcal{L}_{\text{fin}}}{\partial \xi'(u)}\frac{\partial \xi(u)}{\partial \xi_-}\bigg\lvert_{u=u_0} \ .
	\label{finvarren}
\end{equation}
where
\begin{align}
\frac{\partial \mathcal{L}_{\text{fin}}}{\partial \xi'(u)} =\mathcal{N}\ell^{3}\vol{S^{3}}\,\sinh^{3}{(u-c)}\,\sin{\xi(u)}\,&\biggl(-\coth{(u-c)}\,\left[ 1 - \frac{2}{\sinh^{2}{(u-c)}}\right] \,\cos{\xi(u)}\biggr.\nonumber\\
&\qquad\qquad\qquad\qquad\qquad\biggl.+\frac{\xi'(u)}{\sqrt{1 + \xi'(u)^{2}}}\biggr) \ .
\label{varLfin}
\end{align}
The $ u=u_0 $ contribution to \eqref{finvarren} is determined by the interior boundary conditions and is different for the two types of embeddings. For terminating embeddings with $ \xi(u_0) = 0 $, it vanishes by
\begin{equation}
	\frac{\partial \mathcal{L}_{\text{fin}}}{\partial \xi'(u)}\bigg\lvert_{u=u_0} = 0 \ .
\end{equation}
For terminating embeddings $ I_B\lvert_{u=u_0} = 0 $ so that
\begin{equation}
	\frac{\partial I_{\text{D6,on-shell}}^{\text{ren}}}{\partial  \xi_-} = \frac{\partial \mathcal{L}_{\text{fin}}}{\partial \xi'(u)}\frac{\partial \xi(u)}{\partial \xi_-}\bigg\lvert_{u=\Lambda}
	\label{varLambda}
\end{equation}
and all contributions come from $ u=\Lambda $. For space-filling embeddings with $ u_0 = c $ and $ \xi'(c) = 0 $, we instead find
\begin{equation}
	\frac{\partial \mathcal{L}_{\text{fin}}}{\partial \xi'(u)}\bigg\lvert_{u=c} = \mathcal{N}\ell^{3}\vol{S^{3}}\, 2\sin{\xi(c)}\cos{\xi(c)}
\end{equation}
which is exactly canceled by the variation of
\begin{equation}
	-I_{\text{B}}\lvert_{u=c}\, = \mathcal{N}\ell^{3}\vol{S^{3}}\,\sin^{2}{\xi(c)} \ .
\end{equation}
As a result, \eqref{varLambda} holds also for space-filling embeddings as expected.

Now the boundary $ u=\Lambda $ term in \eqref{varLambda} is determined by the on-shell asymptotics \eqref{spherxiasymp} of $ \xi(u) $. Using the expression \eqref{varLfin} and taking the $ \Lambda \rightarrow \infty $ limit of \eqref{varLambda} gives
\begin{equation}
	\frac{\partial I_{\text{D6,on-shell}}^{\text{ren}}}{\partial  \xi_-} = -\mathcal{N}\ell^{3}\vol{S^{3}}\,\alpha^{3}\xi_+
\end{equation}
or equivalently in terms of $ y_{\pm} = \pm\xi_{\pm} $:
\begin{equation}
	\frac{\partial I_{\text{D6,on-shell}}^{\text{ren}}}{\partial  (\alpha y_-)} = \mathcal{N}\ell^{3}\vol{S^{3}}\,\alpha^{2}y_+ \ .
	\label{derexp}
\end{equation}
In field theory variables, the coefficient is explicitly
\begin{equation}
	\mathcal{N}\ell^{3}\vol{S^{3}} = \frac{\pi}{4}N\sqrt{2\lambda} \ .
\end{equation}
Let us relate equation \eqref{derexp} to the integrated 1-point function of the dual operator. Because the gravity action is independent of $ \alpha y_- $, we have
\begin{equation}
	\frac{\partial I_{\text{D6,on-shell}}^{\text{ren}}}{\partial (\alpha y_{-})} =\frac{\partial I^{\text{ren}}_{\text{bulk}}}{\partial (\alpha y_{-})} \ .
	\label{D6bulk}
\end{equation}
In addition, the holographic duality says that
\begin{equation}
	\biggl\langle \exp{\biggl(y_-\int_{S^{3}_{\alpha}}d^{3}\sigma \sqrt{\zeta}\, O^{\text{ren}}(\sigma)\biggr)}\biggr\rangle =e^{-I^{\text{ren}}_{\text{bulk}}} \ ,
\end{equation}
where the expectation value on the left is computed in ABJM theory with massless flavor. Combining with \eqref{derexp}, we obtain
\begin{equation}
	\frac{\pi}{4}N\sqrt{2\lambda}\,\alpha^{2}y_+ = \frac{1}{\alpha}\int_{S^{3}_{\alpha}} d^{3}\sigma\sqrt{\zeta(\sigma)}\,\langle O^{\text{ren}}(\sigma)\rangle_{y_-} \ ,
	\label{integratedvev}
\end{equation}
where the expectation value is computed on $ S^{3}_{\alpha} $ in the presence of the source $ y_- $. This equation shows that $ y_{-} = -\xi_- $ is the source (flavor mass) and $ y_{+} = \xi_+ $ is the (integrated) vev of the fermion bilinear.

By symmetry, the 1-point function on the sphere takes the form
\begin{equation}
	\langle O^{\text{ren}}(x)\rangle_{y_-} = \frac{a_{O}}{\alpha^{2}} \ ,
\end{equation}
where $ a_{O} = a_{O}(\alpha y_-) $ is a dimensionless coefficient. Integrating over $ S^{3}_{\alpha} $ in \eqref{integratedvev} gives the relation
\begin{equation}
	a_{O} = \frac{1}{8\pi}N\sqrt{2\lambda}\, \alpha^{2}y_+ \ .
	\label{aO}
\end{equation}

\section{Invertibility of the vev}\label{app:invvev}

\begin{figure}
	\centering
	\begin{tikzpicture}
	\node (img1)  {\includegraphics{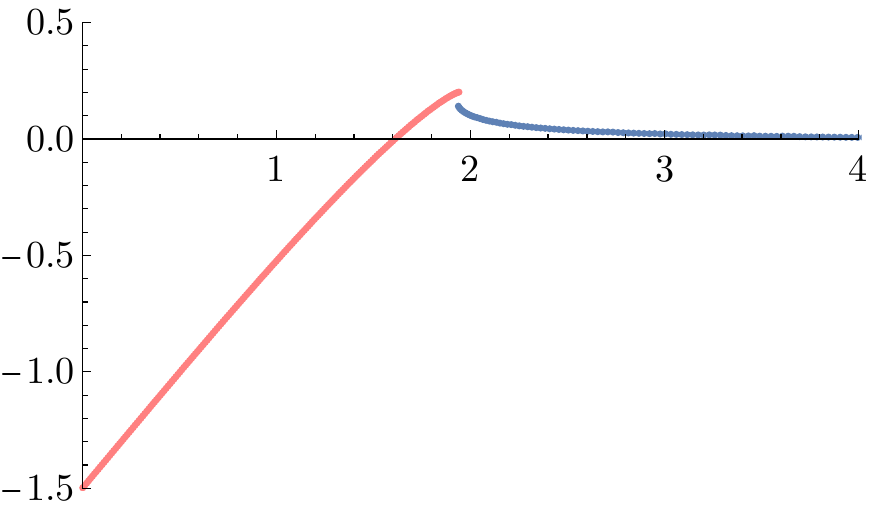}};
	\node[right=of img1, node distance=0cm, yshift=1.2cm, xshift=-1cm] {$ \alpha y_- $};
	\node[above=of img1, node distance=0cm, anchor=center,yshift=-1cm, xshift=-3.7cm] {$ p $};
	\end{tikzpicture}
	\caption{The vev $ p $ as defined by \eqref{papp} in the IR finite scheme $ (b_3,b_1) = (0,0) $.}
	\label{irfinitevev}
\end{figure}

The vev $ p $ is defined as
\begin{equation}
\frac{\partial F^{\text{ren}}_{\text{D6}}}{\partial (\alpha y_-)} = p
\label{papp}
\end{equation}
where $F^{\text{ren}}_{\text{D6}}$ is given by \eqref{generalscheme} and the result is plotted in the IR finite scheme in figure \ref{irfinitevev}. For the purposes of the Legendre transform, we want to find renormalization schemes $ (b_3,b_1) $ where $ p $ is a single-valued function of $ \alpha y_- $. For this, it is useful to introduce the notation
\begin{equation}
F^{\text{ren}}_{\text{D6}} =
\begin{cases}
b_3\,(\alpha y_-)^{3} + b_1\,(\alpha y_-) + b_{-1}\,(\alpha y_-)^{-1} + \ldots, \quad &\alpha y_-\rightarrow \infty\\
1+ c_1\,(\alpha y_-) +c_2\,(\alpha y_-)^{2} + b_3\,(\alpha y_-)^{3} + \ldots, \quad &\alpha y_-\rightarrow 0 \ ,
\end{cases}
\label{genexpF}
\end{equation}
where
\begin{equation}
b_{-1} = -\frac{18\log{2}-11}{8}, \quad c_1 =b_1- \frac{3}{2}, \quad c_2 = \frac{1}{2} \ .
\label{abjmparameters}
\end{equation}
We  require $ p $ to be invertible as a function of $ \alpha y_- $ and we study this in the asymptotic $ \alpha y_-\rightarrow 0,\infty $ limits which gives \textit{necessary} conditions for invertibility of $ p $. This is enough to impose constraints on the renormalization scheme parameter $ b_{3} $. The value of $ b_1 $ does not affect invertibility since it only shifts the value of $ p $ by a constant.

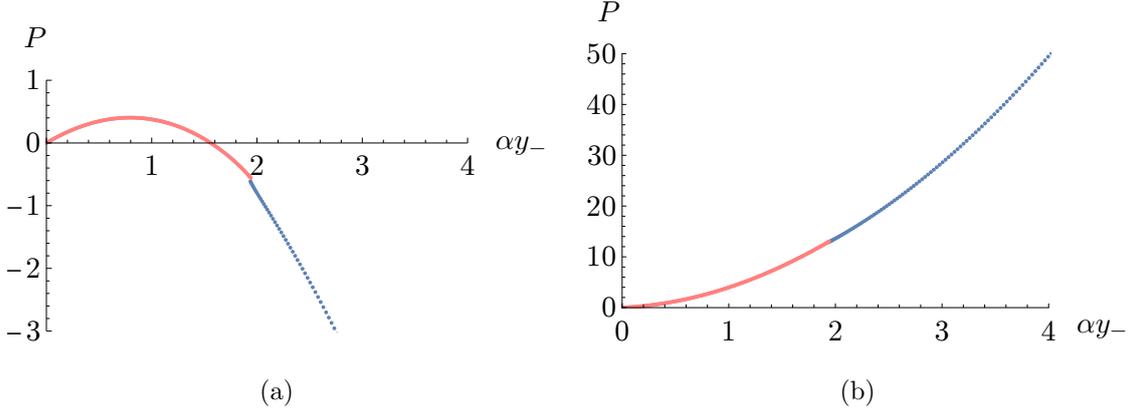
\begin{figure}[t]
	\begin{subfigure}[t]{0.5\textwidth}
		\centering
		\begin{tikzpicture}
			\node (img1)  {\includegraphics{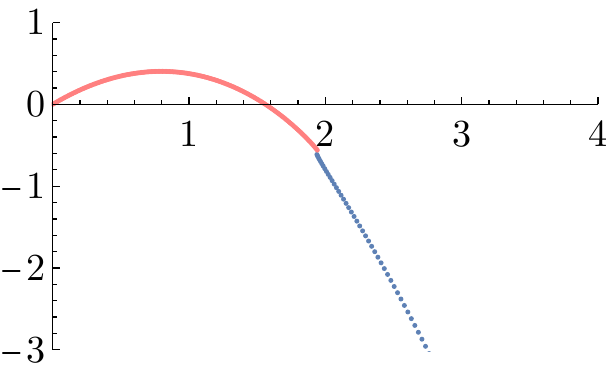}};
			\node[right=of img1, node distance=0cm, yshift=0.8cm, xshift=-1cm] {$ \alpha y_- $};
			\node[above=of img1, node distance=0cm, anchor=center,yshift=-0.8cm, xshift=-2.7cm] {$ P $};
		\end{tikzpicture}
		\subcaption{}
		\label{multivalued}
	\end{subfigure}
	\begin{subfigure}[t]{0.5\textwidth}
		\centering
		\begin{tikzpicture}
			\node (img1)  {\includegraphics{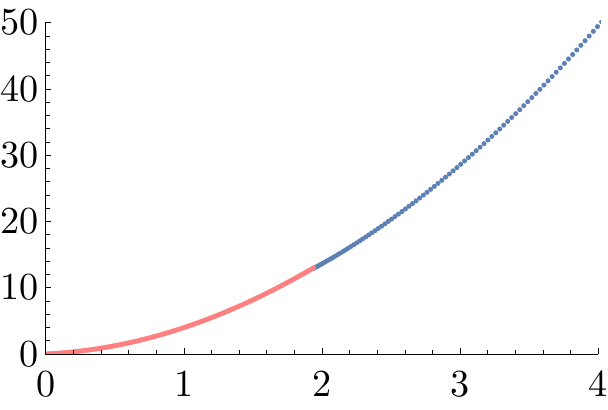}};
			\node[right=of img1, node distance=0cm, yshift=-1.8cm, xshift=-1cm] {$ \alpha y_- $};
			\node[above=of img1, node distance=0cm, anchor=center,yshift=-0.8cm, xshift=-2.8cm] {$ P $};
		\end{tikzpicture}
		\subcaption{}
		\label{singlevalued}
	\end{subfigure}
	\caption{The shifted vev $P = p-c_1$ in two renormalization schemes $ (a) $ a scheme with $ b_3 <0 $ and $ b_1 = \frac{3}{2} $ which leads to the boundary condition $ p(\infty) = -\infty $ $ (b) $ a scheme with $ b_3 > 0 $ and $ b_1 = \frac{3}{2} $ so that $ p(\infty) = \infty $. In the scheme $ (a) $, the vev is multi-valued, while in the scheme $ (b) $, it is single-valued (modulo local effects near the phase transition).}
\end{figure}

At small mass, we obtain
\begin{equation}
p = c_1 + 2\,c_2\,(\alpha y_-) + \ldots, \quad \alpha y_- \rightarrow 0
\end{equation}
so that we have $ p(0) = c_1 $. Taking the derivative gives
\begin{equation}
p'(\alpha y_-) =  2\,c_2 + \ldots, \quad \alpha y_- \rightarrow 0
\end{equation}
which is always positive at $ \alpha y_- = 0 $ since $ c_2  = 1\slash 2 $. Hence there is no scheme in which $ p $ is single-valued if we impose the boundary condition $ p(\infty) < p(0) $, because then the curve turns down (see figure \ref{multivalued}). Thus we must impose $ p(\infty)> p(0) $ and invertibility requires
\begin{equation}
	p'(\alpha y_-) \geq 0, \quad \alpha y_- \in [0,\infty) \ .
	\label{pder}
\end{equation}
At large mass, we obtain instead,
\begin{equation}
	p = 3\,b_{3}\,(\alpha y_-)^{2} + b_{1}- b_{-1}\,(\alpha y_-)^{-2} +\ldots, \quad \alpha y_- \rightarrow \infty \ .
\end{equation}
If $ b_3 < 0 $, we have  $ p(\infty) = -\infty $ which was ruled out above. If $ b_3 = 0 $, we obtain  $ p(\infty) = b_1 > c_1 $, but since $ b_{-1}<0 $ it is approached from above meaning $ p(\alpha y_-) $ has turned around and is not single-valued. Hence invertibility requires
\begin{equation}
	b_3 > 0
	\label{singlecondition}
\end{equation}
so that $ p(\infty) = \infty $. See figure \ref{singlevalued} for a plot of $ P = p - c_1 $ in such a scheme.

In the scheme \eqref{singlecondition}, we can invert $ P $ order by order to obtain
\begin{equation}
	\alpha y_- =
		\frac{1}{\sqrt{3\,b_3}}\,P^{1\slash 2}-\frac{3}{\sqrt{48\,b_3}}\,P^{-1\slash 2} + (-3+16\,b_{3}\,b_{-1})\,P^{-3\slash 2} +\mathcal{O}(P^{-5\slash 2}), \quad P\rightarrow \infty
\end{equation}
and
\begin{equation}
		\alpha y_- = \frac{1}{2c_2}\,P -\frac{3\,b_3}{8\,c_2^{3}}\, P^{2} + \mathcal{O}(P^{3}), \quad P\rightarrow 0 \ ,
\end{equation}
from which the asymptotic expansions of the quantum effective potential \eqref{gammaIR} and \eqref{gammaUV} follow.

\addcontentsline{toc}{section}{References}
\bibliography{abjmv5.bib}
\bibliographystyle{JHEP}

\end{document}